\documentclass[%
 aip,
 amsmath,amssymb,
 reprint,%
]{revtex4-1}

\usepackage{graphicx}% Include figure files
\usepackage{dcolumn}% Align table columns on decimal point
\usepackage{bm}% bold math
\usepackage{enumitem}

\usepackage[utf8]{inputenc}
\usepackage[T1]{fontenc}
\usepackage{mathptmx}
\usepackage{etoolbox}
\usepackage[colorlinks=true,linkcolor=black]{hyperref}%
\usepackage{amsmath}
\usepackage{amssymb}
\usepackage{physics}
\usepackage{subcaption}
\newcommand{\onebraket}[1]{\langle #1 \rangle}
\usepackage{tikz}
\usepackage[normalem]{ulem}
\usetikzlibrary{positioning, shapes.geometric, arrows.meta}
\pdfoutput=1

\makeatletter
\def\@email#1#2{%
 \endgroup
 \patchcmd{\titleblock@produce}
  {\frontmatter@RRAPformat}
  {\frontmatter@RRAPformat{\produce@RRAP{*#1\href{mailto:#2}{#2}}}\frontmatter@RRAPformat}
  {}{}
}%
\makeatother

\newcommand{\MC}[1]{\textcolor{black}{{#1}}}
\newcommand{\MCC}[1]{\textcolor{black}{{#1}}}

\begin{document}

\preprint{AIP/123-QED}

\title{ABACUS: An Electronic Structure Analysis Package for the AI Era}

% Main affiliation
\affiliation{AI for Science Institute, Beijing 100080, P. R. China}
\affiliation{\mbox{HEDPS, CAPT, \MCC{School of Mechanics and Engineering Science and School of Physics}, Peking University, Beijing, 100871, P. R. China}}
\affiliation{\mbox{Institute of Artificial Intelligence, Hefei Comprehensive National Science Center, Hefei 230026, P. R. China}}
\affiliation{\mbox{Key Laboratory of Quantum Information, University of Science and Technology of China, Hefei 230026, P. R. China}}
\affiliation{Institute of Physics, Chinese Academy of Sciences, Beijing 100190, P. R. China}

% Main contributors

\author{Weiqing Zhou}
\affiliation{AI for Science Institute, Beijing 100080, P. R. China}
\affiliation{Key Laboratory of Artificial Micro- and Nano-structures of Ministry of Education and School of Physics and Technology, Wuhan University, Wuhan 430072, P. R. China}

\author{Daye Zheng}
\affiliation{AI for Science Institute, Beijing 100080, P. R. China}

\author{Qianrui Liu}
\affiliation{\mbox{HEDPS, CAPT, \MCC{School of Mechanics and Engineering Science and School of Physics}, Peking University, Beijing, 100871, P. R. China}}

\author{Denghui Lu}
\affiliation{\mbox{HEDPS, CAPT, \MCC{School of Mechanics and Engineering Science and School of Physics}, Peking University, Beijing, 100871, P. R. China}}
\affiliation{AI for Science Institute, Beijing 100080, P. R. China}

\author{Yu Liu}
\affiliation{\mbox{HEDPS, CAPT, \MCC{School of Mechanics and Engineering Science and School of Physics}, Peking University, Beijing, 100871, P. R. China}}

\author{Peize Lin}
\affiliation{\mbox{Institute of Artificial Intelligence, Hefei Comprehensive National Science Center, Hefei 230026, P. R. China}}

% AISI

\author{Yike Huang}
\affiliation{AI for Science Institute, Beijing 100080, P. R. China}

\author{Xingliang Peng}
\affiliation{AI for Science Institute, Beijing 100080, P. R. China}

\author{Jie J. Bao}
\affiliation{AI for Science Institute, Beijing 100080, P. R. China}

\author{Chun Cai}
\affiliation{AI for Science Institute, Beijing 100080, P. R. China}

\author{Zuxin Jin}
\affiliation{AI for Science Institute, Beijing 100080, P. R. China}

\author{Jing Wu}
\affiliation{AI for Science Institute, Beijing 100080, P. R. China}

% USTC

\author{Haochong Zhang}
\affiliation{\mbox{Institute of Artificial Intelligence, Hefei Comprehensive National Science Center, Hefei 230026, P. R. China}}

\author{Gan Jin}
\affiliation{\mbox{Institute of Artificial Intelligence, Hefei Comprehensive National Science Center, Hefei 230026, P. R. China}}

\author{Yuyang Ji}
\affiliation{\mbox{Key Laboratory of Quantum Information, University of Science and Technology of China, Hefei 230026, P. R. China}}

\author{Zhenxiong Shen}
\affiliation{\mbox{Institute of Artificial Intelligence, Hefei Comprehensive National Science Center, Hefei 230026, P. R. China}}
\affiliation{\mbox{Key Laboratory of Quantum Information, University of Science and Technology of China, Hefei 230026, P. R. China}}

\author{Xiaohui Liu}
\affiliation{\mbox{Supercomputing Center, University of Science and Technology of China, Hefei, Anhui 230026, P. R. China}}

% PKU

\author{Liang Sun}
\affiliation{\mbox{HEDPS, CAPT, \MCC{School of Mechanics and Engineering Science and School of Physics}, Peking University, Beijing, 100871, P. R. China}}

\author{Yu Cao}
\affiliation{Institute of Physics, Chinese Academy of Sciences, Beijing 100190, P. R. China}
\affiliation{\mbox{HEDPS, CAPT, \MCC{School of Mechanics and Engineering Science and School of Physics}, Peking University, Beijing, 100871, P. R. China}}

\author{Menglin Sun}
\affiliation{School of Materials Science and Engineering, Peking University, Beijing 100871, P. R. China}

\author{Jianchuan Liu}
\affiliation{\mbox{School of Electrical Engineering and Electronic Information, Xihua University, Chengdu 610039, P. R. China}}

\author{Tao Chen}
\affiliation{\mbox{HEDPS, CAPT, \MCC{School of Mechanics and Engineering Science and School of Physics}, Peking University, Beijing, 100871, P. R. China}}

\author{Renxi Liu}
\affiliation{\mbox{Academy for Advanced Interdisciplinary Studies, Peking University, Beijing, 100871, P. R. China}}
\affiliation{\mbox{HEDPS, CAPT, \MCC{School of Mechanics and Engineering Science and School of Physics}, Peking University, Beijing, 100871, P. R. China}}
\affiliation{AI for Science Institute, Beijing 100080, P. R. China}

\author{Yuanbo Li}
\affiliation{\mbox{HEDPS, CAPT, \MCC{School of Mechanics and Engineering Science and School of Physics}, Peking University, Beijing, 100871, P. R. China}}

\author{Haozhi Han}
\affiliation{\mbox{HEDPS, CAPT, \MCC{School of Mechanics and Engineering Science and School of Physics}, Peking University, Beijing, 100871, P. R. China}}
\affiliation{AI for Science Institute, Beijing 100080, P. R. China}

\author{Xinyuan Liang}
\affiliation{\mbox{Academy for Advanced Interdisciplinary Studies, Peking University, Beijing, 100871, P. R. China}}
\affiliation{\mbox{HEDPS, CAPT, \MCC{School of Mechanics and Engineering Science and School of Physics}, Peking University, Beijing, 100871, P. R. China}}

\author{Taoni Bao}
\affiliation{\mbox{HEDPS, CAPT, \MCC{School of Mechanics and Engineering Science and School of Physics}, Peking University, Beijing, 100871, P. R. China}}

\author{\MCC{Zichao Deng}}
\affiliation{\mbox{HEDPS, CAPT, \MCC{School of Mechanics and Engineering Science and School of Physics}, Peking University, Beijing, 100871, P. R. China}}

\author{\MCC{Tao Liu}}
\affiliation{\mbox{HEDPS, CAPT, \MCC{School of Mechanics and Engineering Science and School of Physics}, Peking University, Beijing, 100871, P. R. China}}

\author{Nuo Chen}
\affiliation{\mbox{HEDPS, CAPT, \MCC{School of Mechanics and Engineering Science and School of Physics}, Peking University, Beijing, 100871, P. R. China}}

\author{Hongxu Ren}
\affiliation{\mbox{HEDPS, CAPT, \MCC{School of Mechanics and Engineering Science and School of Physics}, Peking University, Beijing, 100871, P. R. China}}

\author{Xiaoyang Zhang}
\affiliation{\mbox{HEDPS, CAPT, \MCC{School of Mechanics and Engineering Science and School of Physics}, Peking University, Beijing, 100871, P. R. China}}
\affiliation{AI for Science Institute, Beijing 100080, P. R. China}

\author{Zhaoqing Liu}
\affiliation{\mbox{College of Chemistry and Molecular Engineering, Peking University, Beijing 100871, P. R. China}}

% Xi an jiaotong

\author{Yiwei Fu}
\affiliation{International Research Center for Renewable Energy, State Key Laboratory of Multiphase Flow, Xi’an Jiaotong University, Xi’an, Shaanxi 710049, P. R. China}

\author{Maochang Liu}
\affiliation{International Research Center for Renewable Energy, State Key Laboratory of Multiphase Flow, Xi’an Jiaotong University, Xi’an, Shaanxi 710049, P. R. China}
\affiliation{Suzhou Academy of Xi’an Jiaotong University, Suzhou, Jiangsu 215123, P. R. China}

% DPA: The University of Hong Kong

\author{Zhuoyuan Li}
\affiliation{\mbox{Center for Structural Materials, Department of Mechanical Engineering, The University of Hong Kong, Hong Kong, P. R. China}}
\affiliation{\mbox{Materials Innovation Institute for Life Sciences and Energy (MILES), The University of Hong Kong, Shenzhen, P. R. China}}

\author{Tongqi Wen}
\affiliation{\mbox{Center for Structural Materials, Department of Mechanical Engineering, The University of Hong Kong, Hong Kong, P. R. China}}
\affiliation{\mbox{Materials Innovation Institute for Life Sciences and Energy (MILES), The University of Hong Kong, Shenzhen, P. R. China}}

% Section VII. DeepH

\author{Zechen Tang}
\affiliation{State Key Laboratory of Low Dimensional Quantum Physics and Department of Physics, Tsinghua University, Beijing, 100084, P. R. China}

\author{Yong Xu}
\affiliation{State Key Laboratory of Low Dimensional Quantum Physics and Department of Physics, Tsinghua University, Beijing, 100084, P. R. China}
\affiliation{Frontier Science Center for Quantum Information, Beijing, P. R. China}
\affiliation{RIKEN Center for Emergent Matter Science (CEMS), Wako, Saitama 351-0198, Japan}

\author{Wenhui Duan}
\affiliation{State Key Laboratory of Low Dimensional Quantum Physics and Department of Physics, Tsinghua University, Beijing, 100084, P. R. China}
\affiliation{Institute for Advanced Study, Tsinghua University, Beijing 100084, P. R. China}
\affiliation{Frontier Science Center for Quantum Information, Beijing, P. R. China}

% Section VII. DeePTB

\author{Xiaoyang Wang}
\affiliation{Laboratory of Computational Physics, Institute of Applied Physics and Computational Mathematics, Fenghao
 East Road 2, Beijing 100094, P. R. China}

\author{Qiangqiang Gu}
\affiliation{\mbox{School of Artificial Intelligence and Data Science, University of Science and Technology of China, Hefei 230026, P. R. China}}
\affiliation{AI for Science Institute, Beijing 100080, P. R. China}

\author{Fu-Zhi Dai}
\affiliation{\mbox{School of Materials Science and Engineering, University of Science and Technology Beijing, Beijing 100083, P. R. China}}
\affiliation{AI for Science Institute, Beijing 100080, P. R. China}

\author{Qijing Zheng}
\affiliation{\mbox{Department of Physics, University of Science and Technology of China, Hefei, Anhui 230026, P. R. China}}

% Section HamGNN
\author{\MCC{Yang Zhong}}
\affiliation{Key Laboratory of Computational Physical Sciences (Ministry of Education), Institute of Computational Physical Sciences, State Key Laboratory of Surface Physics, and Department of Physics, Fudan University, Shanghai, 200433, P. R. China}

\author{\MCC{Hongjun Xiang}}
\affiliation{Key Laboratory of Computational Physical Sciences (Ministry of Education), Institute of Computational Physical Sciences, State Key Laboratory of Surface Physics, and Department of Physics, Fudan University, Shanghai, 200433, P. R. China}

\author{\MCC{Xingao Gong}}
\affiliation{Key Laboratory of Computational Physical Sciences (Ministry of Education), Institute of Computational Physical Sciences, State Key Laboratory of Surface Physics, and Department of Physics, Fudan University, Shanghai, 200433, P. R. China}

\author{Jin Zhao}
\affiliation{\mbox{Department of Physics, University of Science and Technology of China, Hefei, Anhui 230026, P. R. China}}

\author{Yuzhi Zhang}
\affiliation{DP Technology, Beijing 100080, P. R. China}
\affiliation{AI for Science Institute, Beijing 100080, P. R. China}

\author{Qi Ou}
\affiliation{\mbox{Basic Research Department, SINOPEC Research Institute of Petroleum Processing Co., Ltd, Beijing 100083, P. R. China}}

\author{Hong Jiang}
\affiliation{\mbox{College of Chemistry and Molecular Engineering, Peking University, Beijing 100871, P. R. China}}

\author{Shi Liu}
\affiliation{\mbox{Department of Physics, School of Science, Westlake University, Hangzhou, Zhejiang 310030, P. R. China}}
\affiliation{\mbox{Institute of Natural Sciences, Westlake Institute for Advanced Study, Hangzhou, Zhejiang 310024, P. R. China}}

\author{Ben Xu}
\affiliation{Graduate School of China Academy of Engineering
 Physics, Beijing 100193, P. R. China}

\author{Shenzhen Xu}
\affiliation{School of Materials Science and Engineering, Peking University, Beijing 100871, P. R. China}
\affiliation{AI for Science Institute, Beijing 100080, P. R. China}

\author{Xinguo Ren}
\email{renxg@iphy.ac.cn}
\affiliation{Institute of Physics, Chinese Academy of Sciences, Beijing 100190, P. R. China}

\author{Lixin He}
\email{helx@ustc.edu.cn}
\affiliation{\mbox{Key Laboratory of Quantum Information, University of Science and Technology of China, Hefei 230026, P. R. China}}
\affiliation{\mbox{Institute of Artificial Intelligence, Hefei Comprehensive National Science Center, Hefei 230026, P. R. China}}
\affiliation{Synergetic Innovation Center of Quantum Information and Quantum Physics, University of Science and Technology of China, Hefei, 230026, P. R. China}

\author{Linfeng Zhang}
\affiliation{DP Technology, Beijing 100080, P. R. China}
\affiliation{AI for Science Institute, Beijing 100080, P. R. China}

\author{Mohan Chen}
\email{mohanchen@pku.edu.cn}
\affiliation{AI for Science Institute, Beijing 100080, P. R. China}
\affiliation{\mbox{HEDPS, CAPT, \MCC{School of Mechanics and Engineering Science and School of Physics}, Peking University, Beijing, 100871, P. R. China}}

\date{\today}

\newcommand{\bR}{\mathbf{R}}
\newcommand{\bk}{\mathbf{k}}
\newcommand{\bz}{\mathbf{0}}
\newcommand{\br}{\mathbf{r}}
\newcommand{\rrp}{\br-\br'}
\newcommand{\Uu}{U\mu}
\newcommand{\Vv}{V\nu}
\newcommand{\Kk}{K\kappa}
\newcommand{\Ll}{L\lambda}
\newcommand{\Ua}{U\alpha}
\newcommand{\Va}{V\alpha}
\newcommand{\Aa}{A\alpha}
\newcommand{\Bb}{B\beta}
\newcommand{\pRU}{\phi^{\bR_U}_{\Uu}}
\newcommand{\pRV}{\phi^{\bR_V}_{\Vv}}
\newcommand{\pRK}{\phi^{\bR_K}_{\Kk}}
\newcommand{\pRL}{\phi^{\bR_L}_{\Ll}}
\newcommand{\PRA}{P^{\bR_A}_{\Aa}}
\newcommand{\PRB}{P^{\bR_B}_{\Bb}}
\newcommand{\tx}[1]{\text{#1}}
\newcommand{\F}[1]{\Vert #1 \Vert}

% abstract
\begin{abstract}
ABACUS (Atomic-orbital Based Ab-initio Computation at USTC) is an open-source software for first-principles electronic structure calculations and molecular dynamics simulations. It mainly features density functional theory (DFT) and molecular dynamics functions and is compatible with both plane-wave basis sets and numerical atomic orbital basis sets. ABACUS serves as a platform that facilitates the integration of various electronic structure methods, such as Kohn-Sham DFT, stochastic DFT, orbital-free DFT, and real-time time-dependent DFT, etc. In addition, with the aid of high-performance computing, ABACUS is designed to perform efficiently and provide massive amounts of first-principles data for generating general-purpose machine learning potentials, such as DPA models. Furthermore, ABACUS serves as an electronic structure platform that interfaces with several AI-assisted algorithms and packages, such as DeePKS-kit, DeePMD, DP-GEN, DeepH, DeePTB, HamGNN, etc.
\end{abstract}

\maketitle

%\vspace*{\fill}
%\clearpage

\tableofcontents

\begin{figure*}
    \centering \includegraphics[width=0.8\linewidth]{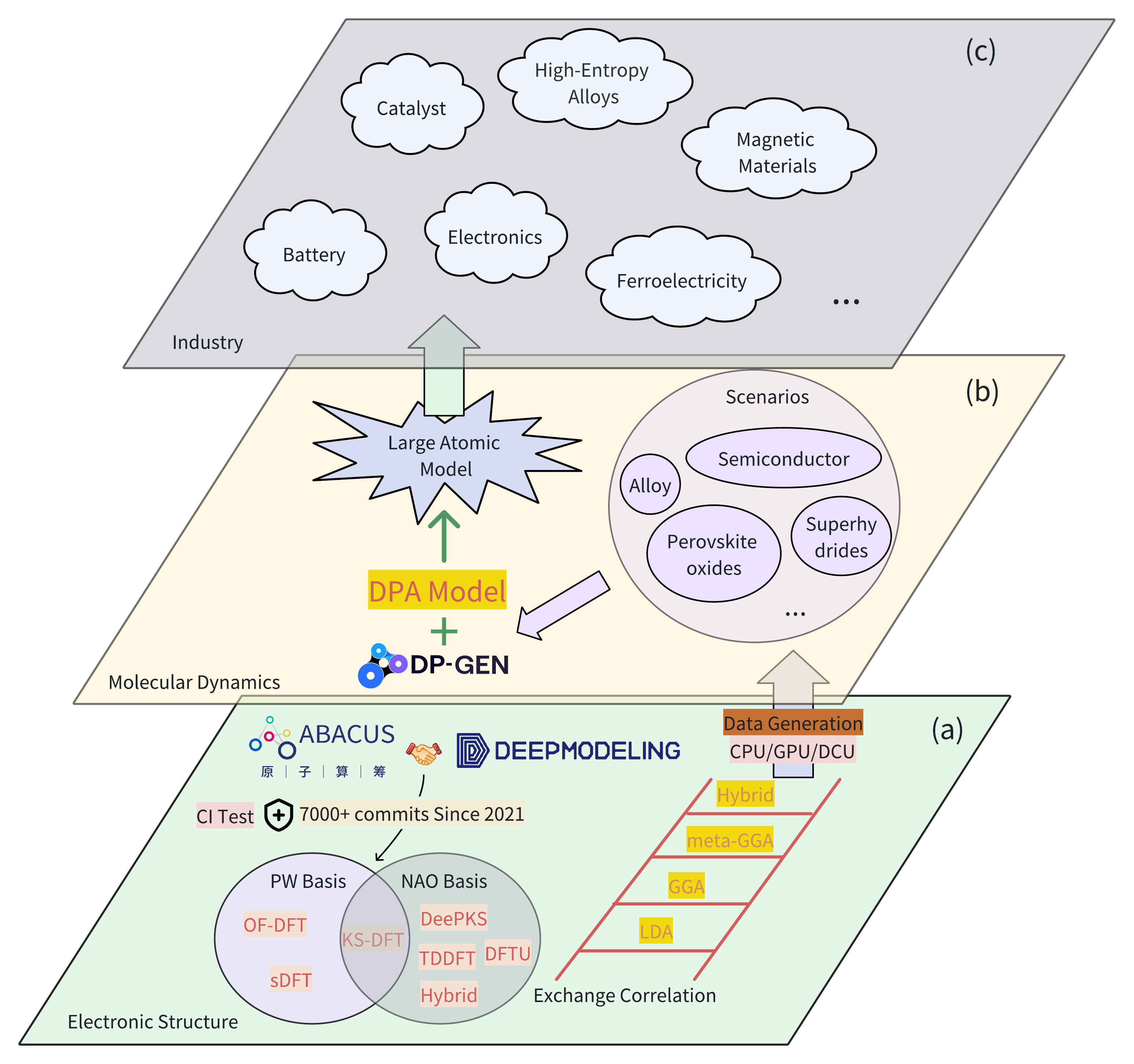}
    \caption{ABACUS is dedicated to building an algorithm platform and data engine for AI4ES (AI for Electronic Structure). Since partnering with the open-source community DeepModeling in 2021, ABACUS has garnered over 7,300 commits under the protection of Continuous Integration (CI) tests. Its adaptable architecture allows developers to incorporate numerous electronic structure algorithms. The AI-assisted functional correction method DeePKS permits ABACUS to achieve precise functional results at a reduced cost. Owing to high-performance implementation at various devices, ABACUS effectively produces extensive first-principles data across multiple sectors. Combining advanced electronic structure algorithms and AI-assisted pre-trained models, one can transfer the precision of quantum mechanics across scales.}
    \label{fig:cover}
\end{figure*}

% introduction
\section{Introduction} \label{introduction}

The theoretical foundations of density-based methodologies can be traced to the Thomas-Fermi (TF) model proposed in 1927. The TF model established the first kinetic energy functional based on the uniform electron gas approximation.~\cite{27-Thomas-local, 27TANL-Fermi-local} Although this orbital-free (OF) framework pioneered the conceptual basis for density functional theory (DFT), its oversimplified treatment of kinetic energy limited chemical accuracy.~\cite{02Carter} In 1964, Hohenberg and Kohn~\cite{1964-hk} established two theorems. First, the external potential of a many-electron system is uniquely determined by its electron density. Second, a universal energy functional exists whose minimum value corresponds to the ground-state electron density. Later in 1965, Kohn and Sham~\cite{1965-ks} developed the Kohn-Sham density functional theory (KS-DFT), whose accuracy is determined by the exchange-correlation functional. This framework allows a non-interacting electron system where the kinetic energy can be computed exactly via electronic wavefunctions. Crucially, the non-interacting system can yield the electron density of the real system via the so-called self-consistent field (SCF) method.

Over the past few decades, driven by developments of the DFT method itself and continuous refinement of numerical algorithms, various popular KS-DFT packages emerge.~\cite{2020-wien2k,2020-BDF,2020-abinit,2020-CONQUEST,2020-cp2k,2020-octopus,2020-onetep,2020-pyscf,2020-qe,2020-siesta,1999-vasp,g16} In addition, with the exponential growth in computing capabilities, the KS-DFT method has become a cornerstone to predict various properties of materials in physics, chemistry, materials science, biology, geophysics, etc. Despite these achievements, challenges still remain. For example, the community increasingly demands DFT implementations that combine enhanced computational efficiency with improved hardware adaptability across different architectures. Recently, with rapid advancements in artificial intelligence (AI) technologies, an important question arises: can these new technologies drive substantial improvements in computational accuracy and efficiency for electronic structure methods such as KS-DFT and OF-DFT?

In this review paper, we introduce the ABACUS package, which was initiated around 2006 and has achieved a series of progress over the past two decades.~\cite{Chen2010,Chen_2011,Li2016,lin_accuracy_2020,2021-zhengdy,lin_efficient_2021,Lin2021,qu2022dftu,2022-quxin,2022-sdft-liu,2022-Dai,2023-metagga-liu,lin_Initio_2024} For instance, the ABACUS package supports both plane-wave (PW) basis and numerical atomic orbitals (NAO) basis for first-principles electronic structure calculations, geometry relaxation and molecular dynamics simulations. In addition, ABACUS offers robust support for highly efficient parallelization of diverse electronic structure methods, leveraging MPI, OpenMP, and CUDA to optimize performance across distributed and accelerated computing architectures. Based on recent advancements in AI technologies for atomistic simulations and electronic structure methods,~\cite{18PRL-Zhang,18CPC-Wang,21JCTC-Chen,22CPC-Chen,2021-dm21,2020-Jia,2019-deepwf,2020-deephf,22MF-Wen} researchers have identified significant potential for AI to transform both electronic structure methods and their corresponding software implementations. In this context, the effort to create an integrated platform that integrates physics-based models, high-performance computing, and AI capabilities has become particularly critical. Since 2021, the ABACUS development team has maintained an ongoing collaboration with the DeepModeling open-source community.~\cite{abacus-develop}

As illustrated in Fig.~\ref{fig:cover}, ABACUS aims to establish an electronic structure algorithmic platform and computational engine tailored for the AI for Science (AI4S) paradigm. This integration features three aspects. First, state-of-the-art electronic structure method developments with AI-assisted algorithms. Second, molecular dynamics simulations to provide data for the OpenLAM project,~\cite{openlam} which is designed to generate a general-purpose machine learning model for the periodic table. Third, the platform can be potentially used in various industry fields such as battery, catalyst, electronics, ferroelectricity, magnetic materials, high-entropy alloys, etc.

In this work, we provide a detailed overview of the latest developments in ABACUS. Sec.~\ref{sec:ksdft} details implementation of general features in ABACUS, such as the pseudopotentials, exchange-correlation functionals, SCF method, geometry relaxation and molecular dynamics, etc. Sec.~\ref{sec:pw} describes the implementation of the KS-DFT, stochastic DFT (sDFT), and OF-DFT methods with the plane wave basis. Moreover, recent advances for AI-assisted kinetic energy density functionals (KEDFs) are introduced for OFDFT. Sec.~\ref{sec:lcao} introduces the linear combination of atomic orbitals (LCAO) calculation methods, where numerical atomic orbitals are served as basis to solve the KS equation and yield forces and stress, as well as newly developed methods such as hybrid functional, AI-assisted electronic structure method DeePKS, DFT+U \MCC{and} real-time time-dependent DFT (RT-TDDFT), all of which are based on the numerical atomic orbital basis. Sec.~\ref{sec:openlam} mainly focuses on machine-learning-based interatomic potentials and describes the role played by ABACUS. Sec.~\ref{sec:interface} presents various software interfaces between ABACUS and other packages. Finally, we conclude in sec.~\ref{sec:summary}.

\begin{figure*}
    \centering
\includegraphics[width=0.8\linewidth]{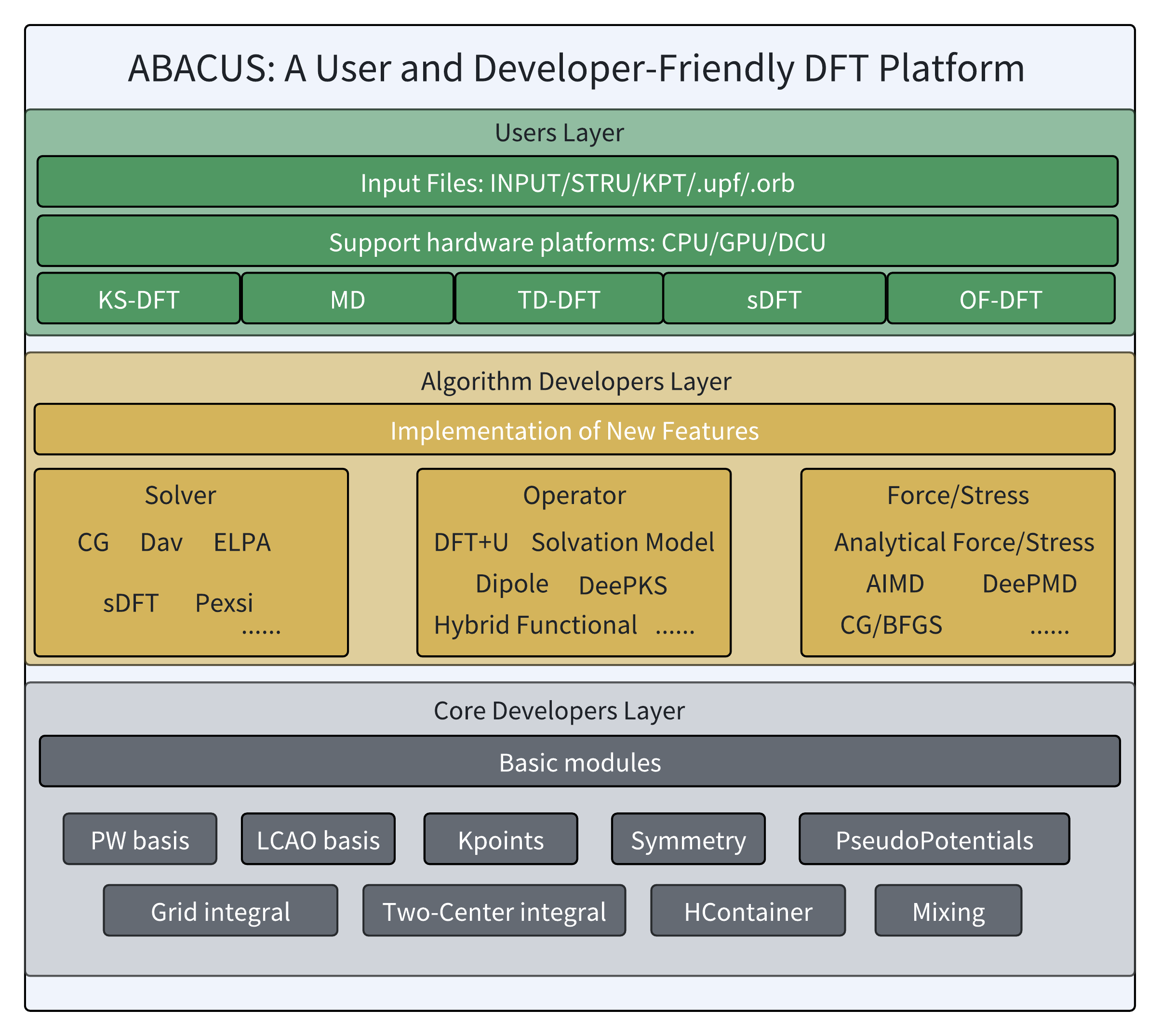}
    \caption{\MC{Code architecture of ABACUS. For users, a few input files need to be prepared in advance, including \textbf{INPUT}, \textbf{STRU}, \textbf{KPT}, pseudopotential files, and orbital files (only for numerical atomic orbitals calculations).
    For developers, the software features a highly modular design that allows for the swift integration of new functionalities. The development team is responsible for formulating and refining key data structures, mathematical routines, and other foundational components in ABACUS. For instance, the {\tt HContainer} module is utilized for storing sparse matrices under the numerical atomic orbital basis set, such as Hamiltonian and density matrices; the {\tt Grid Integral} module is implemented for grid integration.}}
    \label{fig:framework}
\end{figure*}

\section{Kohn-Sham Density Functional Theory} \label{sec:ksdft}

\subsection{Kohn-Sham Equation}

In KS-DFT, the electronic orbitals are obtained by solving the KS equation
\begin{equation}
\left[-\frac{1}{2}\nabla^2+\hat{V}_{\rm KS}\right]\psi_i=\varepsilon_i\psi_i,
\label{eq:ks-equation}
\end{equation}
where $\psi_i$ is the Kohn-Sham wave function with energy $\varepsilon_i$, and $\hat{V}_{\rm KS}$ is an effective potential written as
\begin{equation}
\hat{V}_{\rm KS}=\hat{V}_{\rm ext}+\hat{V}_{\rm H}+\hat{V}_{\rm xc}.
\end{equation}
Here the three terms on the right-hand side are the external potential term, the Hartree term, and the exchange-correlation (XC) term, respectively. Typically, the external potential term includes the ion-electron interactions that can be described via pseudopotentials, as well as external fields like the electric field.

The KS equation can be solved via different types of basis sets, such as plane waves,~\cite{1999-vasp,2020-qe,2020-abinit} numerical atomic orbitals,~\cite{2020-siesta,2011-openmx,2020-CONQUEST,Blum2009} real-space grids,~\cite{1994-Chelikowsky,2020-octopus} Gaussian orbitals,~\cite{g16,2020-cp2k} wavelet basis sets,~\cite{2020-bigDFT} etc. Currently, ABACUS supports both plane waves and NAOs as basis sets. On the one hand, the plane wave basis can accurately and efficiently handle periodic systems with a relatively small number of atoms. On the other hand, the NAOs are centered on atoms and the required number of basis is substantially less than the plane wave basis. Therefore, the NAO basis is suitable to simulate systems containing thousands of atoms.~\cite{2020-cp2k,2020-siesta}

\subsubsection{Pseudopotentials}

The Coulomb potential exhibits a singularity at the atomic nucleus, necessitating an extensive basis set for the precise representation of ``near-core'' wavefunctions, which imposes substantial computational overhead. To overcome this challenge, several pseudopotential methodologies have been developed in the past few decades.~\cite{kleinman1980relativistic,bachelet1982relativistic,1990-rappe-optimized,1990-vanderbilt-soft,1991-martins-efficient,kresse1994norm,1999-vasp,2013-hamann-optimized} These pseudopotentials achieve significant reduction in computational costs while preserving computational accuracy.

ABACUS supports both norm-conserving pseudopotentials (NCPPs)~\cite{1979-Hamann-NC} and ultrasoft pseudopotentials (USPPs).~\cite{1990-vanderbilt-soft} Specifically, the pseudopotential operator $\hat{V}_{\rm ps}$ is decomposed into local and nonlocal components in NCPPs as
\begin{equation}
   \hat{V}_{\rm ps}=\hat{V}_{\rm local}+\hat{V}_{\rm KB},
\end{equation}
where the local pseudopotential $\hat{V}_{\rm local}$ equals the all-electron functions beyond a cutoff radius. Furthermore, $\hat{V}_{\mathrm{KB}}$ is composed of a set of separable nonlocal projectors as proposed by Kleinman and Bylander,~\cite{1982-KB} lated extended by Hamann.~\cite{2013-hamann-optimized} The formula is
\begin{equation}
\begin{aligned}
 \hat{V}_{\mathrm{KB}} = \sum_{ll'mm'} D_{ll^{\prime}} | \beta_{lm} \rangle \langle \beta_{l'm'} |.
\end{aligned}
\end{equation}
Here $\beta_{lm}$ is a projector with angular quantum number $l$ and magnetic quantum number $m$, and its weighting factor is $D_{ll^{\prime}}$. In addition, these projectors $\beta_{lm}$ are expressed as the product of a radial function and a spherical harmonic function. In cases such as transition metals or magnetic systems, the non-linear core corrections~\cite{1982-Louie-Nonlinear} are usually adopted in the pseudopotential file to improve the transferability of pseudopotentials. For the spin-orbital coupling effect, it can be incorporated into the non-local projectors in the $\hat{V}_{\mathrm{KB}}$ term.~\cite{theurich2001self,2005-Corso-spin,2021-Cuadrado-validity}

Compared to NCPPs, the Ultrasoft pseudopotentials~\cite{1990-vanderbilt-soft} offer substantially improved computational efficiency by removing the norm-conservation constraint, which allows greater flexibility in constructing the pseudo-wavefunction within the core region, yielding inherently softer wavefunctions that reduce computational overhead. As a result, USPPs facilitate faster convergence of basis set expansions, though this advantage comes with the trade-off of requiring more sophisticated implementation within KS-DFT packages.

% exchange-correlation functional
\subsubsection{Exchange-Correlation Functionals} \label{sec:xc}

In KS-DFT codes, to describe collinear spin of given systems, the spin-up electron density $\rho^\uparrow(\mathbf{r})$ and spin-down electron density $\rho^\downarrow(\mathbf{r})$ are usually treated separately, and then summed up to the total electron density
\begin{equation}
\rho(\mathbf{r}) = \rho^\uparrow(\mathbf{r}) + \rho^\downarrow(\mathbf{r})
\end{equation}
while the spin density takes the form of 
\begin{equation}
m(\mathbf{r}) = \rho^\uparrow(\mathbf{r}) - \rho^\downarrow(\mathbf{r}).
\end{equation}

In this regard, the exchange-correlation energy $E_{\rm xc}[\rho^\uparrow, \rho^\downarrow]$ exhibits an explicit functional dependence on $\rho^\uparrow$ and $\rho^\downarrow$. The corresponding exchange-correlation potential for spin channel $\sigma$ (where $\sigma \in \{\uparrow, \downarrow\}$) is derived via the functional derivative
\begin{equation}
V_{\text{xc},\sigma}(\mathbf{r}) = \frac{\delta E_{\text{xc}}[\rho^\uparrow, \rho^\downarrow]}{\delta \rho_\sigma(\mathbf{r})}.
\end{equation}

Since the invention of KS-DFT, numerous density functional approximations have been developed. The earliest implementation, the local density approximation (LDA),~\cite{1965-ks} is expressed as a functional of the spin-summed electron density $\rho(\mathbf{r})$. Subsequent refinements introduced the local spin density approximation (LSDA),~\cite{1972-von-local,1980-vosko-accurate} which incorporates spin density components $\rho^\uparrow$ and $\rho^\downarrow$ within a broken-symmetry Slater determinant framework, thereby correctly describing the dissociation limit of H$_2$.

Including density gradients $\nabla\rho(\mathbf{r})$ (or spin density gradients $\nabla\rho^\uparrow$ and $\nabla\rho^\downarrow$) as additional variables led to the development of generalized gradient approximation (GGA) functionals, some of which substantially improved the KS-DFT accuracy compared to LDA/LSDA. For instance, the PBE functional~\cite{1996-pbe} reduces the mean absolute error (MAE) for molecular atomization energies of 20 main-group molecules from 31.4 kcal/mol (LSDA) to 7.9 kcal/mol.

Further advancement was achieved with meta-GGA functionals, which incorporate the kinetic energy density $\tau(\mathbf{r})$. This class of functionals, exemplified by the strongly constrained and appropriately normed (SCAN) functional,~\cite{2015-scan} offers enhanced accuracy over GGAs. Variants of SCAN, such as rSCAN~\cite{2019-rscan} and R2SCAN,~\cite{2020-r2scan} were subsequently developed to improve numerical stability and convergence while largely retaining its accuracy.

One may also include a fraction of exact Hartree-Fock exchange energy into the exchange-correlation term, and the resulting functional is called hybrid functionals, such as the HSE06,~\cite{HSE_heyd_hybrid_2003,2006-erratum} PBE0~\cite{adamo1999toward} and B3LYP functionals.~\cite{stephens1994ab}

Analogous to spin-unpolarized systems,
approximations for $E_{\rm xc}[\rho^\uparrow, \rho^\downarrow]$ are typically constructed hierarchically. The simplest approach, the LSDA,~\cite{1989-lsda} expresses $E_{\rm xc}$ solely as a function of the local spin densities $\rho^\sigma(\mathbf{r})$. More refined treatments, such as the generalized gradient approximation (GGA), introduce additional dependence on the gradients of the spin densities $\nabla \rho^\sigma(\mathbf{r})$. Further sophistication is achieved in meta-GGA functionals through the incorporation of the kinetic energy density, while hybrid functionals explicitly include fractional exact exchange integrals.

The LDA, GGA, and meta-GGA functionals have already been implemented in ABACUS for both plane-wave and numerical atomic orbital basis sets. Both LDA and GGA functionals are accessible through either native implementations or the Libxc library,~\cite{LibXC} whereas the meta-GGA functional is currently restricted to Libxc. Benchmark studies of SCAN, rSCAN,~\cite{2019-rscan} and r2SCAN~\cite{2020-r2scan} within ABACUS are detailed in Ref.~\onlinecite{2023-metagga-liu}. Efficient hybrid functional calculations are available for localized basis sets, utilizing combined capabilities of Libxc and the local resolution of identity (LRI) methodology (see Sec.~\ref{sec:hybrid} for computational details).

% spin-porlarized DFT
%\subsubsection{Spin-Polarized Calculations} \label{spin}
\subsubsection{Non-Collinear Spin}

The non-collinear spin calculation allows electron spins to orient in arbitrary directions rather than being restricted to a single axis. Within the framework of non-collinear spin KS-DFT, the Kohn-Sham electronic wavefunction $i$ can be represented as two-component spinors as
\begin{equation}
\Psi_i(\mathbf{r}) = \begin{pmatrix} \psi_{i}^{\uparrow}(\mathbf{r}) \\ 
\psi_{i}^{ \downarrow}(\mathbf{r}) \end{pmatrix}.
\end{equation}

The spin-resolved density matrix $\rho^{\sigma\sigma'}$ (with $\sigma, \sigma' \in \{\uparrow, \downarrow\}$) contains the full spin structure of the system. It can be further decomposed into charge and spin contributions via the Pauli matrices $\boldsymbol{\sigma} = (\sigma_x, \sigma_y, \sigma_z)$ and the identity matrix $\boldsymbol{I}$ as follows
\begin{equation}
\begin{aligned}
\left(\begin{array}{cc}
    \rho^{\uparrow\uparrow} & \rho^{\uparrow\downarrow} \\
    \rho^{\downarrow\uparrow} & \rho^{\downarrow\downarrow}
\end{array}\right) &= (\rho\bm{I} + \boldsymbol{\sigma}\cdot\bm{m}) \\
&= \frac{1}{2}\left(\begin{array}{cc}
    \rho+m_z & m_x - im_y \\
    m_x+ im_y & \rho - m_z
\end{array}\right).
\end{aligned}
\end{equation}
where $\bm{I}$ is a $2\times2$ identity matrix, $m_x$, $m_y$, and $m_z$ refer to magnetic densities along $x$, $y$, and $z$ axis, respectively.

Next, the exchange-correlation potential $ V_{\rm xc} $ adopts a $2 \times 2$ matrix form acting on two-component spinors
\begin{equation}
V_{\rm xc}(\mathbf{r}) = \begin{pmatrix} V_{\rm xc}^{\uparrow\uparrow}(\mathbf{r}) & V_{\rm xc}^{\uparrow\downarrow}(\mathbf{r}) \\ V_{\rm xc}^{\downarrow\uparrow}(\mathbf{r}) & V_{\rm xc}^{\downarrow\downarrow}(\mathbf{r}) \end{pmatrix},
\end{equation}
where the off-diagonal elements $V_{\rm xc}^{\uparrow\downarrow}$ and $V_{\rm xc}^{\downarrow\uparrow}$ account for spin-flip interactions arising from spin-orbit coupling (SOC) or other magnetic effects. These terms mediate the mixing between spin-up and spin-down states, enabling the description of non-collinear magnetism.

Using the Pauli matrix as basis, the exchange-correlation potential can be converted into
\begin{equation}
\begin{aligned}
v_{\rm xc}(\mathbf{r}) &= \frac{\delta E_{\rm xc}}{\delta \rho(\mathbf{r})}= \frac{1}{2}\Bigl[V_{\rm xc}^{\uparrow\uparrow}(\mathbf{r}) + V_{\rm xc}^{\downarrow\downarrow}(\mathbf{r})\Bigr], \\ 
b^{x}_{\rm xc}(\mathbf{r}) &= -\frac{\delta E_{\rm xc}}{\delta m_x(\mathbf{r})}= \frac{1}{2}\Bigl[V_{\rm xc}^{\uparrow\downarrow}(\mathbf{r}) + V_{\rm xc}^{\downarrow\uparrow}(\mathbf{r})\Bigr], \\ 
b^y_{\rm xc}(\mathbf{r}) &= -\frac{\delta E_{\rm xc}}{\delta m_y(\mathbf{r})}= \frac{i}{2} \left[V_{\rm xc}^{\uparrow\downarrow}(\mathbf{r}) - V_{\rm xc}^{\downarrow\uparrow}(\mathbf{r})\right], \\ 
b^z_{\rm xc}(\mathbf{r}) &= -\frac{\delta E_{\rm xc}}{\delta m_z(\mathbf{r})}= \frac{1}{2}\Bigl[V_{\rm XC}^{\uparrow\uparrow}(\mathbf{r}) - V_{\rm XC}^{\downarrow\downarrow}(\mathbf{r})\Bigr].
\end{aligned}
\end{equation}
Here $\mathbf{b}_{xc}$ is the effective exchange-correlation magnetic field, defined as the functional derivative of the exchange-correlation energy with respect to the magnetization density, and it can be decomposed into three independent components ($\mathbf{b}^x_{xc},\mathbf{b}^y_{xc},\mathbf{b}^z_{xc}$). During Hamiltonian construction, the potential must be reconstructed in its spinor form to operate on the two-component Kohn-Sham wavefunctions.

% self-consistent calculations
\subsection{Self-Consistent Field Method} \label{SCF}
\begin{figure*}
    \centering
    \includegraphics[width=1.0\textwidth]{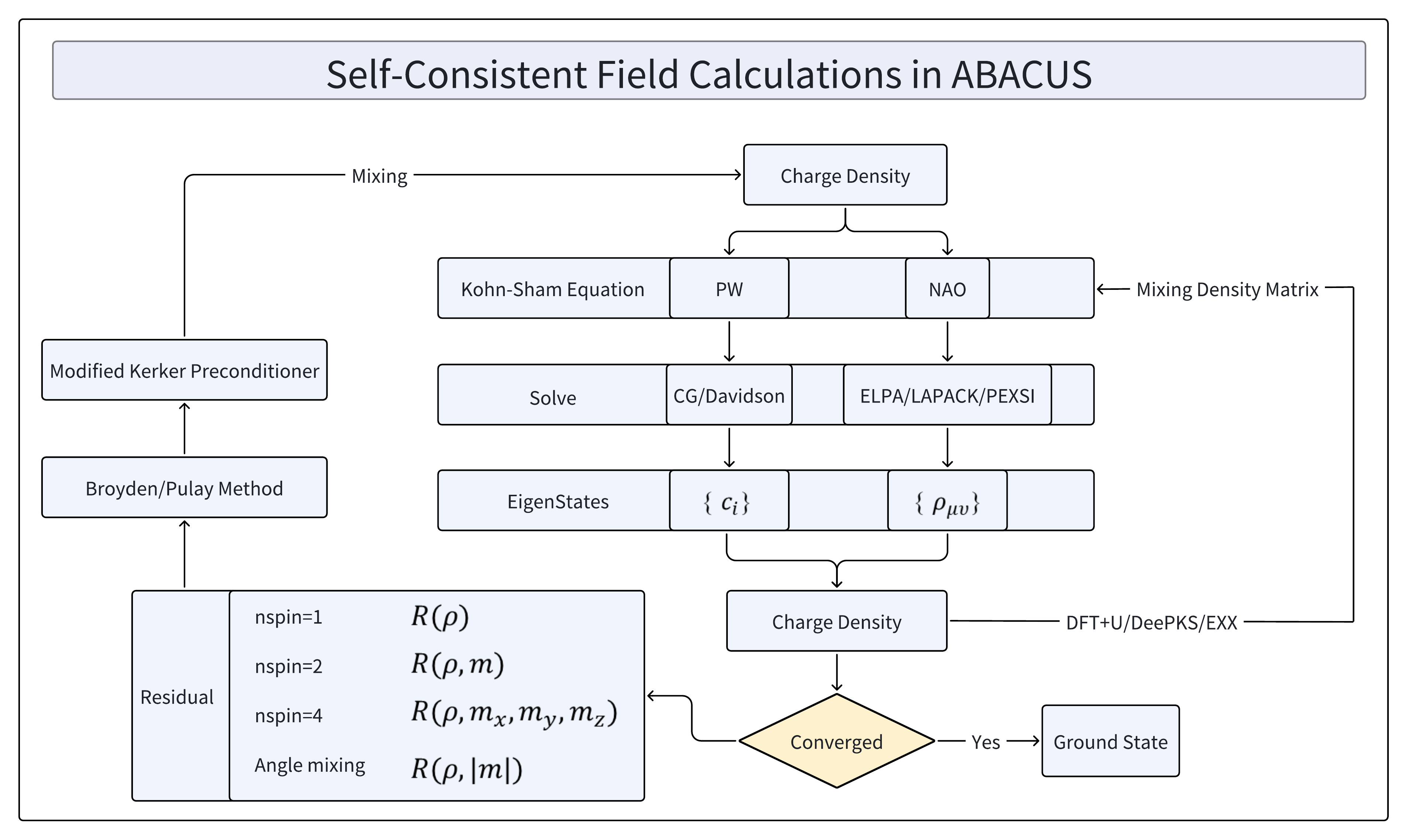}
    \caption{\MC{ABACUS performs self-consistent field (SCF) calculations with two available basis sets: PW and NAOs. The PW basis set employs iterative diagonalization methods like CG and Davidson to solve the Kohn-Sham equations. For NAOs, it uses exact diagonalization or the low-scaling method PEXSI.~\cite{CMS2009,JCPM2013} ABACUS applies different mixing algorithms for various types of calculations, including non-magnetic, collinear, and non-collinear. Additionally, ABACUS supports the density matrix mixing method for DFT+U, hybrid functionals (EXX), and DeePKS calculations with the NAO basis set.}}
    \label{fig:scf}
\end{figure*}

\subsubsection{Mixing of Electron Densities}

The Kohn-Sham equations are typically solved through the SCF method, an iterative procedure where the input electron density $\rho_{\text{in}}$ is systematically updated based on various mixing methods until achieving convergence. The efficiency and reliability of this procedure critically depend on the choice of electron density mixing algorithm, which serves the dual purpose of accelerating convergence toward the ground state while preventing stagnation in local minima. The simplest approach, linear (or ``plain'') mixing, updates the density for non-spin-polarized systems as
\begin{equation}
\rho^{i+1}_{in}=\rho^{i}_{in}+\alpha_{\rho} R[\rho^{i}_{in}],
\label{eq:plain_mixing}
\end{equation}
where $i$ denotes the iteration index. $\alpha_\rho \in [0,1]$ is the mixing parameter, and a residual vector can be defined as
\begin{equation}
R[\rho_{\rm in}]=\rho_{\rm out}-\rho_{\rm in}.
\end{equation}
Besides the linear mixing method, ABACUS implements Pulay's direct inversion in iterative subspace (DIIS) method~\cite{1980-pulay,1982-pulay} and Broyden's quasi-Newton approach.~\cite{1988-johnson} Fig.~\ref{fig:scf} illustrates the electron density mixing strategies implemented in ABACUS to guarantee stable SCF convergence.

% nspin=2
Magnetic calculations usually exhibit significantly poorer convergence behavior compared to their non-magnetic counterparts, frequently encountering two distinct failure cases, i.e., divergence of SCF or convergence to excited magnetic configurations. These issues are due to the complex coupling between charge and magnetization degrees of freedom when solving the Kohn-Sham equations. To address these challenges, ABACUS implements a mixing scheme that explicitly accounts for spin polarization. For collinear cases, the residual vector $R$ is reformulated to simultaneously track both charge and magnetization density variations
\begin{equation}
R(\rho,\bm{m})=R(\rho^{\uparrow }+\rho^{\downarrow},\rho^{\uparrow }-\rho^{\downarrow}).
\label{eq:magnetic_R}
\end{equation}
where $\rho$ represents the total charge density and $m$ denotes the magnetization vector. The mixing procedure employs a history-dependent update with wavevector-dependent preconditioning
\begin{align}
\rho^{i+1}_{in}=\sum_{j=i-n}^{i}p_{j}\left[\rho^{j}_{in}+\alpha_{\rho}\frac{q^{2}}{q^{2}+q^{2}_{\rho}}\left( \rho^{j}_{out} - \rho^{j}_{in} \right)  \right] \nonumber,\\
\bm{m}^{i+1}_{in}=\sum_{j=i-n}^{i}p_{j}\left[\bm{m}^{j}_{in}+\alpha_{m}\frac{q^{2}}{q^{2}+q^{2}_{m}}\left( \bm{m}^{j}_{out} - \bm{m}^{j}_{in} \right)  \right],
\label{eq:magnetic_mix}
\end{align}
where $n$ determines the history length, $p_j$ are Pulay/Broyden coefficients determined according to residual vector (Eq.~\ref{eq:magnetic_R}), and the $q$-dependent terms represents a modified Kerker preconditioner.~\cite{2018-threshold-kerker} The damping parameters $q_\rho$ and $q_m$ control the suppression of long-wavelength oscillations in charge and magnetization densities, respectively.

\begin{figure*}[htbp]
    \centering    \includegraphics[width=0.98\linewidth]{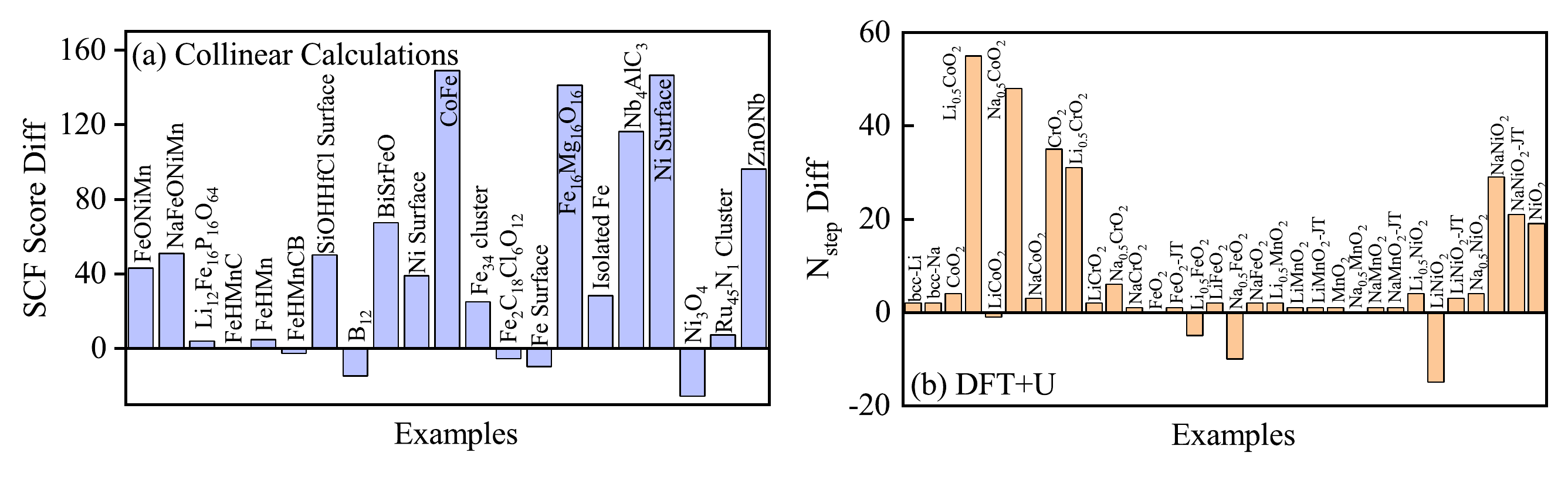}
    \caption{\MC{(a) Differences in SCF performance score ({\it vide infra}) between v3.4 and v3.5 on the selection set of 21 examples. Prior to v3.4, the residual was defined by $\rho_{\uparrow}$ and $\rho_{\downarrow}$, whereas starting from v3.5, the residual is defined by the charge density and the magnetic density, and they are mixed separately. Since some test cases failed to converge in v3.4, we present results with convergence scores rather than directly using convergence steps. The max number of iterations is set to 100 in all calculations, and the SCF convergence threshold {\tt scf\_thr=1e-6}. The SCF score is defined by $|\rm log_{10}(\delta\rho_{\rm last})|\times 10 \times \frac{100}{\rm N_{step}}$, where $\delta \rho_{\rm last}$ is the density difference $R(\rho_{in})$ of the last iteration and $\rm N_{step}$ is the number of convergence iterations. (b) Difference in SCF convergence step ($\rm N_{step}$) for the DFT+U calculation with only mixing charge density and both mixing charge density and density matrix. Here, ``JT'' represents a structure under Jahn–Teller distortion.~\cite{2013-JT-effct} For the sake of reproduction, we have made public all the details of the calculations in the link,~\cite{magnetic-mixing-pr} where one can find the complete report and download all input files.}}
    \label{fig:nspin2+dftu}
\end{figure*}

ABACUS provides several choices to optimize magnetic calculations, beginning with the implementation of separate preconditioning parameters for charge $q_\rho$ and magnetization $q_{m}$ densities to address their distinct physical characteristics and convergence behaviors. By default, the code deactivates magnetic density preconditioning (setting $q_m$ = 0) to prioritize numerical stability in most cases, while still providing users the flexibility to enable magnetic preconditioning when enhanced convergence for challenging magnetic systems is required. This approach combines robust default settings with customizable options, allowing the code to automatically handle routine cases while permitting users to fine-tune the mixing parameters for complex magnetic configurations. We compared the magnetic convergence for several magnetic materials using two versions of the ABACUS (v3.4 and v3.5). All computational details are available online.~\cite{magnetic-mixing-pr} As illustrated in Fig.~\ref{fig:nspin2+dftu}(a), a marked difference in SCF convergence behavior is observed between the two versions. Our findings demonstrate that ABACUS v3.5 achieves significantly enhanced convergence efficiency when compared to the version 3.4 of ABACUS.

% nspin=4
ABACUS generalizes the collinear mixing formalism to non-collinear magnetic calculations through an expanded residual definition that incorporates all magnetization components
\begin{equation}
R=R(\rho,\bm{m}_x,\bm{m}_y,\bm{m}_z),
\label{eq:noncollinear_R}
\end{equation}
where $\bm{m}_x$ denotes the $x$-component of the magnetization density $\bm{m}=\rho^{\uparrow }-\rho^{\downarrow}$, with analogous definitions for the $y$ and $z$ components. While this formulation successfully extends the standard Broyden mixing approach (Eqs.~\ref{eq:magnetic_mix} and \ref{eq:noncollinear_R}) to non-collinear cases, it may fail to converge to the true ground state when magnetic moment directions require optimization. To address this limitation, ABACUS implements an angle mixing scheme based on the work of Ref.~\onlinecite{2013-noncollinear-mixing}. This approach reformulates the residual in terms of a reduced parameter space $R(\rho,|\bm{m}|)$. The residual incorporates the magnitude difference $|\boldsymbol{m}_{\text{in}}^i| - |\boldsymbol{m}_{\text{out}}^i|$, where $|\boldsymbol{m}_{\text{in}}^i|$ is the input spin density of $i$-th iteration and $|\boldsymbol{m}_{\text{out}}^i|$ is the calculated output one. The mixing algorithm proceeds through three key steps at each iteration $i$. First, $|\boldsymbol{m}_{\text{in}}^{i+1}| = |\boldsymbol{m}_{\text{out}}^i|$ is ensured to keep local moment unchanged. Second, for the angle relaxation, the input angle is computed as $\theta_{\text{in}}^{i+1} = \beta \theta_{\text{out}}^i$, where $\theta_{\text{out}}^i$ measures the angular deviation between $\boldsymbol{m}_{\text{in}}^i$ and $\boldsymbol{m}_{\text{out}}^i$, and $\beta$ serves as a relaxation parameter. Third, $\boldsymbol{m}_{\text{in}}^{i+1}$ is restricted to the plane defined by $\boldsymbol{m}_{\text{in}}^i$ and $\boldsymbol{m}_{\text{out}}^i$ to maintain geometric consistency.
This methodology enables efficient exploration of magnetic configuration space while avoiding common convergence pitfalls associated with traditional non-collinear mixing approaches.

% mixing_dmr
Traditional SCF calculations typically focus solely on charge density mixing, neglecting the density matrix that plays a crucial role in methods such as DFT+U. This oversight creates inconsistency, particularly evident when the system is far from its ground state, as the DFT+U method~\cite{cococcioni2005linear, anisimov1993density, anisimov1997first, dudarev1998electron, jiang2010first} explicitly requires the density matrix. Such discrepancy frequently leads to convergence difficulties in strongly correlated systems. To address this discrepancy, ABACUS supports a restart mixing scheme that begins with conventional charge density mixing during initial iterations when the density matrix cannot be reliably estimated from atomic wavefunction coefficients, then automatically switches to simultaneous mixing of both charge density and density matrix once a predefined convergence threshold is achieved.

The charge mixing scheme in DFT+U calculations can be further enhanced by an automated U-Ramping procedure that gradually increases the U parameter during SCF iterations.~\cite{2010-uramping1,2013-uramping2} We performed DFT+U calculations on a series of selected magnetic materials (source files available online~\cite{magnetic-mixing-pr}) using ABACUS v3.6, with particular focus on comparing the SCF convergence behavior between calculations with and without density matrix mixing. Fig.~\ref{fig:nspin2+dftu}(b) demonstrates the reduction of SCF convergence steps after the density matrix mixing method is adopted.

% smearing
\subsubsection{Smearing Methods}

In metallic systems, SCF iterations frequently encounter convergence difficulties stemming from charge density oscillations. These oscillations arise because the electron occupation number drops abruptly to zero at the Fermi energy, with electronic states near the Fermi surface frequently switching between filled and empty electronic states. Furthermore, there exists no variational principle that regulates the convergence of the system's total energy with resepct to $k$-point sampling. Smearing methods mitigate this issue by introducing a controlled degree of partial occupation for Kohn-Sham orbitals in the vicinity of the Fermi energy, effectively damping numerical instabilities and promoting convergence. A few smearing methods that have been implemented in ABACUS are listed below.

First, the electron occupancy can be approximated by a Gaussian distribution
\begin{equation}
    f(\varepsilon)=\frac{1}{w \sqrt{2 \pi}} e^{\frac{\left(\varepsilon-E_{\mathrm{F}}\right)^2}{2 w^2}},
\end{equation}
where $E_\mathrm{F}$ denotes the Fermi energy, $\varepsilon$ represents the single-particle energy levels, and $w$ controls the broadening width.

Second, the Fermi-Dirac smearing scheme is intrinsically aligned with finite-temperature density functional theory,~\cite{1965-mermin-thermal} where the electronic temperature $T$ governs the statistical electron distribution. Crucially, within this framework the smearing width ($k_BT$) arises as a natural consequence of thermal excitations in the electronic system, rather than serving as an empirically tuned broadening parameter. The formula is
\begin{equation}
f(\varepsilon)=\frac{1}{1+\exp(\frac{\varepsilon-E_{\mathrm{F}}}{k_B T})},
\end{equation}
where $k_B$ denotes Boltzmann's constant. However, in comparison to Gaussian smearing, the drawback of the Fermi-Dirac smearing method lies in its long tail, which necessitates the calculation and storage of a large number of partially occupied states to obtain the charge density.

Third, through smooth approximations of the $\delta$ and step functions (Fermi-Dirac-like), the Methfessel-Paxton (MP) method~\cite{1989-mp} enables the Brillouin-zone integration in metals to converge exponentially with the number of sampling points. The approximation formulas for the step function can be written as
\begin{equation}
\begin{aligned}
& S_0(x)=\frac{1}{2}\Bigl[1-\operatorname{erf}(x)\Bigr], \\
& S_N(x)=S_0(x)+\sum_{n=1}^N A_n H_{2 n-1}(x) e^{-x^2},
\end{aligned}
\end{equation}
where $x=\frac{\varepsilon-E_{\mathrm{F}}}{w}$, $\operatorname{erf}(x)$ is the error function, and $H_n$ represents the $N$-th Hermite polynomial with the expansion coefficient $A_n$. Using the formula above, $S_N(x)$ can be used to accurately compute integral $\int S_N(x)F(x)dx$, where $F(x)$ represents physical quantities such as density of states. Although the broadening function of the MP smearing method ensures that the total free energy remains independent of the smearing temperature up to at least third order, thereby making Hellman-Feynman forces consistent with the total free energy, it still introduces nonmonotonic and non-positive-definite occupation functions.

Fourth, to address possible negative occupancies that may lead to negative total electron density during SCF and numerical instability in the MP method, Marzari {\it et al.} introduced the cold-smearing method,~\cite{1999-Marzari-cold} where the broadening function takes the form of
\begin{equation}
\tilde{\delta}(x)=\frac{2}{\sqrt{\pi}}e^{-\left[x-(1/\sqrt{2})\right]^{2}}(2-\sqrt{2}\,x).
\end{equation}
Here $x=\frac{E_\text{F}-\varepsilon}{w}$. In this regard, the occupation numbers $f(\varepsilon_i)$ can be obtained by $\int^{x_i}_{-\infty}\tilde{\delta}(x)dx$. Note that the cold smearing method yields a free energy independent of smearing temperature up to the second order.

\subsection{Geometry Relaxation}

In geometry optimization under fixed cell conditions, the ionic positions are iteratively adjusted to minimize the total energy until the residual atomic forces are smaller than a specified threshold. The force acting on atom $I$ is defined as the negative gradient of the total energy $E_{\rm tot}$ with respect to its atomic position $\mathbf{R}_{I}$, and the formula is
\begin{equation}
    \mathbf{F}_{I} = -\partial E_{\rm tot}/\mathbf{R}_{I}.
    \label{eq:force}
\end{equation}
This minimization problem can be solved using various optimization algorithms, including the steepest descent (SD), the conjugate gradient (CG), the Broyden-Fletcher-Goldfarb-Shanno (BFGS),~\cite{1985-head-broyden} and the Fast Inertial Relaxation Engine (FIRE)~\cite{2006-fire} methods. At each iteration step, the atomic positions are updated according to
\begin{equation}
    \mathbf{R}^{n+1}_{I} = \mathbf{R}^n_{I} + \alpha \mathbf{D}^n_{I},
\end{equation}
where $\mathbf{D}_{I}$ represents the optimization direction, $\alpha$ is the step length, and $n$ denotes the index of geometry relaxation step.

Furthermore, when the cell is allowed to relax, the stress tensor $\sigma_{\alpha\beta}$ is computed as
\begin{equation}
    \sigma_{\alpha\beta}=-\frac{1}{\Omega}\frac{\partial E_{\rm tot}}{\partial \epsilon_{\alpha\beta}},
    \label{eq:stress}
\end{equation}
where $\Omega$ denotes the cell volume and $\epsilon_{\alpha\beta}$ is the strain tensor components. Next, different geometry relaxation strategies can be employed. For example, one approach to perform cell relaxation consists of two sequential optimization stages. First, optimizing atomic positions with lattice vectors held fixed, followed by adjusting lattice vectors while keeping ionic positions constant. An alternative strategy involves optimization of both atomic positions and lattice vectors at the same time. This simultaneous refinement allows for dynamic coupling between atomic rearrangements and lattice deformation, potentially accelerating convergence to the equilibrium structure. The geometry optimization methods implemented in ABACUS have been applied to study alloys,~\cite{24AM-Liu} interfaces,~\cite{21JNM-Liu} slabs,~\cite{22PCCP-Liu} and low-dimensional materials,~\cite{22B-Chen,24-Liu} etc.

%Molecular Dynamics}
\subsection{Molecular Dynamics} \label{sec:md}

By solving Newton’s equations of motion for each atom in the system, molecular dynamics (MD) simulations numerically integrate atomic forces over a small time step to trace the positions of atoms. In this regard, MD simulations offer a robust framework for investigating the temporal evolution of atomic systems, facilitating the study of a wide range of phenomena such as phase transitions, diffusion processes, chemical reactions, and material properties.

Central to MD simulations is the potential energy surface (PES), which describes atomic interactions and can be derived through various methods. Quantum mechanical approaches like density functional theory (DFT) can provide high-accuracy PES calculations but are computationally intensive. Alternatively, machine learning potentials~\cite{18CPC-Wang,23JCP-Zeng} offer a balance between accuracy and efficiency by learning the PES from DFT training data, enabling simulations of larger systems or longer timescales. Although classical MD methods enable rapid PES evaluations for systems, their reliance on empirical force fields (such as Lennard-Jones potential~\cite{24-LJ1,24-LJ2}) results in lower accuracy compared to the other two approaches.

Once the PES is established, atomic forces are computed via gradients of the total energy with respect to atomic positions, which are then integrated over time to predict atomic trajectories. The velocity Verlet algorithm,~\cite{82JCP-Swope} a widely used integration scheme that ensures long-term stability, updates atomic positions $\mathbf{r}(t+\Delta t)$ and velocities $\mathbf{v}(t+\Delta t)$ in discrete time steps $\Delta t$.

MD simulations can be performed under various statistical ensembles, each imposing different constraints to model specific experimental conditions. ABACUS supports the following ensembles. First, the NVE (microcanonical) ensemble conserves particle number $N$, volume $V$, and total energy $E$, simulating isolated systems where energy exchange with the environment is negligible. Second, for simulations requiring constant temperature $T$, the NVT (canonical) ensemble~\cite{96MP-Martyna} employs thermostats such as velocity rescaling, a straightforward but non-equilibrium approach, or the Nos\'{e}-Hoover chain method,~\cite{84JCP-Nose,85A-Hoover,92JCP-Martyna} which rigorously incorporates auxiliary degrees of freedom to maintain temperature via Hamiltonian dynamics. The Anderson thermostat~\cite{80JCP-Andersen} couples the system to a heat bath that imposes the desired temperature to simulate the NVT ensemble. The coupling to a heat bath is represented by a stochastic collision that occasionally acts on randomly selected particles. The Berendsen thermostat~\cite{84JCP-Berendsen} rescales the velocities of atoms at each step to reset the temperature of a group of atoms. In addition, the Langevin dynamics~\cite{78B-Schneider} couples the system to a heat bath through stochastic and frictional forces, ensuring that the system reaches and maintains the desired temperature over time, mimicking conditions where the system exchanges energy with its environment. Third, the NPT (isothermal-isobaric) ensemble maintains constant pressure $P$ and temperature $T$ using barostats like Parrinello-Rahman.

Besides the above methods, ABACUS also supports the multi-scale shock technique (MSST),~\cite{2003-MSST} enabling the simulation of shock wave propagation in materials, thereby extending the range of accessible timescales for investigating shock phenomena within such non-equilibrium frameworks.

\begin{figure}[htp]
    \includegraphics[width=0.8\linewidth]{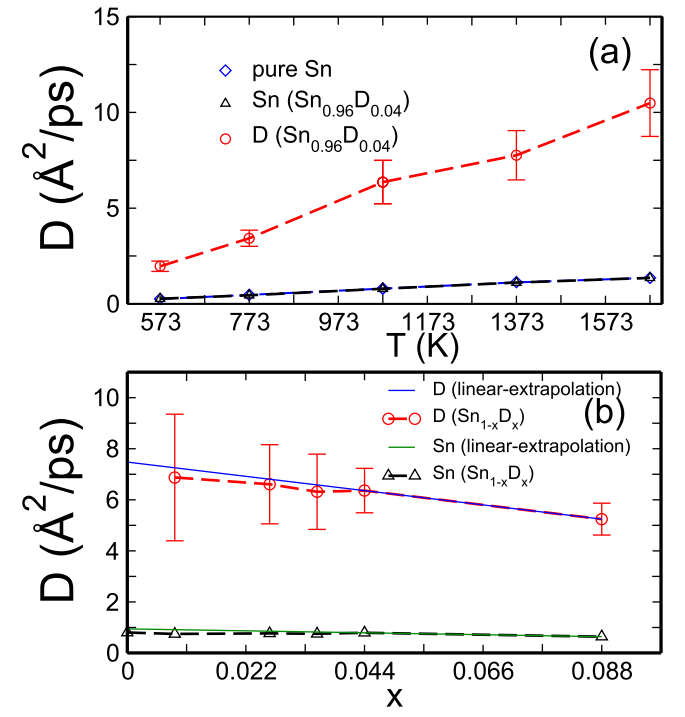}
    \caption{\MC{(a) Diffusion coefficients of D in liquid $\mathrm{Sn_{0.96}D_{0.04}}$, Sn in $\mathrm{Sn_{0.96}D_{0.04}}$, and Sn in pure liquid Sn as a function of temperature. 
    (b) Diffusion coefficients of D and Sn in liquid $\mathrm{Sn_{1-x}D_x}$ at 1073 K with $x$ being 0.009, 0.027, 0.036, 0.044, and 0.085. (Adapted with permission from J. Chem. Phys. 147, 064505 (2017). Copyright 2017 AIP Publishing.)}
 }
    \label{fig::md_jcp}
\end{figure}

Born-Oppenheimer Molecular Dynamics (BOMD) decouples the motion of nuclei and electrons and treats nuclei as classical particles moving on a PES that is instantaneously determined by the electronic structure. As an {\it ab initio} method, BOMD relies on explicit electronic structure calculations and has been implemented in ABACUS. Here we list some applications.~\cite{24MRE-Chen,Liu2017,19A-Wang,18GCA-Liu,21CPB-Wang,21JNM-Zheng} Liu {\it et al.}~\cite{Liu2017} employed BOMD simulations to predict the diffusion coefficients of deuterium in liquid Sn over a temperature range of 573 to 1673 K. Fig.~\ref{fig::md_jcp}(a) shows the diffusion coefficients of deuterium in a liquid Sn cell consisting of 216 atoms at five temperatures between 573 and 1673 K. These simulations reveal that deuterium diffuses through liquid Sn more rapidly than Sn atoms diffuse within themselves. Fig.~\ref{fig::md_jcp}(b) depicts the effects of deuterium concentration on the diffusion rates of both deuterium and Sn at 1073 K. The findings suggest that Sn's structural and dynamic characteristics remain largely unaffected by the presence of deuterium for the tested temperatures and concentrations.

\begin{figure}[htp]
    \includegraphics[width=0.9\linewidth]{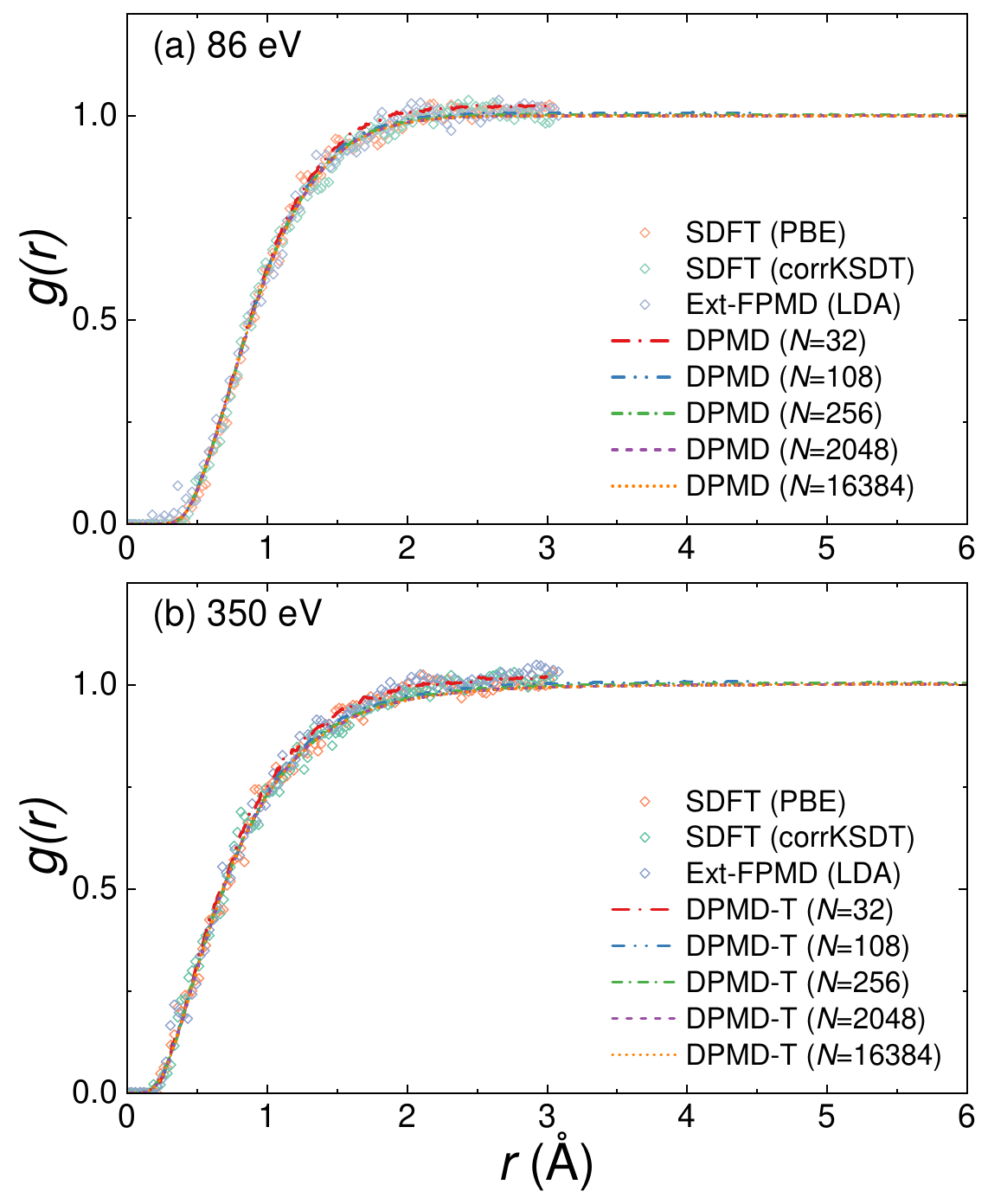}
    \caption{\MC{Radial distribution functions (RDFs) $g(r)$ for B systems at a density of 2.46 $\mathrm{g/cm^3}$ and the temperatures are (a) 86 eV and (b) 350 eV. Some of the $g(r)$ are derived from Ext-FPMD,~\cite{22CPP-Blanchet} as well as the sDFT calculations employed the PBE~\cite{1996-pbe} and corrKSDT~\cite{18L-Karasiev} XC functionals. The number of B atoms is set to 32 in AIMD simulations. In addition, DPMD denotes the model trained by the traditional DP method,~\cite{18CPC-Wang} whereas DPMD-T indicates the TDDP method applied model training as reported by Zhang {\it et al.}~\cite{20PP-Zhang} $N$ is the number of B atoms in a cell. (Adapted with permission from Matter. Radiat. Extremes 9, 015604 (2024). Copyright 2024 AIP Publishing.)}}
    \label{fig::md_mre}
\end{figure}

Chen {\it et al.}~\cite{24MRE-Chen} combined stochastic density functional theory (sDFT) with BOMD to explore warm dense matter systems at temperatures ranging from several tens of eV to 1000 eV. They also trained machine-learning-based interatomic models using the first-principles data and employed these models to examine large systems via BOMD simulations. Furthermore, they evaluated the structural and dynamic characteristics, as well as the transport coefficients, of warm dense matter. Fig.~\ref{fig::md_mre} shows the radial distribution functions $g(r)$ of warm dense B with a density of 2.46 $\mathrm{g/cm^3}$ at 86 and 350 eV, and the sDFT results are in excellent agreement with those obtained from extended first-principles molecular dynamics (Ext-FPMD).~\cite{2016-zhang-extended,2022-blanchet-extended}

Ma {\it et al.}~\cite{2024-ma-prb} combine AIMD and finite-temperature orbital-free DFT (FT-OFDFT) with a nonlocal free energy functional XWM to study a variety of warm dense matter systems such as the Si, Al, H, He, and H-He mixtures. The KS-DFT calculations were performed using ABACUS, while the OF-DFT calculations were carried out with ATLAS.~\cite{2016-atlas} The XWMF functional is expected to be a good choice for the realistic simulations of warm dense matter systems covering a broad range of temperatures and pressures.

% surface chemistry
\subsection{Implicit Solvation Model}

Electrochemical reactions, which refer to potential-driven processes at electrode/solvent interfaces, play a pivotal role in advancing green-energy technologies for clean fuel production. A fundamental understanding of the atomic-scale structure and physicochemical properties of these interfaces is indispensable for the rational design and optimization of electrochemical systems. Atomic-scale computational simulations have emerged as a powerful tool for elucidating the complex nature of electrochemical interfaces. However, theoretical modeling at the molecular level faces substantial challenges, such as solvent layers, electrical double layers, and variations in electron numbers at electrode-solvent interfaces, among others.

\begin{figure}
    \centering
    \includegraphics[width=0.96\linewidth]{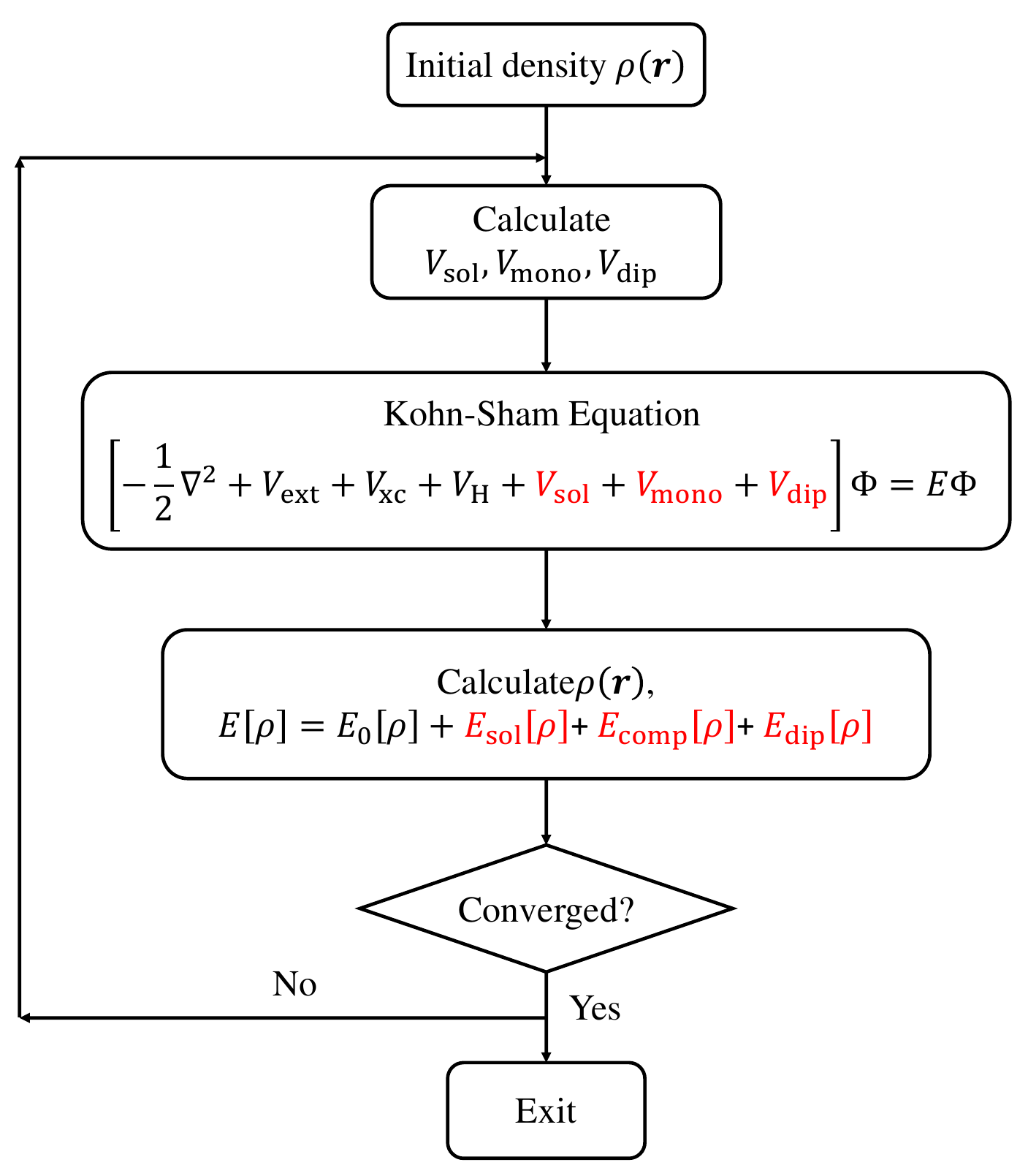}
    \caption{\MC{Implementation of the implicit solvation model in ABACUS. The self-consistent loop of solving the KS equation starts with an initial charge density $\rho(\mathbf{r})$, followed by the evaluation of correction potentials, including the implicit solvent potential $V_{\mathrm{sol}}$, the compensating charge potential $V_{\mathrm{comp}}$, and the dipole correction potential $V_{\mathrm{dip}}$.
    These corrections are subsequently applied to the KS equations for iterative electron density and total energy calculations, where $E_0[\rho]$ denotes the energy without corrections.}}
    \label{fig:surchem_framework}
\end{figure}

To address these challenges, we implement a combined approach incorporating (i) an implicit solvation model to approximate solvent effects, (ii) a dipole correction scheme to mitigate spurious electrostatic interactions, and (iii) a compensating charge plate to maintain charge neutrality. As illustrated in Fig.~\ref{fig:surchem_framework}, these methodological components are integrated into the SCF workflow of ABACUS, where the potential of the implicit solvent $V_{\mathrm{sol}}$, the compensating charge $V_{\mathrm{comp}}$, and the dipole correction $V_{\mathrm{dip}}$ is included in iterative electron density and total energy calculations. This integrated approach enables more realistic simulations of electrochemical systems while maintaining computational efficiency.

Solid-liquid interfaces are ubiquitous and frequently encountered and employed in electrochemical simulations. For accurately modeling such systems, it is important to consider the solvation effect. The implicit solvation model is a well-developed method to deal with solvation effects, widely used in both finite and periodic systems. This approach treats the solvent as a continuous medium instead of individual ``explicit'' solvent molecules, which means that the solute embedded in an implicit solvent and the average over the solvent degrees of freedom becomes implicit in the properties of the solvent bath.

We place the solute in a cavity surrounded by a continuum dielectric medium characterized by the relative permittivity of the solvent, as employed in a previous work.~\cite{2014Implicit} We describe the dielectric response in terms of the solute's electron density, considering the polarization of solvents in response to the electronic structures of solute, the effects of cavitation and dispersion, and the reaction of solute system to the presence of the solvent.

We determine the form of the dielectric cavity in the solvent by assuming a diffuse cavity that is a local functional of the electron density $\rho(\mathbf{r})$ of the solute, which satisfies the following functional dependence
\begin{equation}
\epsilon(\rho)=1+(\varepsilon_{b}-1)S(\rho),
\end{equation}
where $\epsilon_{b}$ is the relative permittivity of the bulk solvent and $S(\rho)$ is the cavity shape function, given by~\cite{2005Joint}
\begin{equation}
S(\rho)=\frac{1}{2}\mathrm{erfc}{\left[\frac{\ln{\left(\frac{\rho}{\rho_\mathrm{c}}\right)}}{\sqrt2b}\right]}.
\end{equation}
The parameter $\rho_{c}$ is the charge density cutoff, defining the electron density at which the dielectric cavity forms. The parameter $b$ determines the width of the diffuse cavity. This assumption leads to a smooth variation of the relative permittivity from $\epsilon(\mathbf{r})$=1 of the solute to $\epsilon_{b}$ in the solvent.

The conjugate gradient method is used to solve the generalized Poisson equation
\begin{equation}
\nabla\cdot\Big[\epsilon(\rho)\nabla\phi(\mathbf{r})\Big]=-4\pi\Bigl[N(\mathbf{r})-\rho(\mathbf{r})\Bigr],
\end{equation}
where $\phi(\mathbf{r})$ is the electrostatic potential due to the electron density $\rho(\mathbf{r})$ and nuclear charge density $N(\mathbf{r})$ of the solute system in a polarizable medium.

The typical Kohn-Sham Hamiltonian consists of two additional terms in the local part of the potential. One of them is the electrostatic correction caused by the induced charge~\cite{2014Implicit}
\begin{equation}
\label{eq:surchem_Vel}
V_{\mathrm{el}}=-\frac{d\epsilon(\rho)}{d\rho}\frac{{|\mathrm{\nabla\phi}|}^2}{8\pi},
\end{equation}
and the other term is the cavity potential, which describes the cavitation, dispersion, and repulsion interaction between the solute and the solvent that is not captured by the electrostatic terms alone~\cite{2014Implicit} 
\begin{equation}
\label{eq:surchem_Vcav}
    V_{\mathrm{cav}}=\tau\frac{d|\mathrm{\nabla S}|}{d\rho},
\end{equation}
where $\tau$ is the effective surface tension parameter. The two corrections (Eqs.~\ref{eq:surchem_Vel} and ~\ref{eq:surchem_Vcav}) are collectively referred to as the implicit solvent-induced potential term $V_{\mathrm{sol}}$, while the energy correction terms are
\begin{equation}
    E_{\mathrm{el}}=-\frac{1}{8\pi}\int{\epsilon\left(\rho\right){|\mathrm{\nabla\varphi}|}^2 \mathrm{d}\mathbf{r}}
\end{equation}
\begin{equation}
E_{\mathrm{cav}}=\tau\int{|\mathrm{\nabla S}|\mathrm{d}\mathbf{r}}.
\end{equation}

\begin{figure}
    \centering
    \includegraphics[width=1.0\linewidth]{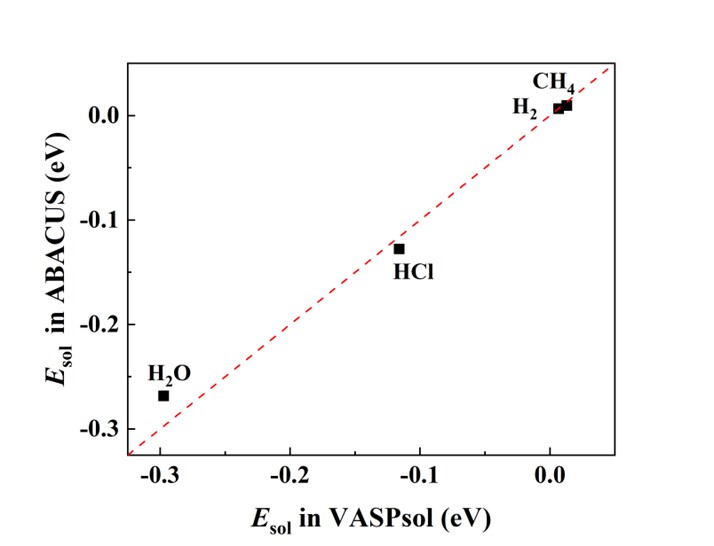}
    \caption{\MC{Solvation energies $E_{sol}$ (the total energy difference between a solvated condition and a vacuum condition) calculated by VASPsol and ABACUS for H$_2$O, HCl, H$_2$, and CH$_4$ systems.}}
    \label{fig:solvation_energy}
\end{figure}

We benchmark the accuracy of the implicit solvation implementation by calculating molecular solvation energies $E_{\mathrm{sol}}$, which is defined as the total energy difference between a solvated condition and a vacuum condition in ABACUS and comparing them against VASPsol-calculated values.~\cite{2019-vaspsol} We can see from Fig.~\ref{fig:solvation_energy} that the solvation energies from ABACUS agree well with those produced by the VASPsol package. The minor discrepancies between the solvation energies computed by the two methods may be attributed to the pseudopotential difference. Note that we use projector-augmented-wave (PAW) potentials in VASPsol and the norm-conserving pseudopotentials in ABACUS.

%Dipole Correction
ABACUS also supports adding dipole correction and compensating charge when modeling surfaces, which were utilized in a recent work.~\cite{2025-sun-probing} The periodic boundary conditions imposed on the electrostatic potential create an artificial electric field across a slab. By introducing an isolated slab-shaped density distribution $\rho(\mathbf{r})$ that is normal to the $z$-axis, a dipole correction~\cite{99B-Bengtsson} is added to the bare ionic potential to compensate for the artificial dipole field within the context of periodic supercell calculations.

Modeling a constant-potential electrochemical surface reaction requires the adjustment of electron numbers in a simulation cell. Simultaneously, we must preserve the supercell’s neutrality under the periodic boundary conditions. Thus, a distribution of compensating charge needs to be implemented in the vacuum region of surface models when extra electrons are added to or extracted from the system. The compensating charge implemented in ABACUS follows the methodology developed by Brumme {\it et al.}~\cite{14B-Brumme} We assume that the monopole with a total charge of $-n_{\mathrm{dop}}$ per unit cell is located at $z_{\mathrm{mono}}$ along the $z$ axis. In this case, the effective potential $V_{\mathrm{mono}}(\mathbf{r})$ is added to the Hamiltonian, and the additional term $E_{\mathrm{mono}}$ is included in the total energy. Since $V_{\mathrm{mono}}(\mathbf{r})$ is independent of the electron density, it is unnecessary to update in a self-consistent manner. We also implemented the correction on the ionic forces induced by the presence of the monopole, allowing for calculating the electronic structure and complete structural relaxation in the field-effect configuration.

\begin{figure}
    \centering
    \includegraphics[width=\linewidth]{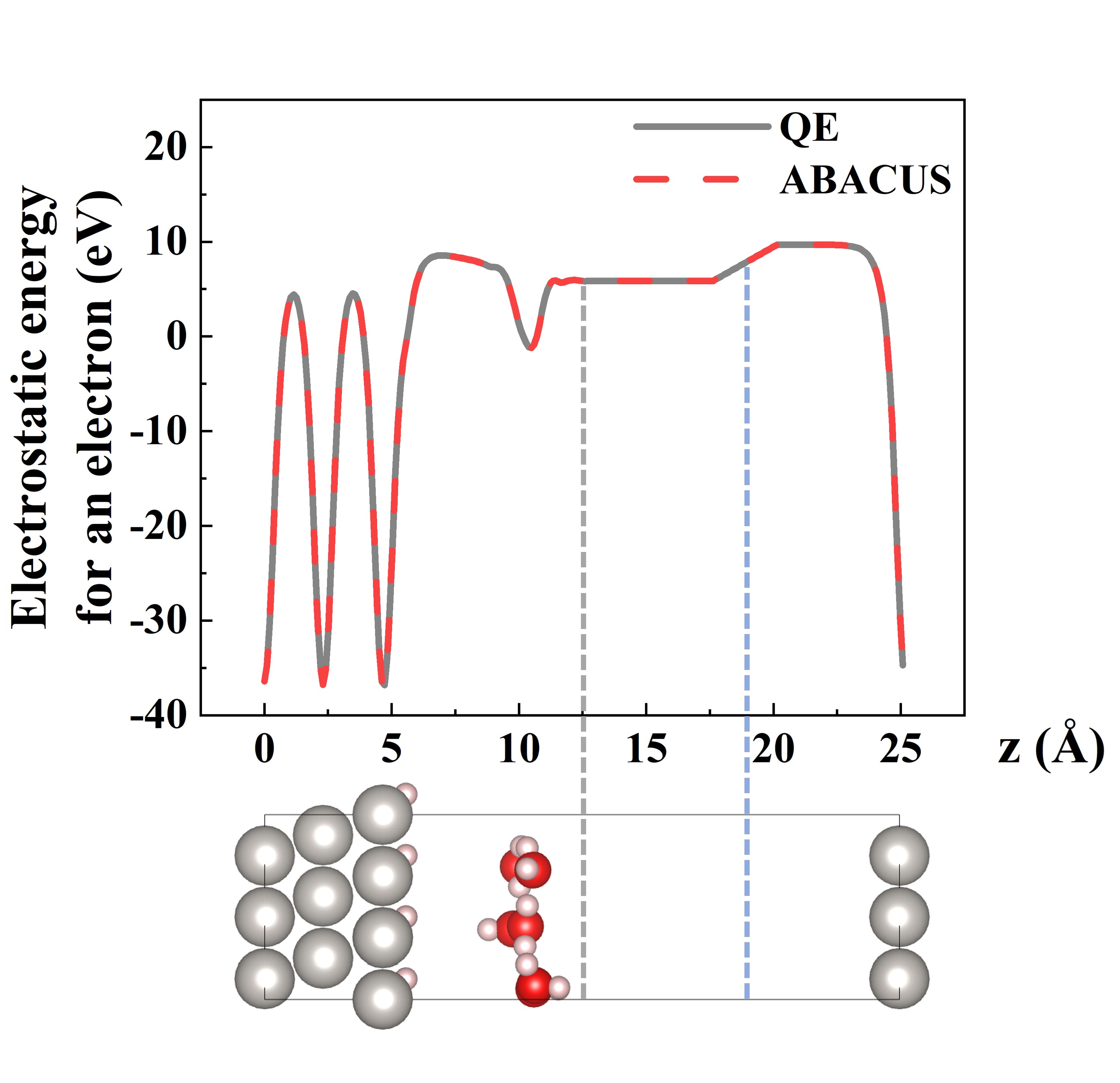}
    \caption{\MC{
        Electrostatic energy of an electron along the $z$ direction perpendicular to the Pt surface calculated by Quantum ESPRESSO and ABACUS.~\cite{2025-sun-probing}
        The gray vertical dashed line represents the position of the compensating charge plate, while the blue dashed line corresponds to the dipole correction. A schematic structure plot of the testing interface model is aligned below the average electrostatic energy curves. Gray, red, and white spheres correspond to platinum, oxygen, and hydrogen atoms, respectively.}
    }
    \label{fig:dipole_correction}
\end{figure}

We output the electrostatic energy of an electron along the $z$ direction perpendicular to the Pt surface for a testing system with an added electron number $N_e^{\mathrm{extra}} = 0.2$ in the simulation cell. The testing system is a (3×3) Pt (111) surface slab composed of three atomic layers, as illustrated in Fig.~\ref{fig:dipole_correction}. The modeled electrode surface contains 27 Pt atoms, with a monolayer of hydrogen coverage and six explicit water molecules. We place a compensating charge plate in the vacuum region above the water layer, and a dipole correction is also included in DFT calculations. Fig.~\ref{fig:dipole_correction} shows that the electrostatic energy results from ABACUS are in perfect agreement with those computed from Quantum ESPRESSO.~\cite{2020-qe}

\section{Methods in Plane Wave Basis} \label{sec:pw}

% pw: energy/force/stress
\subsection{Kohn-Sham Equation in Plane Wave Basis} \label{sec:pw-energy}

In the atomic system with periodic boundary conditions, the electronic wave functions can be expanded with the plane waves, which take the form of
\begin{equation}
    \psi_{n\mathbf{k}}(\mathbf{r}) = \sum_{\mathbf{G}} c_{n\mathbf{k}}(\mathbf{G}) e^{i(\mathbf{k+G})\cdot\mathbf{r}},
\end{equation}
where $\mathbf{G}$ and $\mathbf{k}$ respectively represent the wave vectors of plane waves and sampling points in the Brillouin zone, $n$ denotes the band index, and $\{c_{n\mathbf{k}}(\mathbf{G})\}$ are the expansion coefficients of plane wave basis. The $k$-point sampling method is typically employed as the Monkhorst-Pack scheme.~\cite{76B-Monkhorst} Next, the Kohn-Sham equation described with the plane wave basis can be written as
\begin{equation}
    \label{eq:ks_equation_pw}
    \sum_{\mathbf{G'}} \left[ \frac{1}{2} (\mathbf{k+G})^2 \delta_{\mathbf{GG'}} + \hat{V}_\mathrm{KS}(\mathbf{G-G'}) \right] c_{n\mathbf{k}}(\mathbf{G'}) = \varepsilon_{nk} c_{n\mathbf{k}}(\mathbf{G}),
\end{equation}
where $\hat{V}_\mathrm{KS}(\mathbf{G-G'})$ is the plane wave represenation of the Kohn-Sham potential $\hat{V}_\mathrm{KS}$, $\varepsilon_{nk}$ is the eigenvalue of the Kohn-Sham equation, and $\delta_{\mathbf{GG'}}$ is the Kronecker delta function. In this regard, the total energy can be calculated as
\begin{equation}
\begin{aligned}
    E_\mathrm{tot} = & \sum_{n\mathbf{k}} f(\varepsilon_{n\mathbf{k}}) \varepsilon_{n\mathbf{k}} - \frac{1}{2}\int\int{\frac{\rho(\mathbf{r})\rho(\mathbf{r'})}{|\mathbf{r} -\mathbf{r'}|}\mathrm{d}\mathbf{r}\mathrm{d}\mathbf{r'}} \\
    & -\int{V_\mathrm{xc}(\mathbf{r})\rho(\mathbf{r})\mathrm{d}\mathbf{r}}+E_{\mathrm{xc}}[\rho(\mathbf{r})] + E_{\rm II},
    \label{eq:ksenergy_band}
\end{aligned} 
\end{equation}
where $f(\varepsilon_{n\mathbf{k}})$ is the Fermi-Dirac distribution function, $\rho(\mathbf{r})$ is the electron density, $V_\mathrm{xc}$ is the exchange-correlation potential, $E_{\mathrm{xc}}$ is the exchange-correlation energy, and $E_{\rm II}$ is the ionic energy calculated by the Ewald method.~\cite{electronic-martin}

According to the Hellmann-Feynman theorem,~\cite{35AP-Hellmann,39-Feynman} the force acting on atom $I$ is defined as Eq.~\ref{eq:force} and the stress is defined as Eq.~\ref{eq:stress}. Consequently, when including a non-local part in the norm-conserving pseudopotential, the force acting on atom $I$ of type $\tau$ can be divided into three parts, namely
\begin{equation}
    \mathbf{F}_{I}=\mathbf{F}_{I}^\mathrm{Ewald} + \mathbf{F}_{I}^\mathrm{L} + \mathbf{F}_{I}^\mathrm{NL},
    \label{eq:force_pw}
\end{equation}
where \( \mathbf{F}_{I}^\mathrm{Ewald} \) is the Ewald force. \( \mathbf{F}_{I}^\mathrm{L} \) is the force contributed by the local part of the pseudopotential, and \( \mathbf{F}_{I}^\mathrm{NL} \) is from the non-local part. The local potential term is given by
\begin{equation}
    \mathbf{F}_{I}^\mathrm{L}=-i\Omega\sum_{\mathbf{G}\neq\mathbf{0}}{\mathbf{G}e^{i\mathbf{G}\cdot\mathbf{R}_I}v_\tau^\mathrm{L}(\mathbf{G})\rho^*(\mathbf{G})},
\label{eq:localforce}
\end{equation}
where $\Omega$ is the cell volume and $v_\tau^\mathrm{L}$ is the local potential of atom type $\tau$. Moreover, $\rho(\mathbf{G})$ is the electron density in the plane wave basis, which is computed as
\begin{equation}
    \rho(\mathbf{G}) = \frac{1}{\Omega}\int{\rho (\mathbf r)e^{-i\mathbf G\cdot \mathbf r} \mathrm{d}\mathbf{r}}.
\end{equation}
Furthermore, the non-local potential term is written as
\begin{equation}
    \begin{aligned}
    \mathbf{F}_{I}^\mathrm{NL} = &-2i\sum_{n\mathbf{k}\mathbf{GG'}}f(\epsilon_{n\mathbf{k}})c^*_{n\mathbf{k}}(\mathbf{G})c_{n\mathbf{k}}(\mathbf{G'})\\
    &\times\Big[e^{i(\mathbf{G'-G})\cdot\mathbf{R}_I}(\mathbf{G'-G})v_\tau^\mathrm{NL}(\mathbf{k+G,k+G'})\Big],
    \end{aligned}
\end{equation}
where $v_\tau^\mathrm{NL}$ is the nonlocal potential of atom type $\tau$.

%Stress
The stress tensor is decomposed into Ewald term $\sigma_{\alpha\beta}^\mathrm{Ewald}$, Hartree term $\sigma_{\alpha\beta}^\mathrm{H}$, exchange-correlation term $\sigma_{\alpha\beta}^\mathrm{xc}$, kinetic energy term $\sigma_{\alpha\beta}^\mathrm{T}$, local pseudopotential term $\sigma_{\alpha\beta}^\mathrm{L}$, and non-local pseudopotential term $\sigma_{\alpha\beta}^\mathrm{NL}$, expressed as
\begin{equation}\label{eq:pw_stress}
\sigma_{\alpha\beta}=\sigma_{\alpha\beta}^\mathrm{Ewald}+\sigma_{\alpha\beta}^\mathrm{Hartree}+\sigma_{\alpha\beta}^\mathrm{xc}+\sigma_{\alpha\beta}^\mathrm{T}+\sigma_{\alpha\beta}^\mathrm{L}+\sigma_{\alpha\beta}^\mathrm{NL}.
\end{equation}
The formulas of the terms can be found in Ref.~\onlinecite{22B-Liu}.

% pw: CG and Davidson method
\subsection{Iterative Diagonalization Methods}
\label{sec:pw-diag}

\begin{figure}[htbp]
    \centering
    \setlength{\fboxsep}{1pt} 
    \fbox{
    \parbox{\dimexpr 0.9\columnwidth-2\fboxrule\relax}{
    \begingroup
    \setlength{\abovedisplayskip}{1pt}
    \setlength{\belowdisplayskip}{1pt}
    \begin{align*}
    &\textbf{for} \ i = 1 \ \textbf{to} \ n \ \textbf{do} & \\
    &\quad \psi_i = (I - \sum_{j<i} \ket{\psi_j}\bra{\psi_j}) \psi_i\\
    &\quad \psi_i = \frac{\psi_i}{\Vert \psi_i \Vert}\\
    &\quad p_i = H\psi_i  \\
    &\quad \lambda_i =\langle{\psi_i}|{p_i}\rangle  \\
    &\quad k = 1 \\
    &\quad \textbf{while}~k \le \ k_{maxiter} ~ \textbf{and}~convergence~not~reached~ \textbf{do}\\
    &\qquad g_i = K(p_i - \lambda \psi_i) \\
    &\qquad g_i = (I - \sum_{j\le i} \ket{\psi_j}\bra{\psi_j}) g_i\\
    &\qquad \textbf{if}~ k = 1 ~\textbf{then} &\\
    &\qquad\quad d_i^k = g_i & \\
    &\qquad \textbf{else}&\\
    &\qquad\quad d_i^k = g_i + \gamma\ d_i^{{k-1}}\\
    &\qquad \textbf{end if}&\\
    &\qquad d_i^k = \frac{d_i^k}{\Vert d_i^k \Vert}\\
    &\qquad h_i = H d_i^k \\
    &\qquad \psi_i = \psi_i \cos\theta + d_i^k \sin\theta  \\
    &\qquad p_i = p_i \cos\theta + h_i \sin\theta  \\
    &\qquad \lambda_i =\langle{\psi_i}|{p_i}\rangle\\
    &\quad\textbf{end while}\\
    &\textbf{end for} \\
    \end{align*}
    \endgroup
    } % parbox
    } % fbox
    \caption{\MC{
    Preconditioned conjugate gradient algorithm for $H\psi_i=\lambda \psi_i$, where $i=1,\cdots,n$ is band index.
    Here $\psi_i$ is the $N$-dimensional wave function and $H$ stands for the $N\times N$ Hamiltonian. A preconditioner $K$ is applied to the residual $H\psi_i-\lambda \psi_i$ to obtain a preconditioned gradient $g_i$, which is then used to construct an improved search direction $d_i$ by Polak-Ribi\`{e}re method.~\cite{polak1971computational} Next, the new  $\psi_i$ and $\lambda_i$ are obtained.~\cite{teter1989solution}
    \MCC{Here, $\gamma$ controls the conjugacy of the search directions, leveraging the algorithm's history for optimal convergence. $\theta$ acts as an optimal step size, ensuring maximal energy reduction at each iteration.}
    }
    }
    \label{fig:pcg}
\end{figure}

The Kohn-Sham (KS) equations, when expanded in a plane-wave basis, are typically solved using iterative diagonalization techniques such as the Conjugate Gradient (CG) and Davidson methods. For example, Fig.~\ref{fig:pcg} shows the CG algorithm for solving the Kohn-Sham equations, where the input wave functions $\{\psi_i\}$ are iteratively updated to satisfy orthogonality conditions and to approximate the eigenfunctions of the given Hamiltonian matrix. Regarding the Hamiltonian matrix, it generally includes electron kinetic energy terms, the effective potential, and non-local pseudopotential terms. Importantly, the H$\psi_i$ operation that involves the multiplication of the Hamilotinan matrix with the electronic wave function $\psi_i$, is computationally expensive because mathematical operations such as Fast Fourier transforms (FFTs) and BLAS operations are needed.

In general, diagonalization of the Hamiltonian matrix is the most computationally intensive step in most KS-DFT calculations. Therefore, the development of efficient diagonalization algorithms optimized for high-performance computing (HPC) architectures is essential. In particular, recent advances in heterogeneous computing platforms have substantially accelerated scientific computations, particularly in computational chemistry and materials science.~\cite{jia2013fast, maintz2018strategies, 2020-qe} In some cases, the plane-wave basis set is capable of dealing with systems consisting of thousands of atoms or more.~\cite{2006-large,2017-qeperformance,2019-vasp-11k,2020-abinit-impact,2024-PWDFT}

\begin{figure}[ht]
    \centering
    \includegraphics[width=1.0\linewidth]{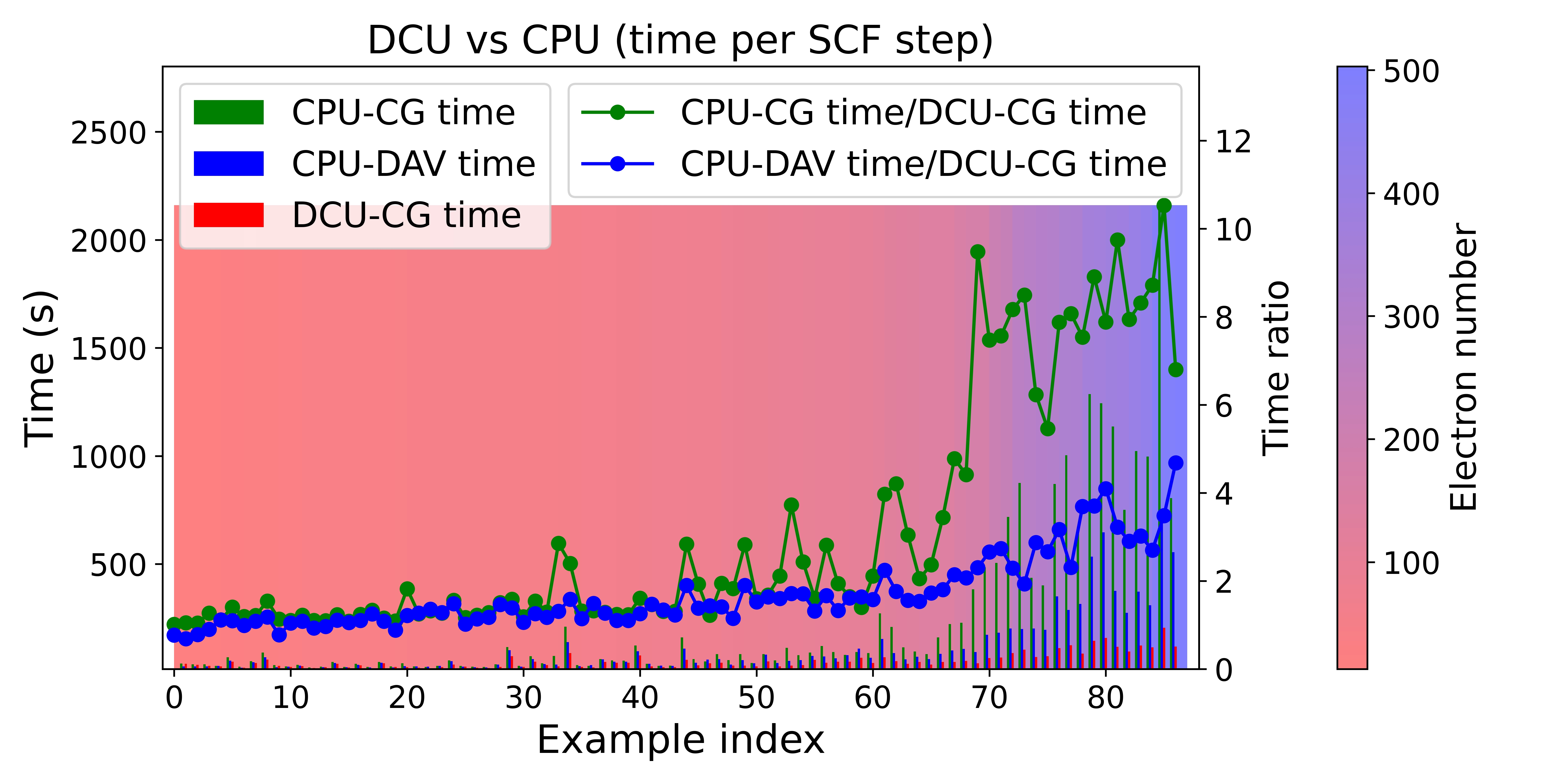}
    \caption{\MC{Computation time per SCF step for a collection of examples (with different numbers of electrons) by using the conjugate gradient (CG) method on DCU (Red), the Davidson (DAV) method on CPU (Blue), and the CG method on CPU (Green), respectively. The line chart shows their ratio. The examples are available from the link.~\cite{openlam-report} 
    %All calculations are performed in the Bohrium platform \cite{bohrium}. 
    The hardware is {\tt c32\_m64} for CPU calculations and {\tt 4*DCU} for DCU calculations.}}
    \label{fig:dcu_cpu}
\end{figure}

ABACUS employs a unified framework that operates seamlessly across diverse hardware architectures. Key linear algebra operations, such as generalized matrix-vector multiplication (\texttt{gemv}) and vector division (\texttt{vecdiv}), are encapsulated within a platform-agnostic interface, supporting execution on platforms such as CPU, GPU, and DCU. Consequently, the algorithm achieves high computational efficiency across multiple platforms. To evaluate the efficiency of heterogeneous acceleration in ABACUS, we conducted SCF calculations using plane-wave basis sets on the Bohrium cloud platform,~\cite{bohrium} comparing CPU and DCU performance at equivalent computational costs. The Davidson method was selected for these benchmarks due to its demonstrated superiority in convergence speed compared to conjugate gradient approaches. Our analysis reveals a clear scaling relationship between system size and acceleration performance, with DCU computations achieving up to 5-fold speedup for systems containing approximately 50 atoms, as illustrated in Fig.~\ref{fig:dcu_cpu}. This performance improvement grows increasingly pronounced in larger systems, underscoring the effectiveness of our heterogeneous computing framework.

% sDFT & MDFT
\subsection{Stochastic DFT}

\subsubsection{Formulas}

The stochastic density functional theory (sDFT)~\cite{13L-Baer} was proposed to circumvent the $O(N^3)$ scaling of KS-DFT calculations. Specifically, by introducing stochastic orbitals and the Chebyshev expansion scheme in the sDFT method, the electron density can be evaluated by tracing operators, bypassing the diagonalization of the Hamiltonian. Since the computational costs of tracing operations scale linearly with system size, this approach significantly enhances computational efficiency for large systems. Later, the finite-temperature sDFT~\cite{18B-Cytter} defines the Fermi-Dirac operator at finite temperature as
\begin{equation}
  \label{eq:fd_operator}
  \hat{f}_H = \frac{1}{1+\exp\left(\frac{\hat{H}-E_\mu}{k_BT}\right)},
\end{equation}
where $\hat{H}$ is the KS Hamiltonian operator and $E_{\mu}$ represents the chemical potential. In fact, the order of expansion decreases with increasing temperature, thereby enhancing the computational efficiency of sDFT at high temperatures.

In sDFT, the electron density is directly computed using the Hamiltonian operator as
\begin{equation}
  \label{eq:trace_density}
  \begin{aligned}
    \rho(\mathbf{r}) &= 2\mathrm{Tr}\left[\hat{f}_H \delta(\hat{\mathbf{r}} - \mathbf{r})\right],
  \end{aligned}
\end{equation}
where the chemical potential in $\hat{f}_H$ is determined by solving the conservation of electron number via
\begin{equation}
  \label{eq:solve_mu}
  N = \mathrm{Tr}\left[\hat{f}_{H\mu}\right].
\end{equation}
By adopting the self-consistent field method, the electron density can be self-consistently computed. The total energy can similarly be computed via
\begin{equation}
\begin{aligned}
  E_\mathrm{tot} = & 2\mathrm{Tr}\left[\hat{f}_H \hat{H}\right] - \frac{1}{2}\iint \frac{\rho(\mathbf{r}) \rho(\mathbf{r'})}{|\mathbf{r} - \mathbf{r'}|}\mathrm{d}\mathbf{r}\mathrm{d}\mathbf{r'} \\
  & - \int V_\mathrm{xc}(\mathbf{r}) \rho(\mathbf{r}) \mathrm{d}\mathbf{r} + E_\mathrm{xc}[\rho(\mathbf{r})]+E_{\rm II}.    
\end{aligned}
\label{eq:trace_energy}
\end{equation}
The entropy is similarly formulated as
\begin{equation}
   \label{eq:trace_entropy}
   S = -2k_B \mathrm{Tr}\left[\hat{f}_H \ln \hat{f}_H + (1 - \hat{f}_H) \ln (1 - \hat{f}_H)\right].
\end{equation}

However, the use of stochastic orbitals inevitably introduces stochastic errors. To achieve higher accuracy, a large number of stochastic orbitals are typically required, which substantially increases the computational time despite its linear scaling property. For instance, fragmentation approaches have been developed to compute molecular systems~\cite{14JCP-Daniel} and covalent materials,~\cite{17JCP-Aronon,19JCP-Ming} while stochastic ``embedding'' methods have been used to calculate p-nitroaniline in water.~\cite{19JCP-Li} Additionally, the energy window method proposed by Chen {\it et al.}~\cite{19JCP-Ming2} decomposes the electron density into different components based on orbital energies to reduce stochastic errors, demonstrating good generality and robustness.

For simulations of materials at extremely high temperatures, mixed stochastic-deterministic Density Functional Theory (mDFT) was proposed.~\cite{20L-White} The method combines the advantages of traditional KS-DFT and sDFT. The mDFT method retains the computational efficiency of sDFT for high-temperature systems while reducing stochastic errors using some of the deterministic Kohn-Sham orbitals.

\subsubsection{Stochastic Orbitals}

In detail, the sDFT defines a set of stochastic orbitals. For any orthogonal complete basis $\{\phi_j\}$, the stochastic orbitals $\{\chi_a\}$ are defined as
\begin{equation}
  \label{eq:initsto1}
  \onebraket{\phi_j|\chi_a} = \frac{1}{\sqrt{N_\chi}}\exp(i2\pi\theta_{j}^a),
\end{equation}
where $\{\theta_{j}^a\}$ are stochastic numbers uniformly distributed over $(0,1)$, and $N_\chi$ is the number of stochastic orbitals. In addition, stochastic orbitals can also be defined~\cite{19JCP-Ming,19JCP-Ming2,89CSSC-Hutchinson,2023-DFPM} as
\begin{equation}
  \label{eq:initsto2}
  \onebraket{\phi_j|\chi_a} = \pm\frac{1}{\sqrt{N_\chi}},
\end{equation}
each with a probability of 1/2. Notably, Baer {\it et al.}~\cite{22ARPC-Baer} demonstrated that both definitions yield the same expected values and similar variance for Hermitian matrices with comparable magnitudes of real and imaginary parts. As $N_\chi \to +\infty$, the stochastic orbitals form a complete basis with the relation of
\begin{equation}
\lim_{N_\chi\to+\infty}\sum_{a=1}^{N_\chi}\onebraket{\phi_i|\chi_a}\onebraket{\chi_a|\phi_j} = \delta_{ij},
\end{equation}
and
\begin{equation}
\lim_{N_\chi\to+\infty}\sum_{a=1}^{N_\chi}|\chi_a\rangle\langle\chi_a| = \hat{I}.
\end{equation}

%\subsubsection{Mixed Stochastic-Deterministic Orbitals}
To enhance the accuracy of stochastic orbitals, mDFT introduces a set of deterministic, orthogonal but not complete Kohn-Sham orbitals ${\varphi_i}$, along with stochastic orbitals ${\tilde{\chi}_a}$, where the stochastic orbitals are required to be orthogonal to the deterministic ones
\begin{equation}
  \label{eq:stoorthogonal}
  \ket{\tilde{\chi}_a} = \ket{\chi_a} - \sum_{i=1}^{N_\varphi}\onebraket{\varphi_i|\chi_a}\ket{\varphi_i},
\end{equation}
where $N_\varphi$ is the number of deterministic orbitals.
The mixed orbitals composed of deterministic and orthogonal stochastic orbitals also form a complete basis set as
\begin{equation}
    \lim_{N_\chi\to + \infty} \sum_{a=1}^{N_\chi} \ket{\tilde{\chi}_a}\bra{\tilde{\chi}_a} + \sum_{i=1}^{N_\varphi}\ket{\varphi_i}\bra{\varphi_i} = \hat{I}.
\end{equation}
Therefore, the trace of any operator $\hat{O}$ is given by
\begin{equation}
  \mathrm{Tr}\left[\hat{O}\right] = \lim_{N_\chi\to + \infty}\sum_{a=1}^{N_\chi} \onebraket{\tilde{\chi}_a|\hat{O}|\tilde{\chi}_a} + \sum_{i=1}^{N_\varphi}\onebraket{\varphi_i|\hat{O}|\varphi_i}.
\end{equation}
For example, the formula for electron density in mDFT is
\begin{equation}
  \label{eq:mdftdensity}
  \rho(\mathbf{r})\approx 2\sum_{a=1}^{N_\chi} \left|\left\langle\tilde{\chi}_a \left|\hat{f}_H^{1/2}\right|\mathbf{r}\right\rangle\right|^2 + 2\sum_{i=1}^{N_\varphi}f(\varepsilon_i)|\psi_i(\mathbf{r})|^2.
\end{equation}
Note that here we use a finite number of stochastic orbitals $N_\chi$, which introduces some stochastic errors.

\subsubsection{Plane-Wave-Based Implementation}

\begin{figure}
    \centering
    \includegraphics[width=\linewidth]{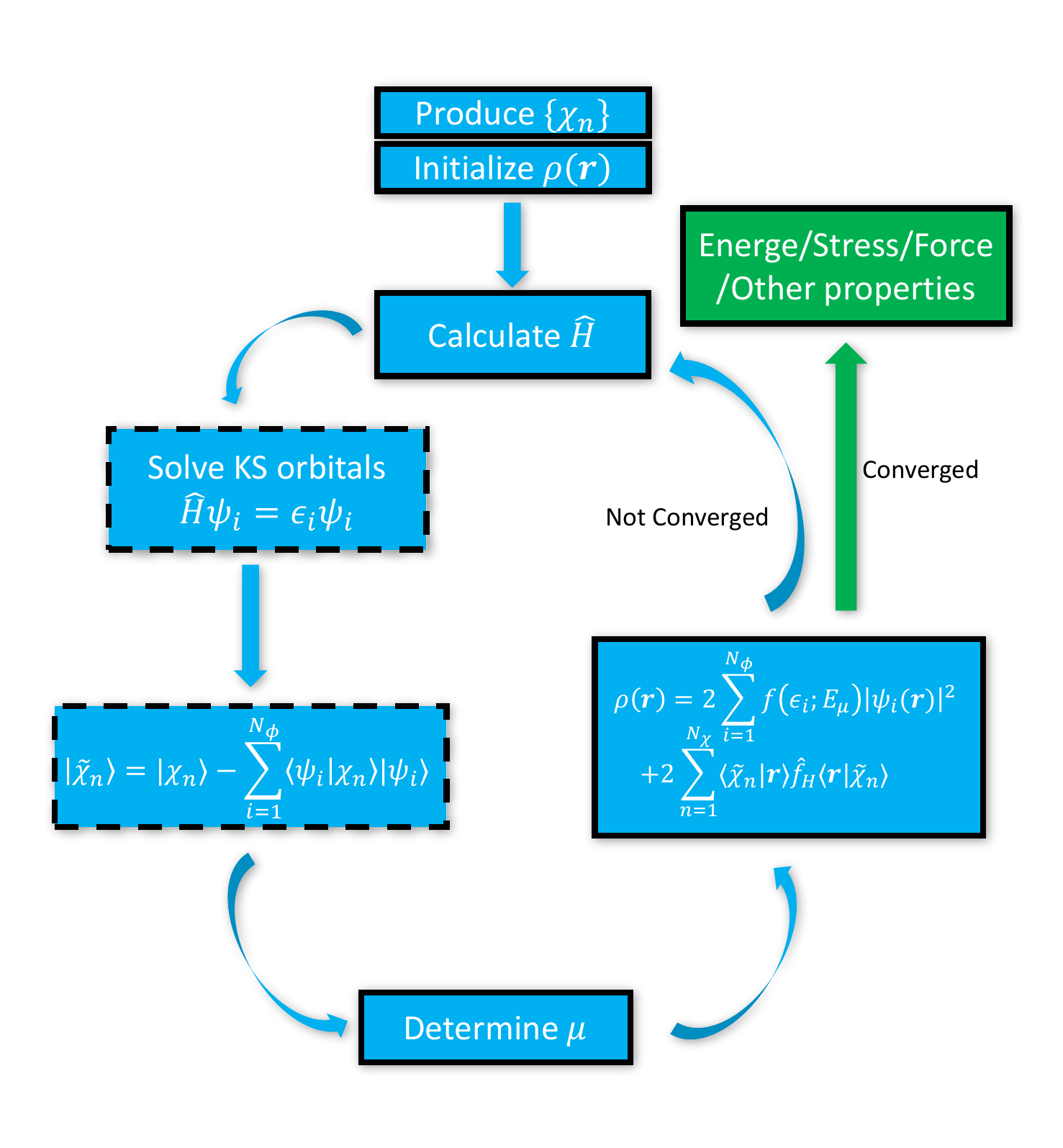}
    \caption{\MC{Flowchart of stochastic DFT in ABACUS. When the number of KS orbitals $N_\phi$ is larger than 0, it is the flowchart of mixed stochastic-deterministic DFT. When the number of KS orbitals $N_\phi$ reduces to 0, indicating that the dashed box does not exist, it degenerates into the flowchart of sDFT. The chemical potential is $\mu$.}}
    \label{fig:sdft_flow}
\end{figure}

We have implemented the sDFT and mDFT methods based on plane wave basis~\cite{22B-Liu} and periodic boundary conditions in ABACUS. Furthermore, both methods can be used with the $k$-point sampling method.
The flowchart of sDFT and mDFT is shown in Fig.~\ref{fig:sdft_flow}.

The stochastic orbitals based on the plane-wave basis can be defined as
\begin{equation}
    |\chi_{a\mathbf{k}}\rangle = \frac{1}{\sqrt{N_\chi}}\sum_{\mathbf{G}}\exp\left(i2\pi\theta_{\mathbf{k,G}}^a\right)|\mathbf{k+G}\rangle,
\end{equation}
where $\theta_{\mathbf{k,G}}^a$ are independent stochastic numbers uniformly distributed in $(0,1)$, and $N_\chi$ is the number of stochastic orbitals. Besides the sum of stochastic orbitals, the sum of $\mathbf{k}$-point should also be done. In plane-wave-based sDFT, the electron density can be computed as
\begin{equation}
\rho(\mathbf{r})=2\sum_{\mathbf{k}}W(\mathbf{k})\sum_{a\mathbf{G}}\left|\left\langle \mathbf{r}  | \mathbf{k+G} \right\rangle  \left\langle \mathbf{k+G} \left| \hat{f}_H^{1/2} \right| \chi_{a\mathbf{k}} \right\rangle\right|^2.
\end{equation}
Let us define
\begin{equation}
    \gamma_{a\mathbf{k}}(\mathbf{G}) = \left\langle \mathbf{k+G} \left| \hat{f}_H^{1/2} \right| \chi_{a\mathbf{k}} \right\rangle,
\end{equation}
so the electron density can be written as
\begin{equation}
\rho(\mathbf{r})=2\sum_{\mathbf{k}}W(\mathbf{k})\sum_{a\mathbf{G}}\left|\gamma_{a\mathbf{k}}(\mathbf{G})e^{i(\mathbf{k+G})\cdot\mathbf{r}}\right|^2.
\end{equation}

For the atomic forces, only the nonlocal potential part is different from the traditional Kohn-Sham method, which is given by
\begin{equation}
\begin{aligned}    \mathbf{F}_{I}^\mathrm{NL}=&2{\rm Tr}\Big[-i\hat{f}_{H}\sum_{\mathbf{GG'}}(\mathbf{G'-G})e^{i(\mathbf{G'-G})\cdot\mathbf{R}_I} \\ 
    & v_\tau^\mathrm{NL}(\mathbf{k+G,k+G'})|\mathbf{k+G}\rangle\langle\mathbf{k+G'}|\Big]\\
=&-2i\sum_\mathbf{k}W(\mathbf{k})\sum_{a{\mathbf{GG'}}}\gamma_{a\mathbf{k}}(\mathbf{G})\gamma_{a\mathbf{k}}(\mathbf{G'})\\
    &\times(\mathbf{G'-G})e^{i(\mathbf{G'-G})\cdot\mathbf{R}_I}v_\tau^\mathrm{NL}(\mathbf{k+G,k+G'}).
  \end{aligned}
\end{equation}
Additionally, the stress tensor for the kinetic part and the nonlocal part is evaluated differently in sDFT and mDFT when compared to the traditional Kohn-Sham method. Specifically, the kinetic part is
\begin{equation}
    \begin{aligned}
      \sigma_{\alpha\beta}^{T}=&\frac{2}{\Omega}{\rm Tr}\Big[\hat{f}_{H}\sum_{\mathbf{GG'}}(\mathbf{k+G})_\alpha(\mathbf{k+G'})_\beta|\mathbf{k+G}\rangle\delta(\mathbf{GG'})\langle\mathbf{k+G'}|\Big]\\
      =&\frac{2}{\Omega}\sum_\mathbf{k}W(\mathbf{k})\sum_{a{\mathbf{GG'}}}\gamma_{a\mathbf{k}}(\mathbf{G})(\mathbf{k+G})_\alpha\delta(\mathbf{GG'})(\mathbf{k+G'})_\beta\gamma_{a\mathbf{k}}(\mathbf{G'}),
    \end{aligned}
\end{equation}
and the nonlocal part takes the form of
\begin{equation}
    \begin{aligned}
        \sigma_{\alpha\beta}^\mathrm{NL}&=-\frac{2}{\Omega}{\rm Tr}\Big[\hat{f}_{H}\sum_{\mathbf{GG'}\tau}S_\tau(\mathbf{G'-G})\\
        &\qquad \quad \frac{\partial v_\tau^\mathrm{NL}(\mathbf{G+k,G'+k})}{\partial \epsilon_{\alpha\beta}}|\mathbf{k+G}\rangle\langle\mathbf{k+G'}|\Big]\\
        & = -\frac{2}{\Omega}\sum_{a{\mathbf{GG'}}\tau}\gamma_{a\mathbf{k}}(\mathbf{G})S_\tau(\mathbf{G'-G})\\
        &\qquad \quad \frac{\partial v_\tau^\mathrm{NL}(\mathbf{G+k,G'+k})}{\partial \epsilon_{\alpha\beta}}\gamma_{a\mathbf{k}}(\mathbf{G'}).
    \end{aligned}
\end{equation}

\begin{figure}
    \centering
    \includegraphics[width=\linewidth]{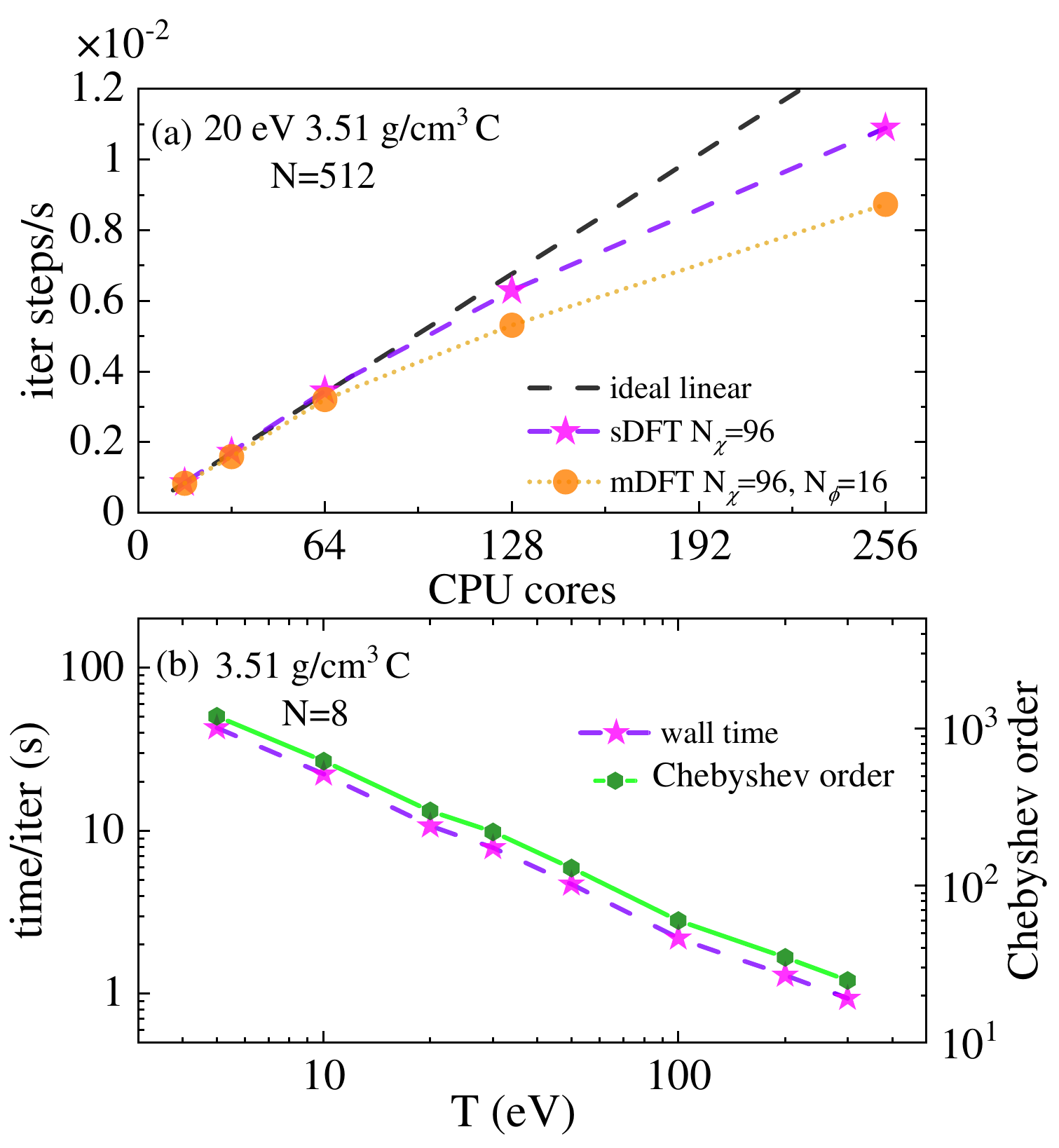}
    \caption{\MC{Efficiency tests of the sDFT and mDFT methods. (a) Parallel efficiency of sDFT and mDFT for a C system with the number of atoms being $N$=512 at the temperature of 20 eV.
    (b) Averaged wall time for an electronic iteration step of the sDFT method when calculating a C system ($N$=8) at a temperature range from 5 to 300 eV. The tested machines are Intel(R) Xeon(R) Platinum 9242 CPU @ 2.30GHz nodes.}}
    \label{fig:sdft_time}
\end{figure}

Fig.~\ref{fig:sdft_time}(a) shows the parallel efficiency of sDFT and mDFT. sDFT demonstrates excellent scalability as all stochastic orbitals are independently evaluated, while mDFT involves calculating KS orbitals and additional data communication among CPU cores, reducing its efficiency compared to sDFT. Fig.~\ref{fig:sdft_time}(b) compares the operational efficiency at different temperatures by recording the average time per electronic iteration for the sDFT method. The temperature ranges from 5 to 300 eV, and 96 stochastic orbitals were adopted. In sDFT, the order of Chebyshev polynomial expansions was selected to ensure an electron error less than $10^{-9}$. Specifically, for temperatures of 5, 10, 20, 30, 50, 100, 200, and 300 eV, Chebyshev orders of 1200, 620, 300, 220, 130, 60, 35, and 25 were used, respectively. As the temperature increases, the number of Chebyshev orders decreases, causing the wall time to drop exponentially. The sDFT and mDFT methods implemented in ABACUS has been successfully applied to study warm dense matter.~\cite{22B-Liu,24MRE-Chen,2025-chen-first}

% pw: ofdft
\subsection{Orbital-Free DFT} \label{OFDFT}

\subsubsection{Formulas}

An alternative method of KS-DFT is Orbital free density functional theory (OF-DFT),~\cite{02Carter, 12CPC-Karasiev} which achieves a more affordable computational complexity of $O(N\ln N)$ or $O(N)$ by calculating the non-interacting kinetic energy directly via charge density instead of Kohn-Sham orbitals. In OF-DFT, once the kinetic energy density functional (KEDF) is defined, the total energy is a pure functional of charge density, taking the form of
\begin{equation}
    {E_{\rm{OF}}[\rho]} = {T_{\rm{s}}}[\rho ] +  E_{\rm{ext}}[\rho] + {E_{{\rm{H}}}}[\rho ]  + {E_{{\rm{xc}}}}[\rho ] + E_{\rm{II}},    
\end{equation}
so that the ground state energy can be obtained by directly minimizing the total energy functional with optimization algorithms,~\cite{04JCTC-Jiang} such as the truncated Newton method~\cite{80MC-Nocedal-tn-of} and the conjugate gradient method.~\cite{05SJ-Hager-cg-of, 92SJ-Gilbert-cg-of}

In practice, to guarantee the conservation of electrons, we define a Lagrangian as
\begin{equation}
    L_{\rm{OF}}[\rho]=E_{\rm{OF}}[\rho]-\mu(\int{\rho(\mathbf{r})\tx{d}\mathbf{r}}-N),
\end{equation}
where $\mu$ is the Lagrangian multiplier and the chemical potential. Then, the minimum of $L_{\rm{OF}}[\rho]$ is found by optimizing $\phi(\mathbf{r}) \equiv \sqrt{\rho(\mathbf{r})}$, which guarantees the non-negativity of $\rho(\mathbf{r})$, and the variation of $L_{\rm{OF}}[\rho]$ to $\phi(\mathbf{r})$ gives
\begin{equation}
    \begin{aligned}
    \frac{\delta L}{\delta \phi}&=\frac{ \delta E_{\rm{OF}}[ \rho ] }{ \delta \phi }-2\mu\phi\\
    &=
    2(V_s + V_{\rm{ext}} + V_{\rm{H}} + V_{\rm{xc}} - \mu) \phi.
    \end{aligned}    
\end{equation}

OF-DFT has been implemented in ABACUS using plane wave basis sets. Up to now, there are five available KEDFs in ABACUS, which are Thomas-Fermi (TF),~\cite{27-Thomas-local, 27TANL-Fermi-local} von Weizsäcker (vW),~\cite{35-vW-semilocal} TF$\lambda$vW,~\cite{83pra-berk-semilocal} Wang-Teter (WT),~\cite{92B-Wang-nonlocal} Luo-Karasiev-Trickey (LKT)~\cite{18B-Luo-semilocal} KEDFs. Besides, due to the absence of Kohn-Sham orbitals, the commonly used norm-conserving pseudopotentials are usually unavailable in the field of OF-DFT unless special treatment is used.~\cite{22NC-Xu-nlps} Therefore, ABACUS employs a local pseudopotential (LPS) instead, supporting types including the bulk-derived (BLPS)~\cite{08PCCP-Huang-BLPS} and a high-quality local pseudopotential.~\cite{24JCTC-Chi-lps}

\subsubsection{Kinetic Energy Density Functional}

Given that $T_s$ is of comparable magnitude to the total energy, the accuracy of OF-DFT is heavily dependent on the form of the KEDF. Nevertheless, the development of an accurate KEDF has remained a significant challenge in the field of OF-DFT for several decades.

Several analytical KEDFs have been proposed over the past few decades,~\cite{12CPC-Karasiev, 18JMR-Witt} and they can be categorized into two main classes. First, the local and semilocal KEDFs are characterized by their kinetic energy density as a function of the charge density, its gradient, the Laplacian of the charge density, or even higher-order derivatives.~\cite{27-Thomas-local, 27TANL-Fermi-local, 35-vW-semilocal, 18B-Luo-semilocal, 18JPCL-Constantin-semilocal, 20Kang-semilocal} Second, the nonlocal KEDFs define the kinetic energy density at each point in real space as a functional of the nonlocal charge density.~\cite{92B-Wang-nonlocal, 99B-Wang-nonlocal, 10B-Huang-nonlocal, 18JCP-Mi-nonlocal, 21B-Shao-nonlocal, 23B-Sun-TKK, 24JCTC-Bhattacharjee-nonlocal} Semilocal KEDFs are generally more computationally efficient, while nonlocal KEDFs tend to provide greater accuracy. However, a universally applicable KEDF that effectively describes both simple metals and semiconductor systems remains largely elusive, and a systematic approach to its development remains to be explored.

Although the exact formula of the non-interacting kinetic energy $T_s$ remains unknown, a rigorous lower bound is provided by the von Weizsäcker (vW) KEDF,~\cite{35-vW-semilocal} which is expressed as
\begin{equation}
T_{\rm{vW}} = \frac{1}{8} \int {\frac{{\left| {\nabla \rho ({\mathbf{r}})} \right|}^2}{\rho \,({\mathbf{r}})} \,\tx{d} {\mathbf{r}}}.
\end{equation}
The remaining part of the non-interacting kinetic energy, known as the Pauli energy,~\cite{88PRA-Levy-pauli} is defined as
\begin{equation}
    T_{\rm{\theta}} = T_{{s}} - T_{\rm{vW}}.
\end{equation}
This Pauli energy can be generally written as
\begin{equation}
    T_{\rm{\theta}} = \int{\tau_{\rm{TF}} F_{\rm{\theta}} {\rm{d}} {\mathbf{r}}},
\end{equation}
where $\tau_{\rm{TF}}$ is the Thomas-Fermi (TF) kinetic energy density,~\cite{27-Thomas-local, 27TANL-Fermi-local} which is accurate for free electron gas (FEG),
\begin{equation}
\tau_{\rm{TF}} = \frac{3}{10}(3\pi^2)^{2/3} \rho^{5/3}.
\end{equation}
Here, $F_{\rm{\theta}}$ represents the enhancement factor.
The corresponding Pauli potential is then given by
\begin{equation}
V_{\rm{\theta}}(\mathbf{r}) = \delta E_{\rm{\theta}}/\delta \rho(\mathbf{r}).
\end{equation}

Notably, in a spin-degenerate system, the Pauli kinetic energy density can be expressed analytically using Kohn-Sham orbitals $\psi_i(\textbf{r})$ and corresponding occupation numbers $f_i$,~\cite{88PRA-Levy-pauli}
\begin{equation}
    \tau_{\theta}^{\rm{KS}} = \sum_{i=1}^M {f_i|\nabla \psi_i|^2} - \frac{|\nabla\rho|^2}{8\rho},
    \label{eq.pauli_e}
\end{equation}
where $i$ indexes the Kohn-Sham orbitals.
The Pauli potential is then defined by
\begin{equation}
    V_{\theta}^{\rm{KS}} = \rho^{-1} \left( \tau_{\theta}^{\rm{KS}} +2 \sum_{i=1}^M {f_i(\varepsilon_M-\varepsilon_i)\psi_i^*\psi_i}\right),
    \label{eq.pauli_p}
\end{equation}
where $\varepsilon_i$ represents the eigenvalue associated with the Kohn-Sham orbital $\psi_i(\textbf{r})$. Additionally, $M$ denotes the highest occupied state, and $\varepsilon_M$ is the eigenvalue of $\psi_M(\textbf{r})$, which corresponds to the chemical potential $E_{\mu}$.

\subsubsection{Machine Learning Based Kinetic Energy Density Functional}

In recent years, machine learning (ML) techniques have breathed new life into the development of KEDF.~\cite{12L-Snyder-mlof, 18TJCP-Seino-mlof, 20JCTC-Meyer-mlof, 21PRR-Imoto-mlof, 22JCTC-Ryczko-mlof, 24NC-Zhang-mlof, 24B-Sun-mlof} For example, Sun {\it et al.} imposed a machine learning based physical-constrained nonlocal (MPN) KEDF and implemented it in ABACUS.~\cite{24B-Sun-mlof} The MPN KEDF is designed to satisfy three exact physical constraints: the scaling law of electron kinetic energy $T_{\rm{\theta}}[\rho_{\lambda}] = \lambda^2 T_{\rm{\theta}}[\rho], \rho_{\lambda}=\lambda^3\rho(\lambda \textbf{r})$, the free electron gas (FEG) limit, and the non-negativity of Pauli energy density.

%-------------
% Figure 1 of OFDFT
%-------------
\begin{figure}[thbp]
	\centering
	\includegraphics[width=\linewidth]{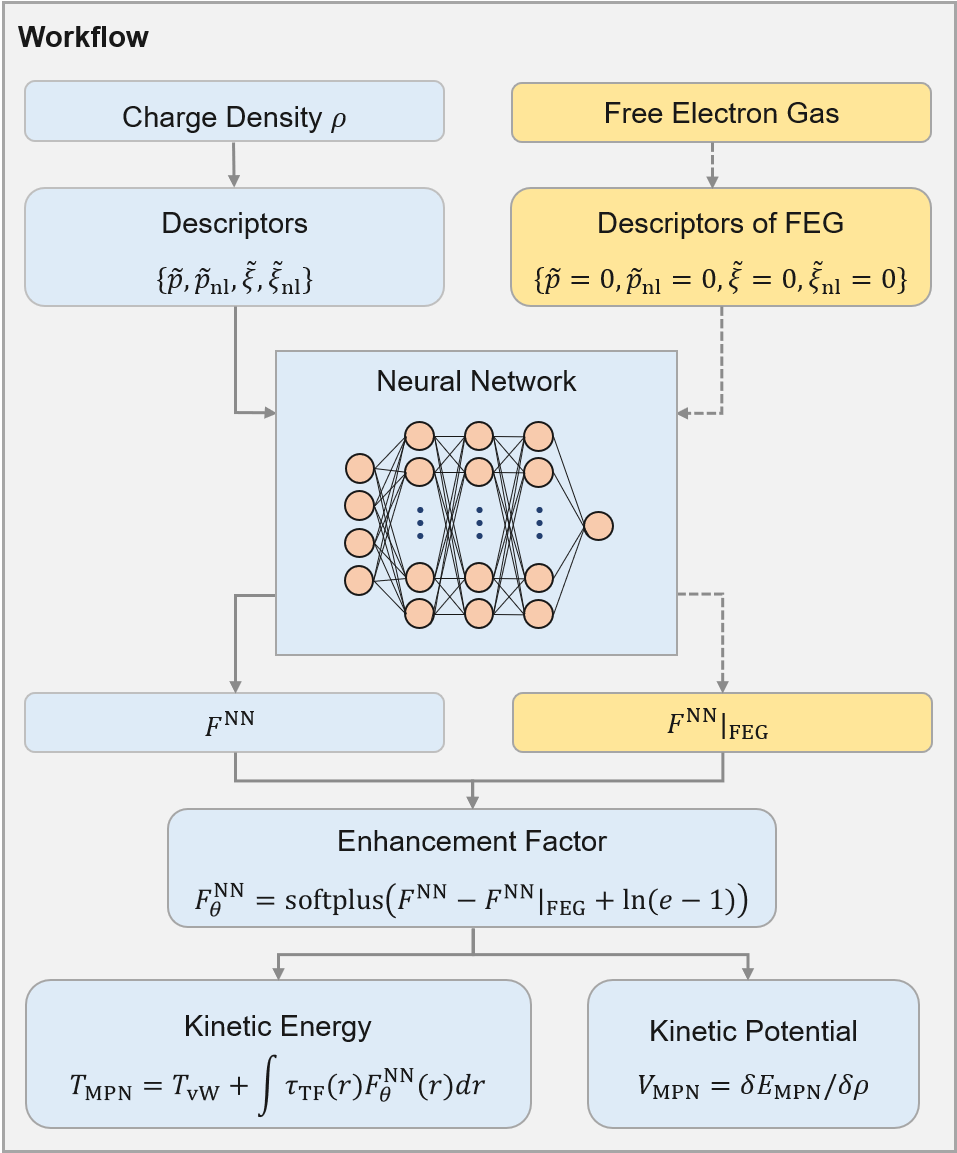}\\
	\caption{Workflow of the MPN kinetic energy density functional within the framework of orbital-free DFT.(Adapted with permission from Phys. Rev. B, 109(11): 115135 (2024). Copyright 2024 American Physical Society.)}
    \label{fig:workflow_ofdft}
\end{figure}

%-------------
% Figure 2 of OFDFT
%-------------
\begin{figure*}[htbp]
    \centering
    \begin{subfigure}{0.49\textwidth}
    \centering
    \includegraphics[width=0.95\linewidth]{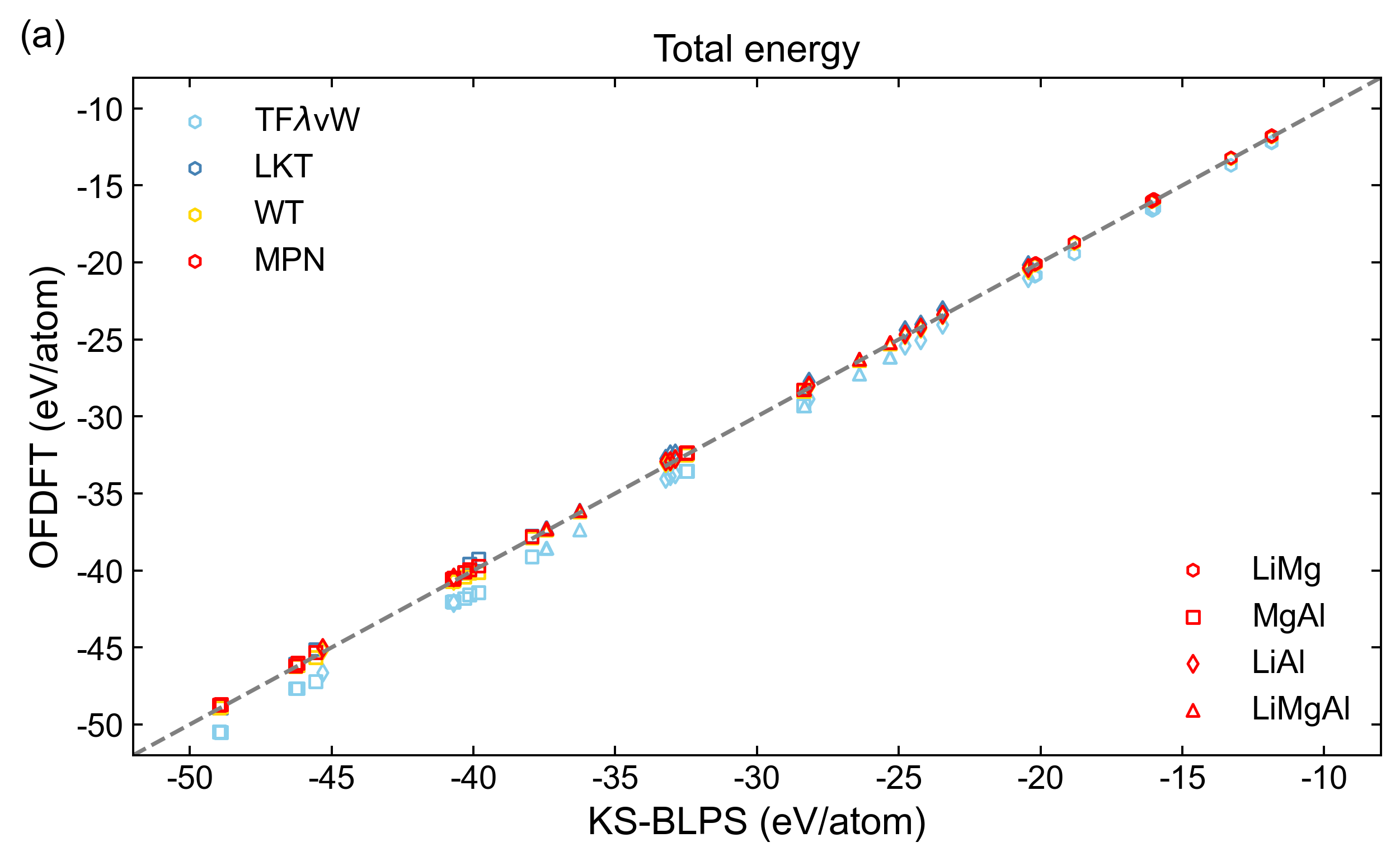}
    \label{fig:etotal}
    \end{subfigure}
    \begin{subfigure}{0.49\textwidth}
    \centering
    \includegraphics[width=0.95\linewidth]{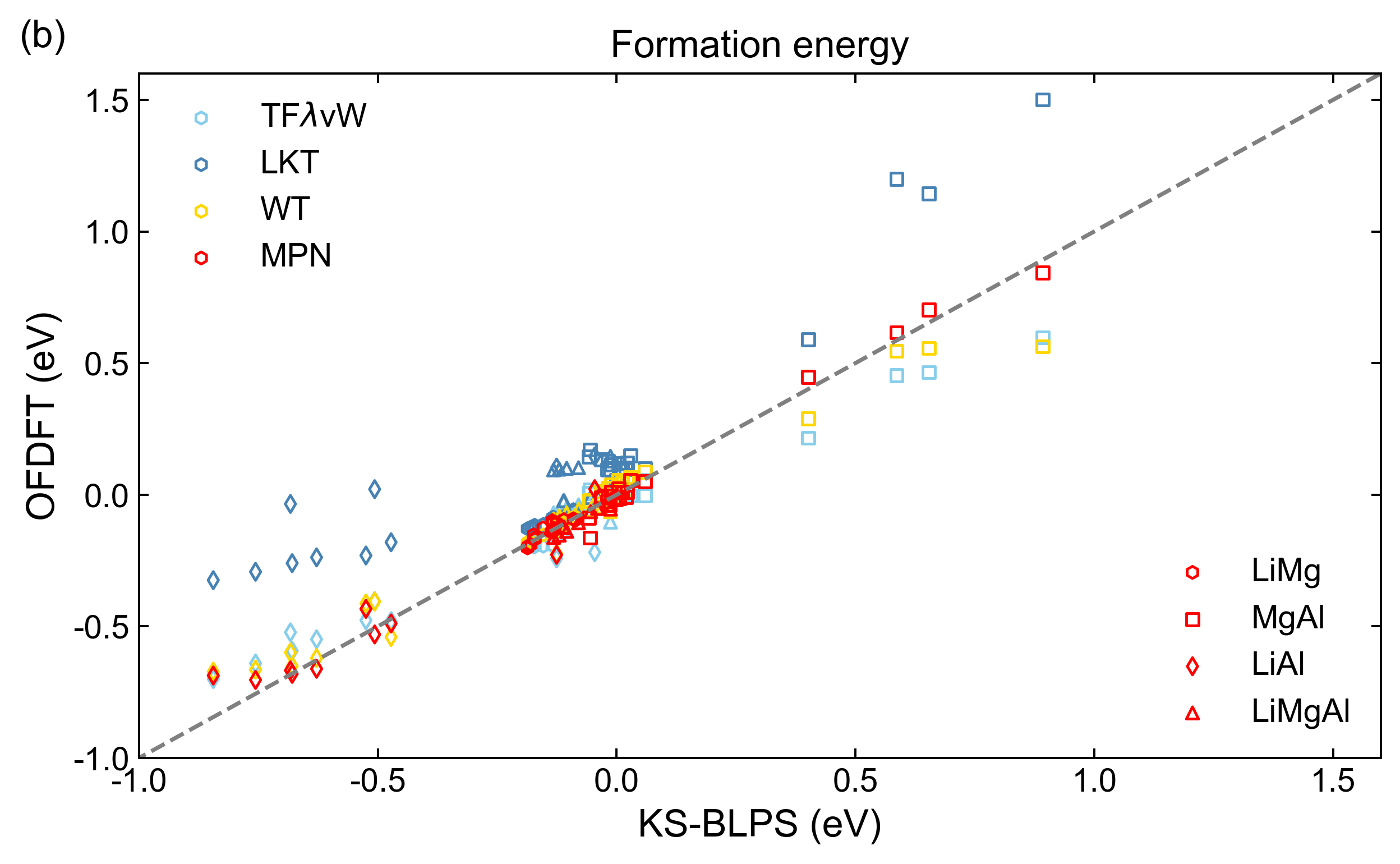}
    \label{fig:eform}
    \end{subfigure}
    \caption{(a) Total energies (in eV/atom) and (b) formation energies (in eV) of 59 alloys, including 20 Li-Mg alloys, 20 Mg-Li alloys, 10 Li-Al alloys, and 9 Li-Mg-Al alloys.
    Different colors indicate the formation energies from different KEDFs (TF$\lambda$vW, LKT, WT, and MPN), while different shapes of markers indicate different alloys. (Adapted with permission from Phys. Rev. B, 109(11): 115135 (2024). Copyright 2024 American Physical Society.)}
    \label{fig:Alloys_ofdft}
\end{figure*}

As illustrated by Fig.~\ref{fig:workflow_ofdft}, the core structure of the MPN KEDF is a neural network (NN). The output of the NN, denoted as $F^{\rm{NN}}(\mathbf{r})$, represents the enhancement factor $F_{\rm{\theta}}$ for each real-space grid point $\mathbf{r}$. To ensure that the calculated Pauli energy and potential adhere to the FEG limit and the non-negativity of the Pauli energy density, the enhancement factor for the Pauli energy is defined as:
\begin{equation}
    F_{\rm{\theta}}^{\rm{NN}} = {\rm{softplus}}\left(F^{\rm{NN}} - F^{\rm{NN}}|_{\rm{FEG}} + \ln{(e-1)}\right),
\end{equation}
where ${\rm{softplus}}(x)=\ln(1+e^x)$ is an activation function commonly used in machine learning, which satisfies ${\rm{softplus}}(x)\geq0$ and ${\rm{softplus}}(x)|_{x=\ln(e-1)}=1.$ By construction, the non-negativity constraint is satisfied
\begin{equation}
    F_{\rm{\theta}}^{\rm{NN}} \geq 0, 
\end{equation}
and the FEG limit, where the enhancement factor should be 1, is also met:
\begin{equation}
    \begin{aligned}
    F_{\rm{\theta}}^{\rm{NN}}|_{\rm{FEG}} &= {\rm{softplus}}\left(F^{\rm{NN}}|_{\rm{FEG}} - F^{\rm{NN}}|_{\rm{FEG}} + \ln{(e-1)}\right)
    \\
    &= 1.
    \end{aligned}
\end{equation}
Furthermore, the selection of the kernel function and descriptors ensures that once the FEG limit of the Pauli energy is satisfied, the FEG limit of the Pauli potential is automatically fulfilled. The scaling law is ensured by the definition of the descriptors, which will be introduced subsequently.

As displayed in Fig.~\ref{fig:workflow_ofdft}, the NN for the MPN KEDF uses four descriptors $\{\tilde{p}, \tilde{p}_{\rm{nl}}, \tilde{\xi}, \tilde{\xi}_{\rm{nl}}\}$ as inputs. First, the semilocal descriptor $\tilde{p}$ is defined as the normalized dimensionless gradient of the charge density
\begin{equation}
 \tilde{p}(\mathbf{r}) = \tanh{\Big(0.2 p(\mathbf{r})\Big)}, 
\end{equation}
where the parameter $p(\mathbf{r})$ is given by:
\begin{equation}
 p(\mathbf{r}) = |\nabla \rho(\mathbf{r})|^2 / \Big[2(3\pi^2)^{1/3} \rho^{4/3}(\mathbf{r})\Big]^2.
\end{equation}
The corresponding nonlocal descriptor $\tilde{p}_{nl}$ is defined as
\begin{equation}
    \tilde{p}_{\rm{nl}}(\mathbf{r}) = \int{w(\mathbf{r}-\mathbf{r}')\tilde{p}(\mathbf{r}'){\rm{d}} \mathbf{r}'},
\end{equation}
where $w(\mathbf{r}-\mathbf{r}')$ is a kernel function similar to the WT kernel function.~\cite{92B-Wang-nonlocal} This kernel function is defined in reciprocal space as
\begin{equation}
    w(\eta) = {{\left( {\frac{1}{2} + \frac{1-\eta^2}{4\eta}\ln \left| {\frac{1 + \eta}{1 - \eta}} \right|} \right)}^{ - 1}} - 3\eta^2 - 1,
\end{equation}
where $\eta = \frac{k}{2k_{\rm{F}}}$ is a dimensionless reciprocal space vector, and $k_{\rm{F}} = (3\pi^2\rho_0)^{1/3}$ is the Fermi wave vector, with $\rho_0$ representing the average charge density.

The third and fourth nonlocal descriptors $\tilde{\xi}$ and $\tilde{\xi}_{\rm{nl}}$ represent the distribution of charge density. These descriptors are defined as
\begin{equation}
    \tilde{\xi}(\mathbf{r}) = \tanh{\left(\frac{\int{w(\mathbf{r}-\mathbf{r}')\rho^{1/3}(\mathbf{r}'){\rm{d}} \mathbf{r}'}}{\rho^{1/3}(\mathbf{r})}\right)},\\
\end{equation}
and
\begin{equation}
    \tilde{\xi}_{\rm{nl}}(\mathbf{r}) = \int{w(\mathbf{r}-\mathbf{r}')\tilde{\xi}(\mathbf{r}'){\rm{d}} \mathbf{r}'}.
\end{equation}
Here, $w(\mathbf{r}-\mathbf{r}')$ is the same kernel function used in the definition of the other nonlocal descriptors.

The loss function for the MPN KEDF is defined as
\begin{equation}
    \begin{aligned}
        L=&\frac{1}{N}\sum_{\mathbf{r}}{\left[ \left(\frac{F_\theta^{\rm{NN}}- F^{\rm{KS}}_{\theta}}{\bar{F}^{\rm{KS}}_{\theta}}\right)^2 +
      \left(\frac{V_\theta^{\rm{MPN}} - V^{\rm{KS}}_{\theta}}{\bar{V}^{\rm{KS}}_{\theta}}\right)^2 \right]}\\
      &+ \left[F^{\rm{NN}}|_{\rm{FEG}}-\ln(e-1)\right]^2,
    \end{aligned}
\end{equation}
where $N$ is the total number of grid points, and $\bar{F}^{\rm{KS}}_{\theta}$ ($\bar{V}^{\rm{KS}}_{\theta}$) represents the mean value of $F^{\rm{KS}}_{\theta}$ ($V^{\rm{KS}}_{\theta}$). The first term in the loss function accounts for the discrepancy between the predicted Pauli energy enhancement factor $F_\theta^{\rm{NN}}$ and the reference KS Pauli energy enhancement factor $F^{\rm{KS}}_{\theta}$. The second term ensures that the predicted Pauli potential $V_\theta^{\rm{MPN}}$ closely matches the KS Pauli potential $V^{\rm{KS}}_{\theta}$. This term is crucial because the Pauli potential plays a significant role in determining the optimization direction and step size during OF-DFT calculations. The final term is a penalty term designed to minimize the magnitude of the FEG correction, thereby enhancing the stability of the MPN KEDF.

The training set for the MPN KEDF includes eight metallic structures, specifically bcc Li, fcc Mg, fcc Al, as well as five alloys: $\rm{Li_3 Mg}$ (mp-976254), LiMg (mp-1094889), $\rm{Mg_3 Al}$ (mp-978271), $\beta''$ $\rm{MgAl_3}$,~\cite{03MSMSE-Carling-mgal} $\rm{LiAl_3}$ (mp-10890).
The numbers in parentheses correspond to the Materials Project IDs.~\cite{13APL-Jain-MP} To evaluate the precision and transferability of the MPN KEDF, a test set was constructed using 59 alloys from the Materials Project database.~\cite{13APL-Jain-MP} This testing set includes 20 Li-Mg alloys, 20 Mg-Li alloys, 10 Li-Al alloys, and 9 Li-Mg-Al alloys. The total energies and formation energies of 59 alloys as calculated by various KEDFs in OF-DFT are presented in Fig.~\ref{fig:Alloys_ofdft}. As depicted in Fig.~\ref{fig:Alloys_ofdft}(a), the TF$\lambda$vW KEDF systematically underestimates the total energies compared to the results from KS-DFT, leading to a substantial mean absolute error (MAE) of 0.934 eV/atom. In contrast, the LKT KEDF demonstrates improved performance with a reduced MAE of 0.145 eV/atom. The nonlocal WT KEDF further enhances accuracy, achieving an MAE of 0.043 eV/atom. Although the MPN KEDF has a higher MAE of 0.123 eV/atom compared to the WT KEDF, it still outperforms both the TF$\lambda$vW and LKT KEDFs. Fig.~\ref{fig:Alloys_ofdft}(b) illustrates the formation energies. The LKT KEDF consistently overestimates the values compared to those obtained by KS-DFT and yields a high MAE of 0.166 eV. This is significantly larger than the MAEs achieved by the TF$\lambda$vW KEDF (0.051 eV) and the WT KEDF (0.035 eV). Notably, the MPN KEDF demonstrates superior performance with an even lower MAE of 0.028 eV, outperforming the WT KEDF.

While the MPN KEDF has shown promising results for simple metals and their alloys,~\cite{24B-Sun-mlof} its performance remains limited for semiconductors. To address this, Sun {\it et al.} developed a multi-channel MPN KEDF~\cite{2024-Sun-multi} that incorporates information from multiple real-space length scales.

\section{Methods in Numerical Atomic Orbital Basis}
\label{sec:lcao}
% NAOs
\subsection{Numerical Atomic Orbitals} \label{sec:nao}

The efficiency and accuracy of first-principles methods are largely determined by the basis sets. As mentioned in Sec.~\ref{sec:pw}, although the accuracy of the plane wave basis in solving Kohn-Sham equation can be systematically improved by increasing the kinetic energy cutoff, its computational costs are generally formidable for systems consisting of hundreds or thousands of atoms. On the other hand, atomic orbital basis sets are generally more efficient than plane wave basis sets, at the cost of a slight loss in accuracy. In particular, numerical atomic orbitals (NAOs) have become a competitive option for basis set types, especially for calculating large systems consisting of hundreds or even thousands of atoms. The NAO basis set also supports the linear scaling algorithms due to the strict locality feature. Over the past several decades, several methods have been proposed~\cite{Talman1978,Sharafeddin1992,Toyoda2010,Cerioni2012,Soler2002,Blum2009} to perform efficient numerical integrations for NAOs.

Unlike the plane-wave basis, there is no unique way to construct NAOs. For example, a popular method involves solving isolated atoms subject to confining potentials.~\cite{Sankey1989,Porezag1995,Horsfield1997,Junquera2001,Soler2002} Based on confining potentials, Blum \textit{et al.}~\cite{Blum2009} proposed to generate NAOs by iteratively picking up basis functions one by one from a pool of predefined candidates, to seek the best improvement of a target energy. Alternatively, Ozaki~\cite{Ozaki2003,Ozaki2004} suggested that eigenfunctions of isolated atoms in confining potentials serve as ``primitive orbitals'' in terms of which NAOs are expanded, and expansion coefficients can be optimized with self-consistent calculations. Apart from the above energy-based methods, another class of methods constructs NAOs towards some reference states. For example, Sanchez-Portal \textit{et al.}~\cite{Sanchez-Portal1995,Sanchez-Portal1996} proposed to optimize the following ``spillage'' function defined as
\begin{align}
    \mathcal{S} &= \sum_{n\mathbf{k}} \ev{(1-\hat{P}(\mathbf{k})}{\psi_n(\mathbf{k})}, 
\end{align}
where $\{\psi_n(\mathbf{k})\}$ denotes reference states from plane-wave calculations, and $\hat{P}(\mathbf{k})$ is a projection operator defined as
\begin{align}
    \hat{P}({\mathbf{k}}) =\sum_{\mu\nu} \ket{\phi_{\mu}(\mathbf{k})} S_{\mu\nu}^{-1}(\mathbf{k})\bra{\phi_{\nu}(\mathbf{k})}.
\end{align}
Here $S_{\mu\nu}(\mathbf{k})$ is the overlap matrix. In the original work, NAOs are chosen to be combinations of pseudo-atomic orbitals (eigenfunctions of isolated atoms with pseudo-potentials) or Slater-type orbitals. Based on the spillage formalism, Chen {\it et al.}~\cite{Chen2010} proposed to construct NAOs with localized spherical Bessel functions as basis,~\cite{Haynes1997} and reference systems are chosen to be a series of isolated dimers/trimers of variable bond lengths. Recently, Lin {\it et al.}~\cite{Lin2021} further improved the basis by introducing a gradient term to the original spillage
\begin{align}
    \mathcal{S}^{\text{LRH}} = \mathcal{S} + \sum_{n\mathbf{k}} \left\|\hat{p}(1-\hat{P}(\mathbf{k}))\ket{\psi_n(\mathbf{k})} \right\|^2,
\end{align}
where $\hat{p}$ is the momentum operator. To use the NAO basis set in ABACUS, the collection of NAO basis sets paired with SG15 pseudopotentials~\cite{Schlipf2015} is publicly available.~\cite{NAOlink}

% LCAO integration techniques
\subsection{Kohn-Sham Equation in NAO Basis} \label{sec:lcao:efs} 

With the usage of localized basis set, the Kohn-Sham equation for a given $\mathbf{k}$ point in the Brillouin zone becomes a generalized eigenvalue problem that takes the form of
\begin{equation}
\boldsymbol H(\mathbf{k})\boldsymbol C(\mathbf{k})=\boldsymbol S(\mathbf{k})\boldsymbol C(\mathbf{k})\boldsymbol E(\mathbf{k}),
\label{eq:general-eigenproblem}
\end{equation}
in which $\boldsymbol H(\mathbf{k})$, $\boldsymbol C(\mathbf{k})$, and $\boldsymbol S(\mathbf{k})$ are the Hamiltonian matrix, the electronic wave function coefficients of NAOs, and the overlap matrix, respectively. The $\boldsymbol E(\mathbf{k})$ is a diagonal matrix with KS eigenvalues. The overlap matrix element $S_{\mu\nu}(\mathbf{k})$ is given by
\begin{equation}
S_{\mu\nu}(\mathbf{k}) = \sum_{\mathbf{R}} \langle \phi_{\mu0} | \phi_{\nu\mathbf{R}} \rangle e^{i\mathbf{k}\cdot\mathbf{R}} , 
\label{eq:overlap_k}
\end{equation}
where $\phi_{\nu\mathbf{R}}(\mathbf{r})$ is a NAO centered in the cell with lattice vector $\mathbf{R}$.

In general, the terms in Hamiltonian matrix $H(\mathbf{k})$ are constructed in two ways, i.e., the two-center integrals and the grid integral techniques. Given an operator $\hat{O}$, the two-center integral calculates
\begin{align}
O_{\mu\nu}(\mathbf{R})&=\int\phi_{\mu}(\mathbf{r})\hat{O}\phi_{\nu}(\mathbf{r}-\mathbf{R})\mathrm{d}\mathbf{r},
\end{align}
where functions $\phi$ (basis or any "projector") centered at atoms spaced by $\mathbf{R}$ are distinguished by $\mu$ and $\nu$. The overlap matrix $S_{\mu\nu}(\mathbf{R})$ and the kinetic energy matrix $T_{\mu\nu}(\mathbf{R})$ are evaluated directly with this form, while the representation of the non-local part of pseudopotential requires the calculation of
\begin{equation}
    V^\mathrm{NL}_{\mu\nu}(\mathbf{R})=\sum_{Iij}D_{ij}^I\langle\phi_{\mu\mathbf{0}}|\beta_i^I\rangle \langle\beta_j^I|\phi_{\nu\mathbf{R}}\rangle.
    \label{eq:vnl-lcao}
\end{equation}
Here, indices $i$ and $j$ run over all projects for atom $I$. Equation \ref{eq:general-eigenproblem} is solved independently for each $\mathbf{k}$ point. 

In the context of LCAO-based DFT calculations within the ABACUS package, real-space uniform grid integrals hold significant importance as they are key to multiple critical computational steps. For example, the grid integration is used to construct the local potential term of the Hamiltonian and the electron density in real space, as well as the Pulay force term when calculating the atomic forces. To optimize these grid integrals, we partition real-space grids across MPI processes and use parallel matrix operations, which substantially accelerate the overall computation.~\cite{zhang2024gpu}

The representation of Hamiltonian or any operator within the momentum space is obtained via a "folding" operation
\begin{equation}
    H_{\mu\nu}(\mathbf{k}) = \sum_{\mathbf{R}}{H_{\mu\nu}(\mathbf{R})}e^{i\mathbf{k}\cdot\mathbf{R}},
\end{equation}
in which $\mathbf{R}$ always runs over all valid neighboring cells. After diagonalization of the Hamiltonian matrix, the eigenvalues and corresponding Kohn-Sham wave functions can be obtained.

To solve the generalized eigenvalue problem in Eq.~\ref{eq:general-eigenproblem}, ABACUS supports numerical libraries such as ScaLAPACK~\cite{scalapack} and the Eigenvalue SoLvers for Petaflop Applications (ELPA).~\cite{elpa} ABACUS also provides the pole expansion and selected inversion (PEXSI) method as a low-scaling alternative (see Sec.~\ref{sec:pexsi} for details). Additionally, we have implemented GPU acceleration through specialized numerical libraries, such as the GPU-enabled ELPA implementation.~\cite{yu2021gpu} After diagonalization, the density matrix can be constructed via the formula
\begin{equation}
\rho_{\mu\nu}(\mathbf{R})=\frac{1}{N_\mathbf{k}}\sum_{n\mathbf{k}}
f_{n\mathbf{k}}c_{n\mu,\mathbf{k}}c_{n\nu,\mathbf{k}}^{\ast}e^{-i\mathbf{k}\cdot\mathbf{R}},
\label{eq:density_matrix}
\end{equation}
where $f_{n\mathbf{k}}$ is the occupation, and the electronic wave functions are $\{c_{n\mu,\mathbf{k}}\}$. Subsequently, the electron charge density $\rho(\mathbf{r})$ can be evaluated through real-space grids.

\begin{figure}
    \centering
    \begin{subfigure}{\linewidth}
        \includegraphics[width=1.05\linewidth]{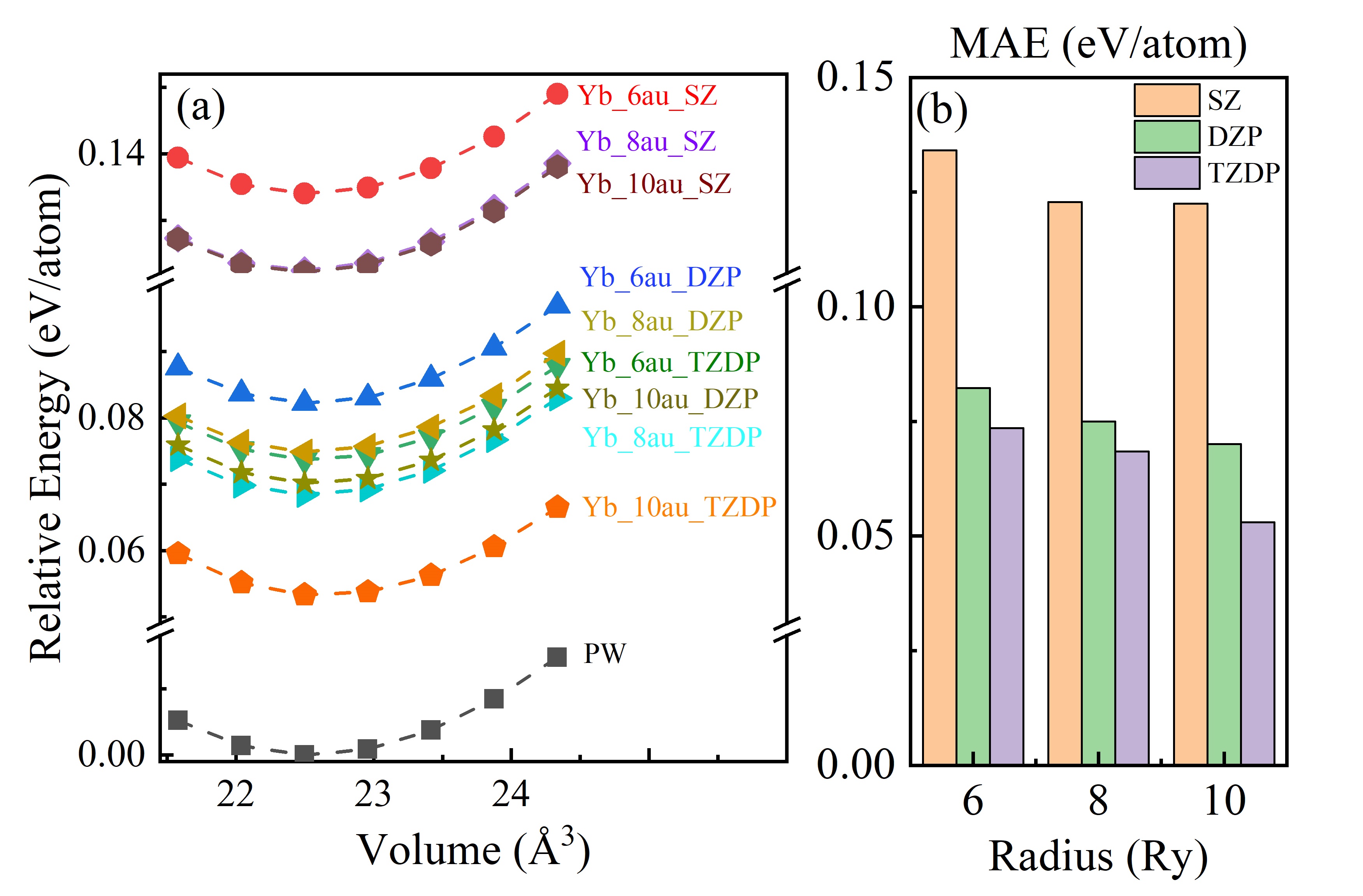}
    \end{subfigure}%
    \caption{\MC{(a) Comparison on energies calculated with NAO and PW basis sets for the $\rm Li_{12} \rm Yb_4 \rm Cl_{24}$ system calculated using Yb SZ/DZP/TZDP orbitals with different orbital cutoffs. (b) Mean absolute error (MAE, in eV/atom) of the energies between NAO and PW basis sets for different volumes.}}
    \label{fig:LCAO-orb-EOS-Yb}
\end{figure}

% benchmarking
To assess the accuracy of the NAO basis set, we conducted a systematic comparison with PW results. A series of NAO basis sets—including Single-$\zeta$ (SZ), Double-$\zeta$ plus polarization functions (DZP), and Triple-$\zeta$ plus double polarization functions (TZDP)—were generated with varying cutoff radii based on the Pseudo-Dojo v0.4 (3plus) pseudopotential. We then performed SCF calculations using these basis sets, and the resulting total energies are presented in Fig.~\ref{fig:LCAO-orb-EOS-Yb}. Our analysis demonstrates that incorporating a larger number of atomic orbitals and employing higher cutoff values significantly reduces the discrepancy between NAO and PW results. Notably, the TZDP basis set with a cutoff of 10 a.u. yields the closest agreement with the PW results, exhibiting a minimal deviation of approximately 0.0053 eV/atom.

\subsection{Forces and Stress}

To obtain the analytical expression for atomic forces and stress, one must start with the analytical expression for the total energy
\begin{equation}
\begin{aligned}
E_{\rm tot} & = \sum_{\mu\nu\mathbf{R}}\rho_{\nu\mu}(\mathbf{R})\Big[T_{\mu\nu}(\mathbf{R})+V^{\rm NL}_{\mu\nu}(\mathbf{R})\Big]+\int V_{\rm local}(\mathbf{r})\rho(\mathbf{r})\mathrm{d} \mathbf{r} \\
&+\frac{1}{2}\iint \frac{\rho(\mathbf{r}) \rho\left(\mathbf{r}^{\prime}\right)}{\left|\mathbf{r}-\mathbf{r}^{\prime}\right|} \mathrm{d} \mathbf{r} \mathrm{d} \mathbf{r}^{\prime}
+E_{\mathrm{xc}}[\rho(\mathbf{r})] +E_{\rm II},
%\frac{1}{2} \sum_{I \neq J} \frac{Z_I Z_J e^2}{\left|R_I-R_J\right|},
\label{eq:lcao_etot2}
\end{aligned}
\end{equation}
where the kinetic and nonlocal pseudopotential matrices are respectively labeled as $T_{\mu\nu}(\mathbf{R})$ and $V_{\mu\nu}^{\rm NL}(\mathbf{R})$, while the local pseudopotential is $V_{\rm local}$ and the ionic energy $E_{\rm II}$ is given by the Ewald method.~\cite{electronic-martin} It should be noted that ABACUS uses Eq.~\ref{eq:ksenergy_band} to calculate the total energy in practical computations. For more information about evaluating force and stress terms in ABACUS, we refer the readers to a recent review work in Ref.~\onlinecite{lin_Initio_2024}.

The atomic forces are obtained by direct differentiation of the total energy $E_{\rm tot}$ with respect to atomic positions
\begin{align}
\bm{F}_{I} &= -\sum_{\mu\nu\mathbf{R}}\rho_{\nu\mu}(\mathbf{R})\frac{\partial H_{\mu\nu}(\mathbf{R})}{\partial \mathbf{R}_I} - \sum_{\mu\nu}\frac{\partial\rho_{\nu\mu}(\mathbf{R})}{\partial \mathbf{R}_I}H_{\mu\nu}(\mathbf{R}) \nonumber \\
&= \bm{F}_I^{\rm FH} + \bm{F}_I^{\rm Pulay} + \bm{F}_I^{\rm Orth} + \bm{F}_I^{\rm Ewald}.
\label{eq:force_lcao}
\end{align}
Here we utilize the Feynman-Hellmann theorem~\cite{35AP-Hellmann,39-Feynman} and divide the total forces into four components. First, the Feynman-Hellmann force $\bm{F}_I^{\rm FH}$ describes the contribution of the partial derivatives of the Hamiltonian operator. In this term, the kinetic and nonlocal pseudopotential terms are evaluated through the two-center integration technique, while the local pseudopotential contribution is evaluated through the plane wave basis as shown in Eq.~\ref{eq:localforce}. Second, the Pulay force $\bm{F}_I^{\rm Pulay}$ describes the partial derivatives of the NAO basis set with respect to ionic positions, as unlike the plane wave basis set, the NAO basis set changes with atomic positions. Third, the orthogonal force $\bm{F}_I^{\rm Orth}$ arises from the nonorthogonality of the NAO basis set, and an energy density matrix must be evaluated to yield this term. Finally, the contribution of ionic interactions to atomic forces is included in $\bm{F}_I^{\rm Ewald}$. The stress tensor defined in Eq.~\ref{eq:stress} can also be derived using the NAO basis set, the formulas are similar to the above force formulas. For more details, please refer to Ref.~\onlinecite{lin_Initio_2024}.

\begin{figure}[ht]
    \includegraphics[width=1.0\linewidth]{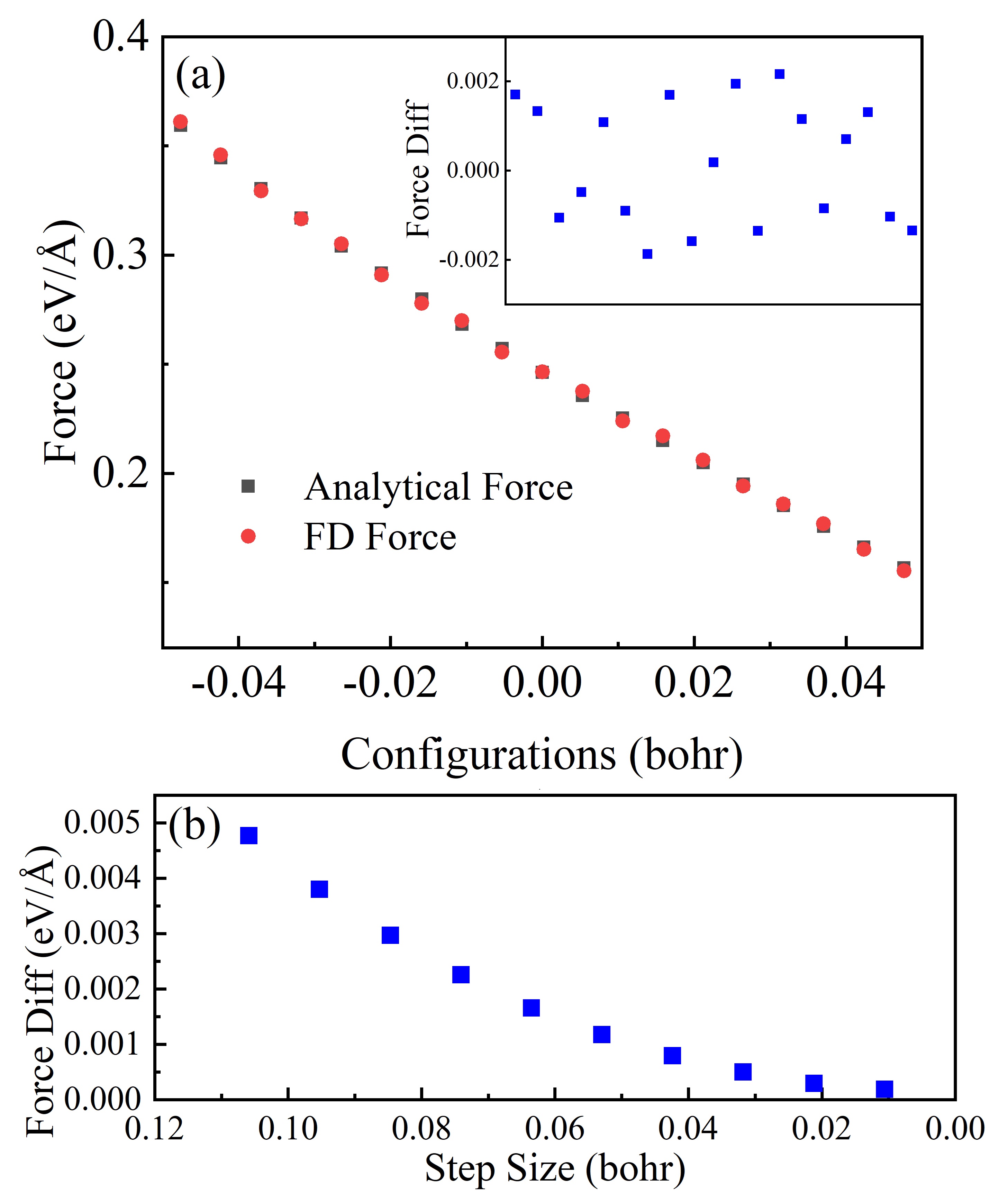}
    \caption{\MC{By using the NAO basis set, the finite difference (FD) tests of atomic forces on the $x$-axis of BCC Fe. (a) Analytical forces and the forces calculated by the FD method along the $x$-direction for different configurations. The zoom-in figure shows the difference between them. (b) Deviations between FD values and analytical values along the Fe $x$-direction for different step sizes used in the FD method.}}
    \label{fig:LCAO-FD}
\end{figure}

To evaluate the consistency of atomic forces using the NAO basis, we performed finite-difference tests on a body-centered cubic (BCC) iron (Fe) system. Specifically, a single Fe atom was displaced along the $x$-axis in both positive and negative directions by a defined step size, and the corresponding energy changes were computed. Fig.~\ref{fig:LCAO-FD}(a) compares the atomic forces derived from the finite-difference method (using a step size of 0.02 Bohr) with analytically calculated forces at varying atomic positions. The inset illustrates the residual differences between the two methods. For the Fe system, the observed force discrepancy is on the order of 0.002 eV/\AA. This deviation is influenced by several factors, including the precision of the finite-difference step size and the numerical accuracy of the SCF calculations. Further investigation of the step-size dependence (Fig.~\ref{fig:LCAO-FD}(b)) reveals that the discrepancy diminishes as the step size decreases, asymptotically approaching zero. This behavior confirms the consistency between the NAO-implemented forces and the energy landscape in ABACUS.

% LCAO hybrid
\subsection{Hybrid Functional}\label{sec:hybrid}

Hybrid density functionals (HDFs) formulated within the generalized KS framework~\cite{seidl_generalized_1996} belongs to the fourth rung of the Jacob's ladder.~\cite{perdew_jacobs_2001} It overcomes the drawbacks of the local and semi-local XC functionals, namely the self-interaction errors~\cite{perdew_self-interaction_1981,mori-sanchez_localization_2008} related to the underestimation of the band gaps in solids. The ABACUS package supports pure exact exchange (EXX) calculations, specifically the Hartree-Fock (HF) method. Building upon this capability, we have implemented several widely-used hybrid functionals, including the Heyd-Scuseria-Ernzerhof (HSE),~\cite{HSE_heyd_hybrid_2003} PBE0,~\cite{PBE01_ernzerhof_assessment_1999,PBE02_adamo_toward_1999} and SCAN0.~\cite{SCAN0_hui_scan-based_2016} The local (semilocal) components of these functionals are computed using the LibXC library,~\cite{2018_libxc} while the exact (Fock) exchange energy is evaluated through two-electron integrals of the Kohn-Sham orbitals.

The exact exchange Hamiltonian under the representation of NAOs basis is given by
\begin{equation}\label{eq:H}
\begin{aligned}
    H^\text{EXX}_{\mu\nu}(\mathbf{R})= & \sum_{\lambda\sigma}\sum_{\mathbf{R}_1}\sum_{\mathbf{R}_2} 
    \rho_{\lambda\sigma}(\mathbf{R}_2-\mathbf{R}_1)(\phi^\mathbf{0}_{\mu}\phi^{\mathbf{R}_1}_{\lambda}|\phi^{\mathbf{R}_2}_{\sigma}\phi^{\mathbf{R}}_{\nu}),
\end{aligned}
\end{equation}
in which the real space density matrix $\rho_{\lambda\sigma}(\mathbf{R})$ is defined in Eq.~\ref{eq:density_matrix} and the 4-index integrals are defined as
\begin{equation}
\begin{aligned}
    (\phi^\mathbf{0}_{\mu}\phi^{\mathbf{R}_1}_{\lambda}&|\phi^{\mathbf{R}_2}_{\sigma}\phi^{\mathbf{R}}_{\nu}) \\
    & = \iint \phi_{\mu}^{\mathbf{0}}(\mathbf{r})\phi_{\lambda}^{\mathbf{R}_1}(\mathbf{r})v(\mathbf{r-r'})\phi_{\sigma}^{\mathbf{R}_2}(\mathbf{r'})\phi_{\nu}^{\mathbf{R}}(\mathbf{r'})\mathrm{d}\mathbf{r}\mathrm{d}\mathbf{r'},    
\end{aligned}
\end{equation}
with $v(\rrp)=1/|\rrp|$ for the bare Coulomb in PBE0 and HF, while $v(\rrp)=\mathrm{erfc(\omega|\rrp|)}/|\rrp|$ for short-range EXX in HSE.

In fact, the above EXX effective potential exhibits an unfavorable $O(N^4)$ scaling behavior, making it significantly more computationally demanding than local functionals in terms of both time and memory requirements. To mitigate this challenge, several algorithmic approaches have been developed, most notably the density fitting (DF) technique~\cite{DF_whitten_coulombic_1973} and its specialized implementation for localized atomic orbitals known as resolution of identity (RI).~\cite{RI_FEYEREISEN1993359,RI_VAHTRAS1993514,ren_resolution_identity_2012} These methods reduce the four-center integrals to combinations of three- and two-center integrals by expanding the orbital products using auxiliary basis functions (ABFs).

For HDF calculations in large periodic systems, the local resolution of identity (LRI) approach~\cite{levchenko_hybrid_2015,ihrig_accurate_2015,lin_accuracy_2020} enables linear-scaling EXX calculations by exploiting the inherent locality of both atomic orbitals (AOs) and ABFs. Extensive benchmarking has demonstrated that with properly chosen ABFs, the LRI approximation achieves sufficient accuracy for HDF calculations.~\cite{lin_accuracy_2020} Residual discrepancies in cohesive properties and band gaps primarily originate from differences in core-valence interaction treatments (pseudopotentials versus all-electron methods) rather than basis set limitations.~\cite{ji_reproducibility_2022}

ABACUS employs two distinct types of ABFs: the ``on-site'' ABFs designed to fit products of orbitals centered on the same atom, and ``opt'' ABFs developed to improve the fitting accuracy for interatomic orbital products. Both types maintain the same functional form as AOs, consisting of a radial function multiplied by spherical harmonics $P_{A\alpha=\{nlm\}}(\br)=g_{nl}(r)Y_{lm}(\hat{\br})$. The radial components of ``on-site'' ABFs are constructed through pairwise multiplication of AO radial functions on the same atom $g_{nl}(r)=f_{n_1l_1}(r)f_{n_2l_2}(r)$, with the angular momentum constraint $|l_1-l_2|\leq l\leq l_1+l_2$, followed by orthogonalization and selection via principal component analysis (PCA). The ``opt'' ABFs are generated in the same way as AO-generation.~\cite{chen_systematically_2010, chen_electronic_2011}

\begin{figure}[ht]
    \centering
    \includegraphics[width=0.95\linewidth]{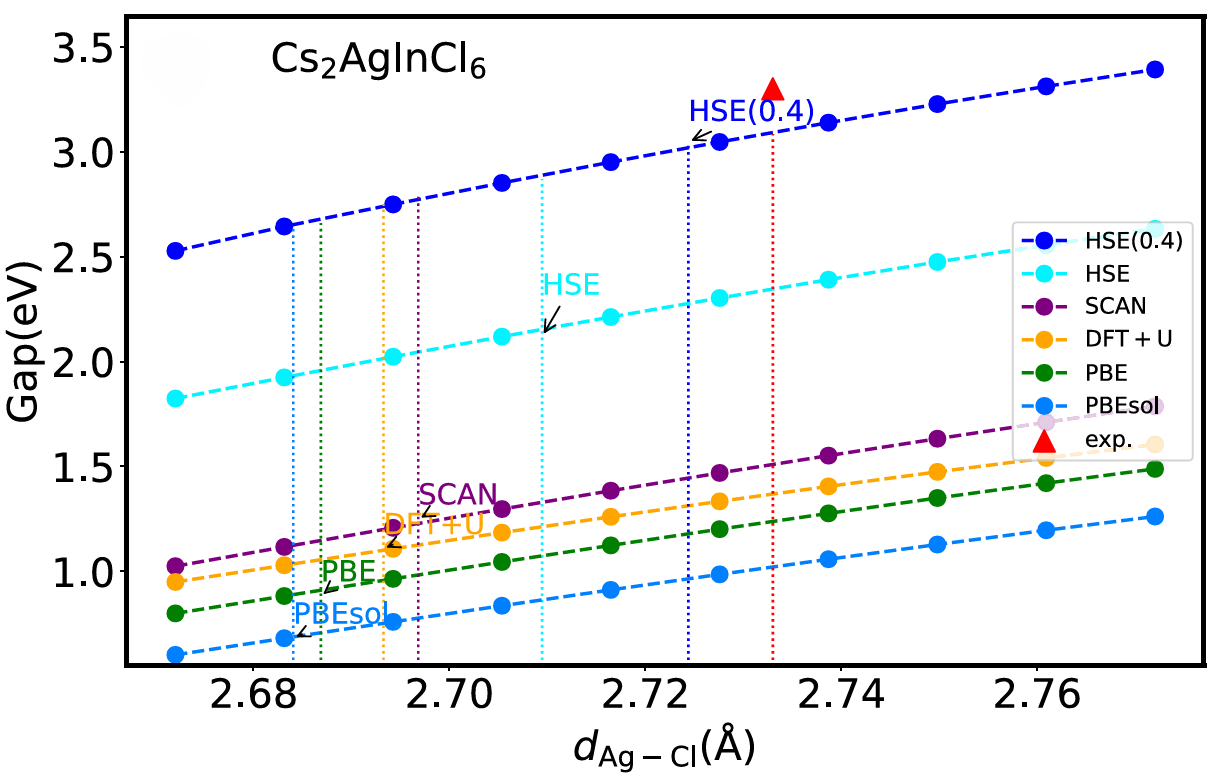}
    \caption{\MC{Band gaps of Cs$_2$AgInCl$_6$ with a lattice constant of 10.481~\AA, as computed by HSE (0.4), HSE, SCAN, DFT+U, PBE ,and
   PBEsol as a function of Ag-Cl bond length. (Adapted with permission from Phys. Rev. Research 6, 033172 (2024). Copyright 2024 American Physical Society.)}}
    \label{fig:EXX2}
\end{figure}

\begin{figure*}[ht]
    \centering
    \includegraphics[width=0.8\textwidth]{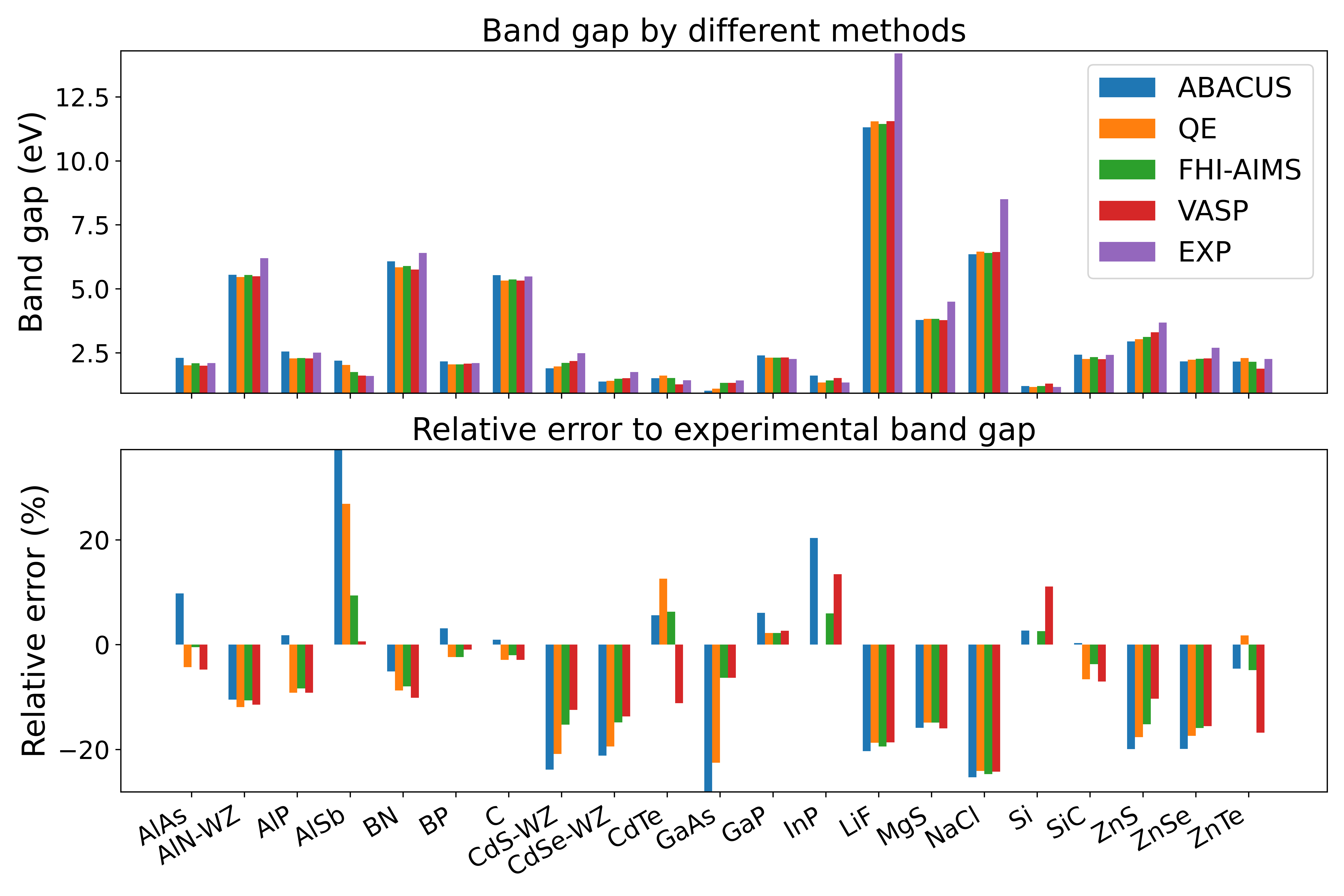}
    \caption{\MC{Band gaps of 21 different semiconductors, calculated by ABACUS+LibRI, along with the reference computed values from other software (QE/FHI-AIMS/VASP)~\cite{lin_accuracy_2020} and experimental values.~\cite{2018-cui}} 
    }
    \label{fig:exxtest}
\end{figure*}

The computational efficiency is substantially enhanced by exploiting the inherent locality of both AOs and the density matrix, which enables effective prescreening of negligible submatrices below a specified threshold. For parallel computation of four-center integrals, we have implemented two complementary distribution strategies: a load-balancing greedy algorithm for multi-processor scheduling, and a spatially optimized scheme based on K-means clustering of ABF atom pairs. The latter minimizes both inter-processor communication overhead and memory requirements by ensuring geometrically proximate pairs are assigned to the same processor, with memory consumption scaling linearly with the union of their neighbor lists.~\cite{lin_efficient_2021}

Notably, the LRI-based four-center integration has been modularized into the standalone LibRI library,~\cite{libri} which provides a portable and efficient implementation that can be readily integrated into various electronic structure codes. This includes both conventional DFT packages and beyond-DFT methods such as the random phase approximation (RPA) and GW calculations,~\cite{librpa} offering substantial acceleration for two-electron Coulomb repulsion integral evaluations across different theoretical frameworks.

In 2025, Lin {\it et al.} developed a linear-scaling algorithm for exact-exchange forces and stresses in hybrid DFT, applicable to both molecules and periodic systems. Leveraging numerical atomic orbitals and localized resolution-of-identity, the massively parallel implementation enables structural relaxation of thousand-atom systems in ABACUS.~\cite{2025-lin-efficient}

The EXX module of ABACUS has already been applied to several studies.~\cite{lin_accuracy_2020,2024-Ji-EXX,Tang2024} In 2024, Lin {\it et al.} investigated the effect of exact exchange on some of the lead-free halide double perovskites (HDPs) Cs$_2BB'X_6$ ($B$=Ag$^+$, Na$^+$; $B$=In$^{3+}$, Bi$^{3+}$; X=Cl$^-$, Br$^-$).~\cite{2024-Ji-EXX} They found some local exchange-correlation functionals fail to capture the geometric and electronic structures of Cs$_2BB'X_6$, which can be traced back to the so-called delocalization error. To show the differences between these functionals, they calculated the band gap as a function of the $B$-$X$ bond length using the functionals: PBE, PBEsol, SCAN, HSE, and HSE~(0.4). Taking Cs$_2$AgInCl$_6$ as an example in Fig.~\ref{fig:EXX2}, one can see the band gap obtained at Ag-Cl bond length (2.724 \AA) is in much better agreement with the experimental value (3.3 eV), marked by the red triangle.

\begin{figure}[ht]
    \centering
    \includegraphics[width=0.95\linewidth]{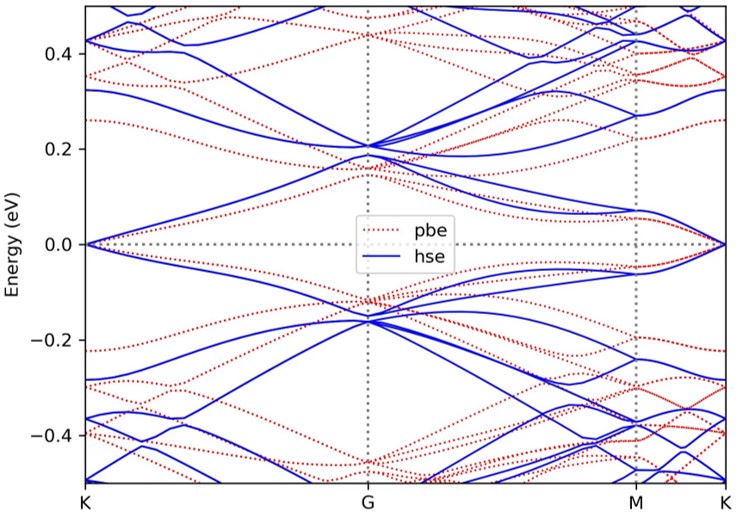}
    \caption{\MC{Band structures of twisted bilayer graphene with $\theta \approx 2.0046^{\circ}$, 3268 atoms/cell computed by ABACUS using the PBE and HSE06 functionals. (Adapted with permission from Nat Commun 15, 8815 (2024). Copyright 2024 Springer Nature.)}}
    \label{fig:graphene_HSE}
\end{figure}

Generally speaking, hybrid functional calculations can only process systems of limited size due to the large computational demand of building the exact exchange Hamiltonian. By combining NAOs basis and LRI techniques, Fig.~\ref{fig:exxtest} shows the time consumption of this part increases almost linearly with the system size, enabling ABACUS to perform hybrid functional calculations for large systems with thousands of atoms. In 2024, Tang {\it et al.} used hybrid functional data from ABACUS and trained a deep equivariant neural network approach for efficient hybrid density functional calculations.~\cite{Tang2024} Taking twisted bilayer graphene (TBG) as an example in Fig.~\ref{fig:graphene_HSE}, they performed the HSE band structure calculations of (17, 16) TBG (twist angle $\theta\approx 2.0046^{\circ}$, 3,268 atoms/cell).~\cite{Tang2024}

We also performed HSE calculations to obtain the band gaps of several semiconductors and compare them with the results from other DFT packages (QE/VASP/FHI-AIMS) and the experimental values.~\cite{lin_accuracy_2020,2018-cui} In these tests, ABACUS and QE employ the SG15-type norm-conserving PBE pseudopotentials (except for In is PSlibrary norm-conserving), VASP employs the PBE PAW potentials, whereas FHI-AMIS performs all-electron calculations.~\cite{1990-delley-allelectron}
Fig.~\ref{fig:exxtest} shows that the band gaps obtained by ABACUS HSE calculations are consistent with the results from other DFT packages, demonstrating satisfactory results when compared to the experimental values.

% LCAO deepks
\subsection{DeePKS} \label{sec:deepks}

The exchange-correlation functional is critical to determining the accuracy of DFT. However, when different levels of XC functionals are selected, there exists a trade-off between accuracy and efficiency. Lower-level XC functionals on Jacob's ladder~\cite{01AIP-Perdew} typically offer higher computational efficiency but lower accuracy, whereas higher-level ones exhibit the opposite trend. With the rapid progress in increasing computational resources and advanced algorithms, AI-assisted methods hold the potential to mitigate this issue.~\cite{22ES-Kulik,23SCI-Huang}

Proposed in 2020, DeeP Kohn-Sham method (DeePKS)\cite{21JCTC-Chen} employs a computationally efficient neural network-based functional model to represent the energy and force difference between a lower-level XC functional and a higher-level XC functional. The resulting model maintains translational, rotational, and permutational invariance and can be used in self-consistent field calculations. Together with DeePKS-kit~\cite{22CPC-Chen} software, ABACUS supports the iterative training of DeePKS models.~\cite{22JPCA-Li} Specifically, DeePKS models can be trained in molecular or periodic systems to learn properties such as energy, force, stress, and band gap. Based on the DeePKS model, ABACUS can achieve accuracy similar to that of the high-level XC functional on these specified properties, while the computational efficiency is similar to that of the low-level XC functional.

In the DeePKS method, we divide the total energy functional into two parts
\begin{align}
    E_\text{DeePKS}[\{\psi_i\}|\omega]=E_\text{baseline}[\{\psi_i\}]+E_\delta[\{\psi_i\}|\omega] ,
\end{align}
where $\{\psi_i\}$ are single-particle orbitals to yield the baseline energy $E_\text{baseline}$, and $E_{\delta}$ is constructed as a neural network model with parameters $\omega$. The model input is constructed based on the projected density matrix for each atom $I$, which takes the form
\begin{align}
D_{nlmm^{\prime}}^I=\sum_{\mu\nu}\langle\alpha_{nlm^{\prime}}^I|\phi_{\nu}\rangle\rho_{\mu\nu}\langle\phi_{\mu}|\alpha_{nlm}^I\rangle.
\end{align}
Here $\rho_{\mu\nu}$ represents the density matrix, and $|\alpha_{nlm}^{I}\rangle$ is a set of localized orbitals centered on atoms, identified by atomic index $I$, and quantum numbers $nlm$. The atomic-centered basis functions $|\alpha_{nlm}^{I}\rangle$ ensure the translational invariance. To maintain rotational invariance, we proceed to extract the eigenvalues of projected density matrix blocks with the same indices $I$, $n$, and $l$, resulting in a set of descriptors that are obtained as eigenvalues of the projected density matrix
\begin{align}
    \mathbf{d}_{nlm}^I=\mathrm{Eig}(D_{nlmm^{\prime}}^I).
\end{align}
The descriptors are then grouped into vectors based on the atomic index $I$, and $E_{\delta}$ is calculated as the sum of atomic contributions as
\begin{align}
    E_\delta=\sum_IF_{\mathrm{NN}}(\mathbf{d}^I|\boldsymbol{\omega}),
\end{align}
where $F_{\mathrm{NN}}$ is the deep neural network. Next, the Hamiltonian operator can be written as
\begin{align}
    H=H_\text{baseline}+\hat{V}^\delta,
\end{align}
where the $H_\text{baseline}$ term is the baseline Hamiltonian, and the correction potential can be evaluated via
\begin{equation}
\hat{V}_{\mu\nu}^{\delta} =\frac{\partial E_\delta}{\partial\rho_{\mu\nu}}.
\end{equation}
Finally, the corresponding force and stress terms can also be obtained for the correction term.~\cite{22JPCA-Li}

\begin{figure}
    \centering
    \includegraphics[width=0.8\linewidth]{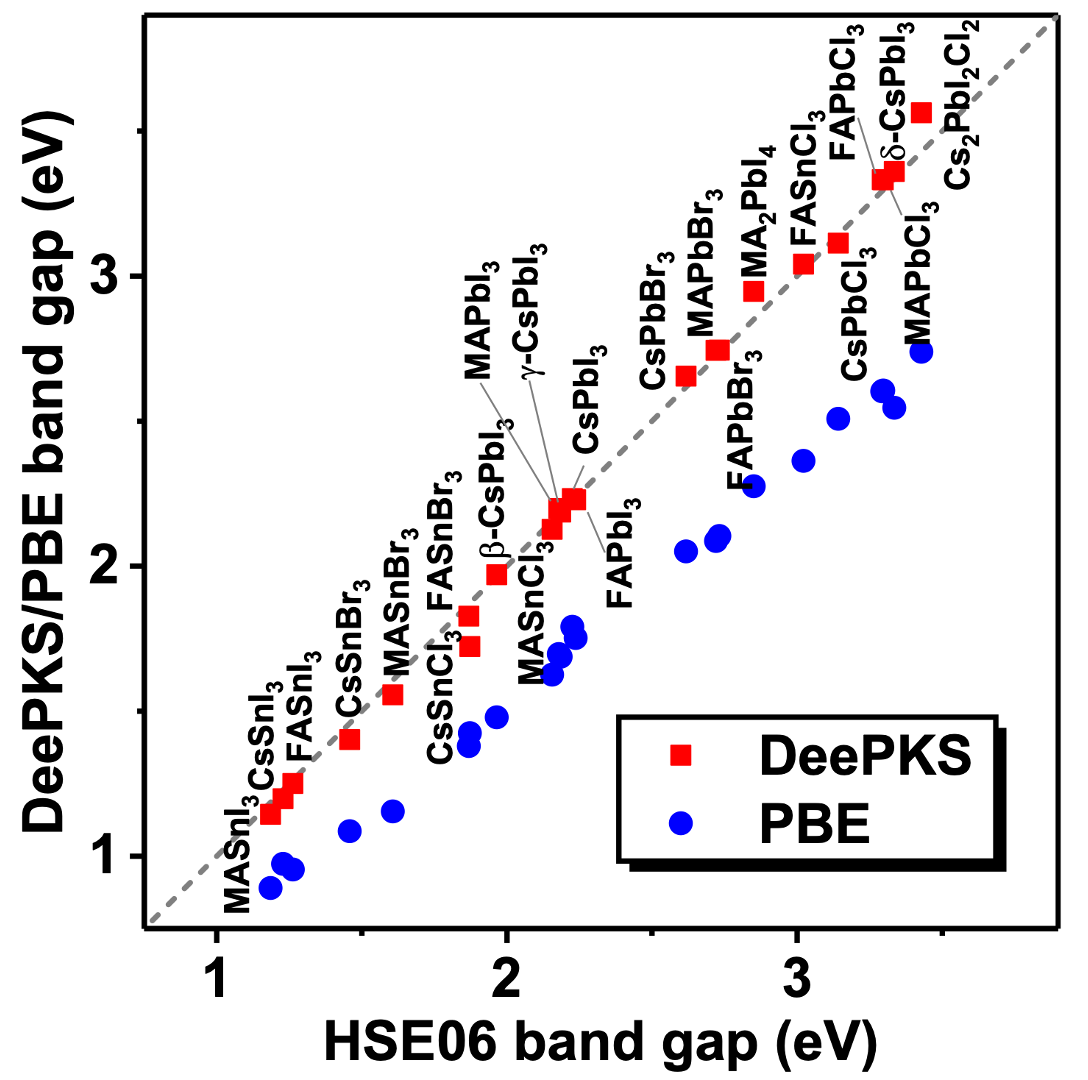}
    \caption{\MC{Band gaps predicted by DeePKS and PBE with respect to the HSE06 results for all tested perovskites.\cite{23JPCC-Ou} All tested non-hybrid perovskites are cubic phase except for those indicated by Greek letters. (Adapted with permission from J. Phys. Chem. C 127, 18755–18764 (2023). Copyright 2023 American Chemical Society.)}}
    \label{fig:bandgap}
\end{figure}

Here we list a few applications using the DeePKS method.
For instance, halide perovskites (ABX$_3$, X=halogen anion) have shown great promise as a cost-effective alternative to current commercial photovoltaic technologies. Designing effective photovoltaic systems requires a precise yet efficient description of the electronic structure of halide perovskites. One key benefit of halide perovskites is adjusting the absorption edge wavelength (band gap) by changing the ratio of different halide ions. Ou {\it et al.}~\cite{23JPCC-Ou} constructed a general DeePKS model that can be utilized for halide perovskites, including different combinations of ABX$_3$ (A=FA, MA, Cs; B=Sn, Pb; X=Cl, Br, I), the organic-inorganic hybrid alternatives, and the Ruddlesden-Popper (RP) perovskites. In detail, the authors built an extensive DeePKS model upon 460 configurations spanning seven types of halide perovskites, with HSE06 accuracy and satisfactory predictions for the band gap. Based on an iterative training process with ABACUS and DeePKS-kit, they showed that the resulting DeePKS model can accurately replicate forces, stress, band gaps, and density of states (DOS) near the Fermi energy for all types of halide perovskites, including RP structures, and hybrid compositions, when compared to HSE06. They further showed the band gaps predicted by DeePKS and PBE, with respect to those by HSE06, over 30 tested systems. As depicted in Fig.~\ref{fig:bandgap}, the DeePKS model demonstrated precise predictions of band gaps for all perovskites examined, closely aligning with the HSE06 findings and yielding an average absolute error (MAE) of 0.0350 eV. On the other hand, PBE significantly underestimated the band gap values, with a large MAE of 0.5222 eV.

\begin{figure}
    \centering
    \includegraphics[width=0.8\linewidth]{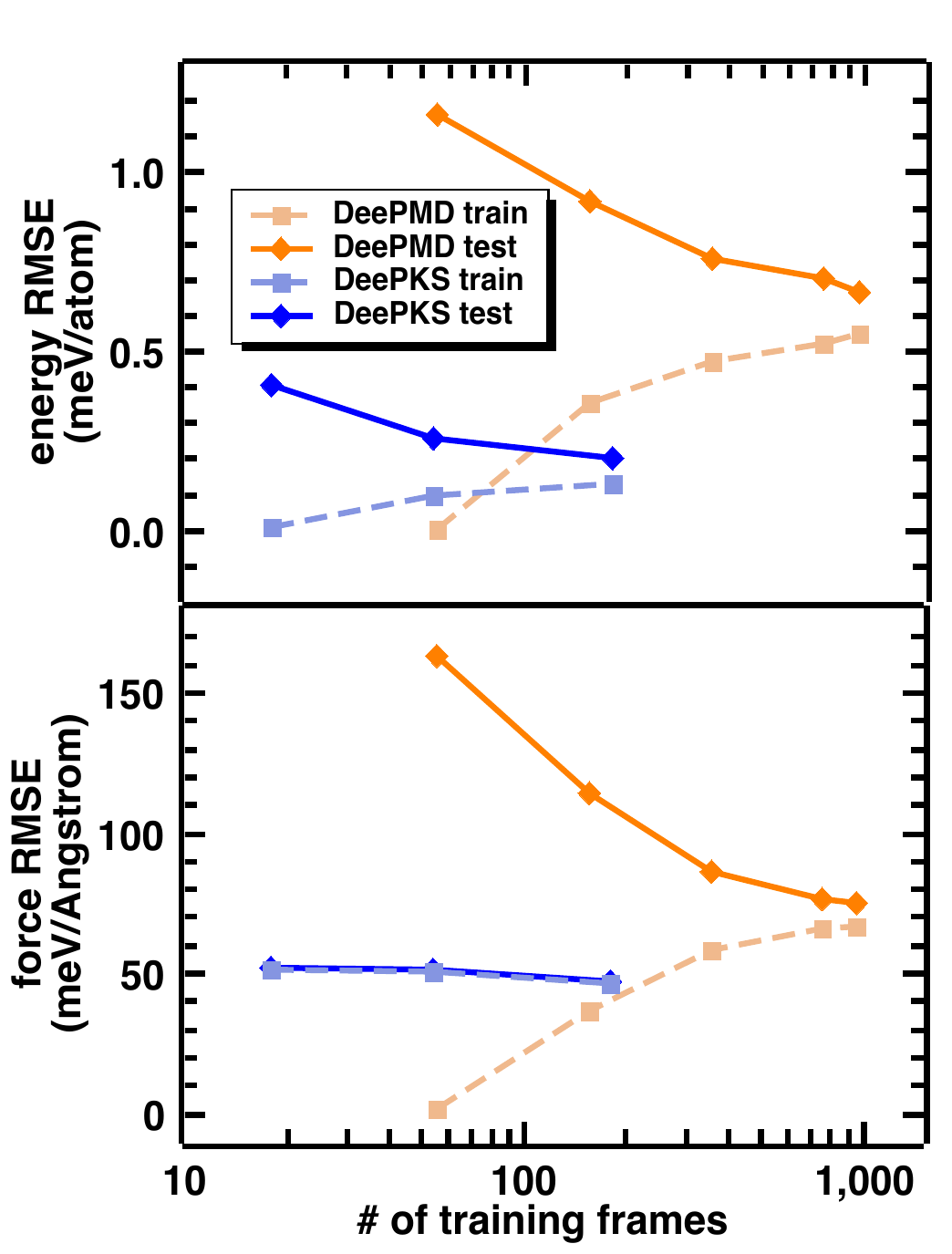}
    \caption{\MC{Learning curves for energy (upper panel) and force (lower panel) given by DeePMD (orange) and DeePKS (blue) with respect to the number of training frames.~\cite{22JPCA-Li} Dashed line with squares indicates train set error, while solid line with diamonds indicates test set error. (Adapted with permission from J. Phys. Chem. A 2022, 126, 49, 9154–9164. Copyright 2022 American Chemical Society.)}}
    \label{fig::learning_curve}
\end{figure}

\begin{figure*}[htp]
    \includegraphics[width=1.0\linewidth]{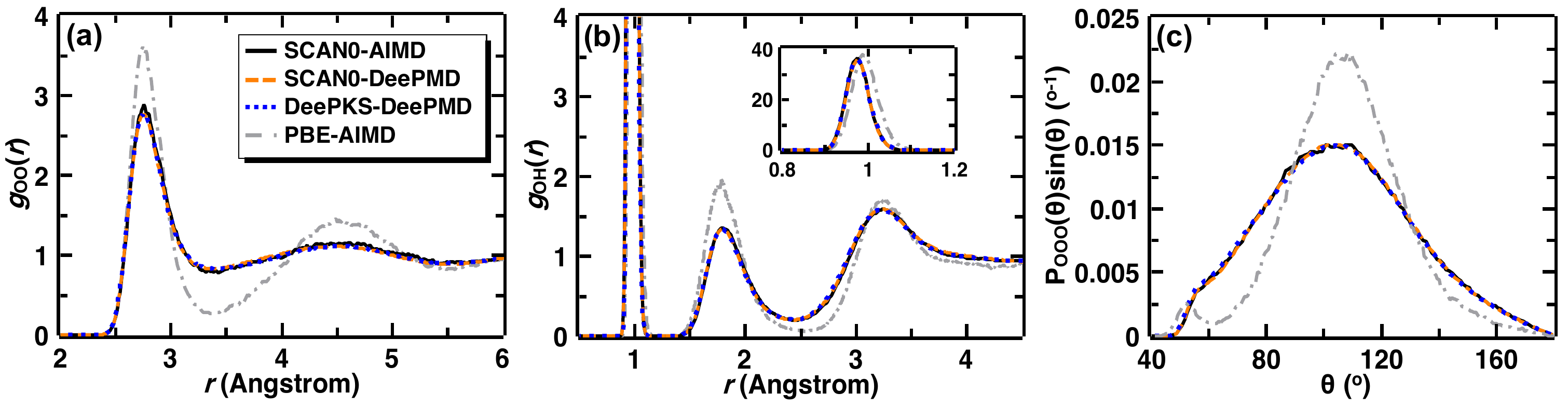}
    \caption{\MC{Radial distribution functions (RDFs) of (a) $g_{\textrm{OO}}(r)$, (b) $g_{\textrm{OH}}(r)$, and (c) bond angle distribution $P_{\textrm{OO}}(\theta)$ given by DeePKS-DeePMD (blue dotted line), SCAN0-AIMD (black solid line), SCAN0-DeePMD (orange dashed line), and PBE-AIMD (gray dotted-dashed line).(Adapted with permission from J. Phys. Chem. A 2022, 126, 49, 9154–9164. Copyright 2022 American Chemical Society.)}}
    \label{fig::water_scan0}
\end{figure*}

In addition to providing an accurate yet efficient description of the electronic structure, DeePKS also serves as a ``bridge'' between expensive quantum mechanical (QM) models and ML-based potentials. While the ML-based potentials such as the Deep Potential Molecular Dynamics (DeePMD)~\cite{18PRL-Zhang,18CPC-Wang} have emerged as powerful tools for mitigating the high computational costs associated with ab initio molecular dynamics (AIMD), training these potentials demands a significant number of QM-labeled frames. DeePKS offers a solution to further save the computational cost by reducing the required QM-labeled frames, owing to its significantly better transferability as compared to DeePMD. Li {\it et al.}~\cite{22JPCA-Li} examined DeePKS and DeePMD's training curves with respect to the number of training samples in systems with 64 water molecules at the accuracy of hybrid functional SCAN0. The DeePKS model outperforms the DeePMD model with fewer frames shown in Fig.~\ref{fig::learning_curve}. Additionally, the DeePKS model has a smaller generalization gap than the DeePMD model. The work~\cite{22JPCA-Li} also showed that excellent agreement can be achieved between the ABACUS-DeePKS-DeePMD results and the ones from SCAN or SCAN0-based DeePMD simulations. By labeling fewer than 200 frames in the training set with SCAN0 or SCAN functionals, the GGA-based DeePKS model can effectively reproduce the energies and forces for pure and salt water systems, with considerable time savings. SCF calculations with trained DeePKS models were carried out and utilized as labels for DeePMD training. Structural properties of liquid water, such as radial distribution function (RDF), bulk density, hydrogen (H) bonds, and dynamic properties like diffusion coefficient, were found to be excellently matched with those obtained by the SCAN0 AIMD and DeePMD methods. For example, for systems consisting of 64 water molecules, they trained a DeePKS model with 180 training samples at the accuracy of SCAN0. Then, the authors used it to efficiently generate 1000 data to train DeePMD potential and performed MD simulations of 512 water molecules. As shown in Fig.~\ref{fig::water_scan0}, various RDFs derived from DeePKS-DeePMD simulations match well with both SCAN0-AIMD and SCAN0-DeePMD results, including a marked decrease in overstructured peaks.

Zhang {\it et al.}~\cite{24JCIM-Zhang} investigated the tautomeric equilibria of glycine in water with the DeePMD model at the accuracy of M06-2X.~\cite{08TCA-Zhao} To avoid expensive computational cost for directly generating dataset at M06-2X level, they utilized the DeePKS method as implemented in ABACUS and DeePKS-kit, together with on-the-fly probability enhanced sampling (OPES)~\cite{20JPCL-Invernizzi} method, to construct the dataset for training the DeePMD model. With this DeePMD potential and OPES, they performed MD simulations and obtained a converged free energy surface (FES). They observed that glycine can undergo tautomerism, transitioning between its neutral and zwitterionic forms through intramolecular and intermolecular proton transfers. Zhang {\it et al.}~\cite{24-Zhang} adopted a similar strategy to study the propensity of water self-ions at air(oil)-water interface. They found that the trained DeePKS model can decrease calculation time by about nine times compared to regular M06-2X calculations. With the resulting DeePMD model for an efficient MD process, they demonstrated the stable ionic double-layer distribution near the interface for both air-water and oil-water interface systems.

Liang {\it et al.}~\cite{25JCTC-Liang} developed an enhanced version of the DeePKS approach, known as DeePKS-ES (electronic structure), to further improve its prediction of electronic structure properties. The DeePKS-ES method modifies the loss function by integrating the Hamiltonian matrix along with its eigenvalues (energy levels) and eigenvectors (wave function coefficients). This adjustment allows DeePKS-ES to reliably predict key electronic structure characteristics like band gaps, density of states (DOS), and Hamiltonian matrices, in addition to total energy and atomic forces. When tested on various water systems, including monomers, clusters, and liquid phases, DeePKS-ES achieves a level of accuracy similar to that of the hybrid functional HSE06 while maintaining the computational efficiency of the generalized gradient approximation PBE. Particularly, the universal DeePKS-ES model shows strong adaptability across different water systems, connecting quantum-mechanical precision with scalable computation.

% LCAO DFT+U
\subsection{DFT+U} \label{sec:dftu}

While local and semilocal density functional approximations (LDA and GGAs) can often predict ground-state properties of various systems with reasonable accuracy, they exhibit significant limitations when applied to strongly correlated materials, such as transition metal oxides and rare-earth compounds. In these systems, the simplified exchange-correlation functionals fail to adequately capture complex electron-electron interactions, leading to substantial inaccuracies in predicted electronic properties, including but not limited to energy band gaps, magnetic moments, and orbital polarization. The DFT+$U$ method~\cite{cococcioni2005linear, anisimov1993density, anisimov1997first, dudarev1998electron, jiang2010first} addresses these limitations by incorporating a Hubbard-$U$ correction term, which significantly improves the description of strongly correlated electronic systems while maintaining computational costs comparable to standard LDA or GGA calculations.

The DFT+$U$ method is implemented through the following total energy functional
\begin{equation}
E_{\mathrm{DFA}+U}[\rho]=E_{\mathrm{DFA}}[\rho]+E_{\mathrm{Hub}}-E_{\mathrm{dc}},    
\end{equation}
where $E_{\text{DFA}}$ denotes the baseline energy. In addition, the Hubbard term and double-counting terms are labeled as $E_{\mathrm{Hub}}$ and $E_{\mathrm{dc}}$, respectively. In ABACUS, we introduce the fully localized limit (FLL) implementation of DFT+$U$ method,~\cite{dudarev1998electron} which assumes that the on-site Coulomb interactions of localized electrons are fully accounted for by the Hubbard $U$ term, while a mean-field average is subtracted to avoid overestimating these interactions. The Hubbard-corrected energy functional is given by
\begin{equation}
E_U\left[\left\{n_{m m^{\prime}}^{I \sigma}\right\}\right] 
=\frac{U}{2} \sum_I \sum_{m\sigma}\Big(n_{m m}^{I \sigma}-\sum_{m^{\prime}} n_{m m^{\prime}}^{I \sigma} n_{m^{\prime} m}^{I \sigma}\Big).
\end{equation}
Here, the Hubbard parameter $U$ represents an empirically adjusted Coulomb penalty specifically applied to localized electrons, designed to correct the systematic underestimation of electron-electron repulsion inherent in standard DFT calculations. Additionally, the central quantity is the on-site density occupancy matrix $n_{m m^{\prime}}^{I \sigma}$ with $I$, $\sigma$, and $m$ being the atom, spin and angular momentum indices, respectively. In general, this matrix can be expressed using a local projection operator 
\begin{equation}
\hat{P}^\sigma_{mm'} = |\alpha^{I\sigma}_m \rangle\langle \alpha^{I\sigma}_{m'}|, 
\end{equation}
where $\alpha^{I\sigma}_{m}$ denotes the projection orbital, and $\hat{\rho}$ represents the density matrix
\begin{equation}
n^{I\sigma}_{mm'} = \langle \alpha^{I\sigma}_m \sigma|\hat{\rho}|\alpha^{I\sigma}_{m'} \sigma\rangle = \mathrm{Tr}(\hat{\rho}\hat{P}_{ mm'}^{I\sigma}).
\end{equation}

Two distinct implementations of the DFT+$U$ method are available in ABACUS when using NAOs as basis sets. These implementations differ primarily in their approach to projecting the electron density onto localized states, a crucial step for accurately modeling strong electron-electron interactions within specific atomic orbitals. The first approach is termed the ``dual projection'' method, while the second is referred to as the ``full projection'' method.

First, the dual projection method~\cite{qu2022dftu,han2006dftu} employs a Mulliken charge projector to construct the on-site density occupancy matrix. This approach facilitates the transformation of Kohn-Sham orbitals into a localized correlated subspace using NAOs, particularly for $d$ or $f$ orbitals. The Mulliken projector satisfies the charge sum rule, ensuring conservation of total electronic charge across all projected channels. This methodology is alternatively termed as the ``dual orbitals'' approach~\cite{han2006dftu} due to its fundamental reliance on dual orbitals that maintain mutual orthogonality. Mathematically, these dual orbitals are derived from the original atomic orbitals. The spin-dependent on-site density occupancy matrix is explicitly defined as
\begin{equation}
\begin{aligned}
n^{I\sigma}_{mm'} &= \frac{1}{4N_k}   \sum_{\mathbf{k}\beta mm'\mu} \Big(  S^{I}_{\beta m,\mu}(\mathbf{k}) \rho^{I\sigma}_{\mu,\beta m'}(\mathbf{k}) + \rho^{I\sigma}_{\beta m,\mu}(\mathbf{k}) S^I_{\mu,\beta m'}(\mathbf{k})\\
&  + S^I_{\beta m',\mu}(\mathbf{k}) \rho^{I\sigma}_{\mu,\beta m}(\mathbf{k}) + \rho^{I\sigma}_{\beta m',\mu}(\mathbf{k}) S^I_{\mu,\beta m}(\mathbf{k}) \Big),
\end{aligned}
\end{equation}
where $\rho^{I\sigma}_{\beta m,\mu}(\mathbf{k})$ denotes the density matrix, and $S^I_{\beta m,\mu}(\mathbf{k})$ is the overlap matrix of projected dual orbital $|\beta^I_m\rangle$ and numeric atomic orbital $|\phi_\mu\rangle$ (Eq.~\ref{eq:overlap_k}). More details are provided in Ref.~\onlinecite{qu2022dftu}.

Second, while the dual projection method offers a robust and efficient DFT+$U$ implementation, it relies on a subset of NAOs as projection operators introduces notable basis set dependence for the Hubbard $U$ parameter. This dependence becomes particularly evident when employing large radial cutoffs, which may affect the description of localized electron characteristics. Therefore, an alternative approach, implemented in ABACUS since version 3.6, involves the modulation of NAOs to construct localized projection operators. This method employs spherical truncation through empirical atomic radius settings, systematically transforming the problem into orbital modulation while satisfying three criteria. First, close correspondence with the original numerical orbitals must be ensured. Second, the orbitals should satisfy the normalization condition. Third, sufficient smoothness at the cutoff of the atomic radius must be ensured.

To be specific, the radial function $\chi(r)$ of the original NAOs undergoes direct truncation followed by normalization operations to yield the truncated localized orbital
\begin{equation}
\alpha(r)=\left.\frac{\chi(r)g(r;\gamma)}{\langle\chi(r)g(r;\gamma)|\chi(r)g(r;\gamma)\rangle} \right|_{\frac{\partial \langle\alpha|\chi\rangle}{\partial \gamma} = 0},
\end{equation}
where the smooth function has the form
\begin{equation}
g(r;\gamma)=\left\{\begin{matrix}
1-\exp\left(-\frac{(r-r_c)^2}{2\gamma^2}\right) & r < r_c \\
0&r\geq{}r_c
\end{matrix}\right. 
.
\end{equation}
Here the parameter $\gamma$ controls the smoothing interval and the optimal $\gamma$ is determined iteratively to maximize two-center integrals between modulated $\alpha(r)$ and the original $\chi(r)$ orbitals, yielding the target localized orbitals. The full projection method constructs the on-site density occupancy matrix via the projection operator in real space~\cite{han2006dftu}
\begin{equation}
\mathrm{n}_{m m^{\prime}}^{I\sigma}= \sum_{\mathbf{R}\mathbf{R}'}\sum_{\mu\nu}\rho^{\mathbf{R}\sigma}_{\mu\nu}\langle \phi^{0}_{\mu}|\alpha_m^{I\mathbf{R}'}\rangle \langle \alpha_{m'}^{I\mathbf{R}'}|\phi^{\mathbf{R}}_{\nu}\rangle.  
\end{equation}

\begin{figure*}[htbp]
    \centering
    \includegraphics[width=0.9\linewidth]{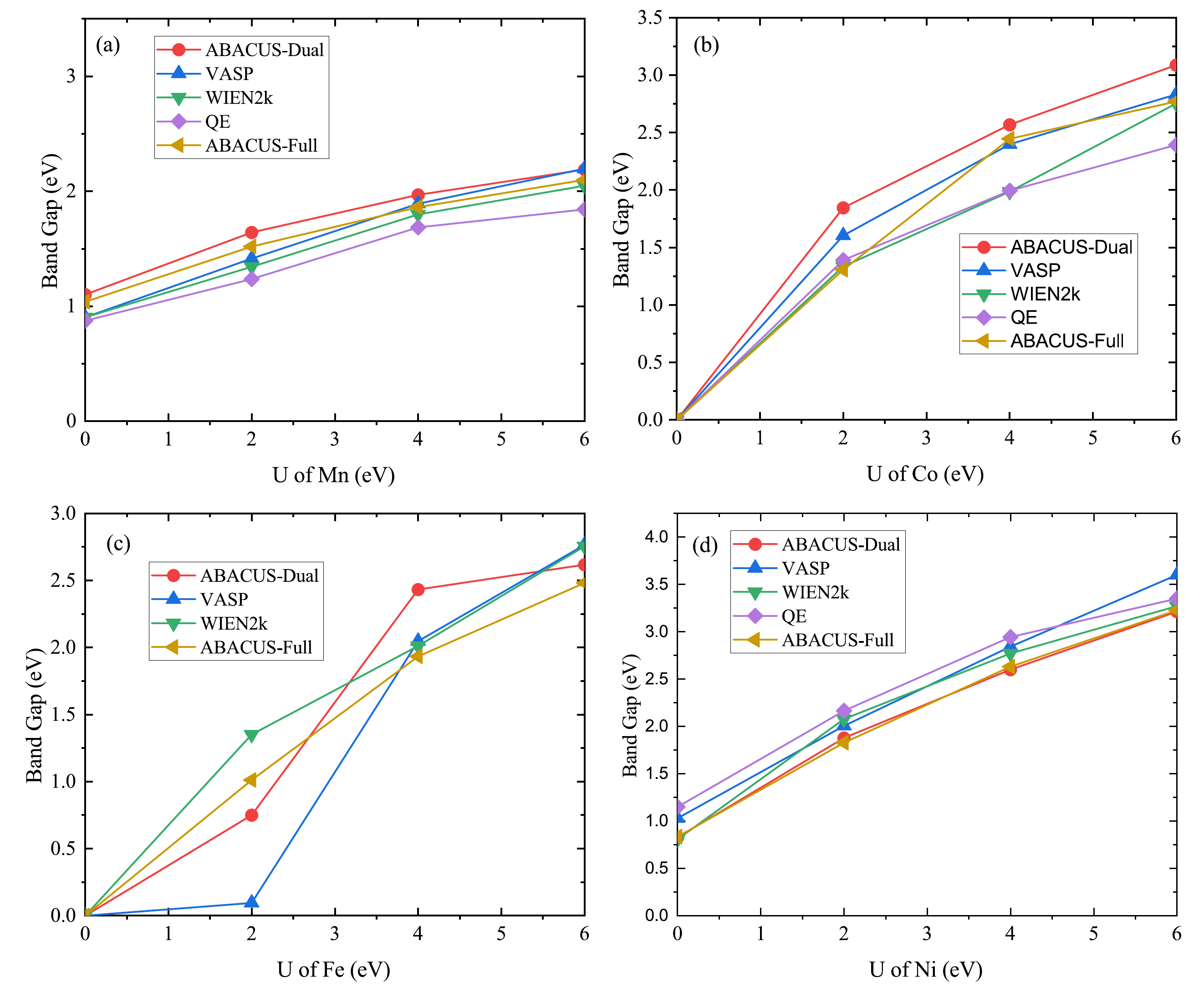}
    \caption{\MC{Band gaps of transition metal monoxide (a) MnO, (b) FeO, (c) CoO, and (d) NiO as influenced by the Hubbard $U$ parameter, utilizing a rhombohedral cell with a type-II antiferromagnetic structure. ``ABACUS-Full'' presents results from the full projection method. For comparative analysis, PBE+$U$ results for other computational methods are from Ref.~\onlinecite{qu2022dftu}.}}
    \label{fig:dftu-bandgap-benchmark}
\end{figure*}

We present systematic benchmarks of the full projection DFT+U method for MnO, CoO, FeO, and NiO systems using ABACUS v3.8. All calculations employ the optimized norm-conserving Vanderbilt (ONCV) pseudopotentials with SG15-v1.0 version. The Brillouin zone discretization is 0.15 $\text{bohr}^{-1}$ and the energy cutoff is 100 Ry. We use the PBE functional~\cite{1996-pbe} with the Hubbard U correction, while spin-orbit coupling effects are excluded from consideration. In addition, we adopt a double-$\zeta$ plus polarization (DZP) atomic basis with a radial cutoff of 9.0 bohr for transition metal (TM) elements and 7.0 bohr for O atoms. 

\begin{table*}
\centering
\caption{Band gaps (in eV) and atomic magnetism (in $\mu_B$) with format ``value of band gap (value of atomic magnetic of Mn/Co/Fe/Ni)'' of MnO, FeO, CoO, and NiO as a function of effective on-site Coulomb energy \( \bar{U} \) (in eV). The experimental values are presented in the last row.}
\hspace{-5mm}
\centering
\begin{ruledtabular}
\begin{tabular}{ccccc}
\( \bar{U} \) (eV) & MnO & CoO & FeO & NiO \\
\hline
0.0 & 0.00 (4.64) & 1.04 (2.46) & 0.00 (3.54) & 0.83 (1.33) \\
1.0 & 0.54 (4.71) & 1.30 (2.56) & 0.00 (3.61) & 1.37 (1.44) \\
2.0 & 1.31 (4.76) & 1.52 (2.62) & 1.01 (3.68) & 1.83 (1.52) \\
3.0 & 2.01 (4.80) & 1.71 (2.67) & 1.38 (3.74) & 2.25 (1.58) \\
4.0 & 2.45 (4.84) & 1.86 (2.72) & 1.93 (3.78) & 2.63 (1.63) \\
5.0 & 2.63 (4.87) & 1.99 (2.76) & 2.28 (3.82) & 3.01 (1.67) \\
6.0 & 2.77 (4.90) & 2.10 (2.79) & 2.48 (3.86) & 3.23 (1.71) \\
\hline
Exp. & $3.6$-$3.8$~\cite{messick1972direct} ($4.58$~\cite{cheetham1983magnetic}) & $2.4$~\cite{bowen1975electrical} ($3.8$~\cite{fender1968covalency}) & $2.4$~\cite{powell1970optical} ($3.32$~\cite{roth1958magnetic}) & $4.0$~\cite{sawatzky1984magnitude}/$4.3$~\cite{hufner1984photoemission} ($1.90$~\cite{khan1970magnetic}) \\
\end{tabular}
\end{ruledtabular}
\begin{flushleft}

\end{flushleft}
\label{tab:dftu-bg-mag}
\end{table*}

Table~\ref{tab:dftu-bg-mag} presents the calculated band gaps and atomic magnetic moments for transition metal oxides (TMOs) with the Hubbard U parameter ranging from 0 to 6 eV. All of the four investigated TMOs (MnO, FeO, CoO, and NiO) are in the rhombohedral structure with a type-II antiferromagnetic (AFM) configuration. The lattice parameters, adopted from Ref.~\onlinecite{tran2006hybrid}, are 4.445, 4.334, 4.254, and 4.171~\AA~for MnO, FeO, CoO, and NiO, respectively, with the AFM ordering aligned along the [111] crystallographic direction. Fig.~\ref{fig:dftu-bandgap-benchmark} further shows the comparison of band gaps for MnO, FeO, CoO, and NiO systems. The results obtained from ABACUS match well with those from other DFT packages such as VASP, Wien2k, and Quantum ESPRESSO.~\cite{qu2022dftu}

% LCAO real-time TDDFT
\subsection{Real-Time Time-Dependent DFT}

DFT is effective in predicting ground-state properties but faces challenges when dealing with excited states. A more accurate theoretical framework for simulating excited states was proposed by Runge and Gross,~\cite{84PRL-Runge} known as time-dependent density functional theory (TDDFT). Real-time TDDFT (rt-TDDFT) enables the investigation of electron dynamics, including applications in optical absorption spectra,~\cite{ren2010optical} stopping power,~\cite{maliyov2018electronic} photocatalysis,~\cite{tddft_application_5_Ren_2024} and field-induced transitions. However, its computational cost presents a significant hurdle for large-scale systems. In this context, the use of numerical atomic orbital (NAO) basis sets helps reduce computational expenses while still permitting detailed analysis of excited-state phenomena and their complex behaviors.~\cite{meng2008real}

The propagation of electrons obeys the time-dependent Kohn-Sham (TD-KS) equation
\begin{equation}
    \mathrm{i}\frac{\partial}{\partial t}\psi_j(\mathbf{r},t)=H\psi_j(\mathbf{r},t).
\end{equation}
Here the Hamiltonian $H$ depends only on the electron density at time $t$ and we adopt the adiabatic approximation. For numerical atomic orbitals, the TD-KS equation can be expressed in the matrix form~\cite{18ATS-SMeng}
\begin{equation}
    \mathrm{i}\frac{\partial \boldsymbol c}{\partial t}=\boldsymbol S^{-1}\boldsymbol H \boldsymbol c.
\end{equation}
where $\boldsymbol S$ represents the overlap matrix, and $\boldsymbol c$ is the column vector of coefficients for the local basis. For simplicity, the subscripts denoting the band index and $k$-points have been omitted. We adopt the first-order Crank-Nicolson method to approximate the propagator
\begin{equation}
    \boldsymbol{C}(t+\mathrm{\Delta}t)=\frac{\boldsymbol S(t+\mathrm{\Delta}t/2)-\mathrm{i}\boldsymbol H(t+\mathrm{\Delta}t/2)\mathrm{\Delta}t/2}{\boldsymbol S(t+\mathrm{\Delta}t/2)+\mathrm{i}\boldsymbol H(t+\mathrm{\Delta}t/2)\mathrm{\Delta}t/2}\boldsymbol{C}(t),
\end{equation}
where $S(t+\mathrm{\Delta}t/2)$ and $H(t+\mathrm{\Delta}t/2)$ are computed using a linear approximation. Since $H(t+\mathrm{\Delta}t/2)$ depends on $\mathbf{C}(t+\mathrm{\Delta}t)$, a self-consistent procedure is needed to perform the propagation. The approach to approximating the propagator ensures its time-reversal invariance, conserves the total energy effectively, and preserves the orthogonality of the propagating wave functions. In addition, the Ehrenfest dynamics is adopted for ion-electron coupled systems and the Verlet algorithm is employed to calculate the ionic velocities and positions at each time step.

The time-dependent electric field $\mathbf{E}(t)$ is introduced into the Hamiltonian to simulate the interactions between laser field and materials, but there are different ways to implement the electric field. First, within the length gauge, the electric field potential term in Hamiltonian takes the form of
\begin{equation}
    V_{\mathrm{efield}}(\mathbf{r},t)=\mathbf{E}(t)\cdot \mathbf{r}.
\end{equation}
However, the periodic nature of the cell breaks the translational invariance of the electric potential. To address this, a sawtooth field in the spatial domain takes the form of
\begin{equation}
    E_{\mu}(x_{\mu},t)=
    \begin{cases}
        E_{\mu}(t),  & \epsilon<x_{\mu}<L_{\mu}-\epsilon,   \\
        -E_{\mu}(t)(L_{\mu}-2\epsilon)/2\epsilon, & -\epsilon<x_{\mu}<\epsilon,
    \end{cases}
\end{equation}
where $\mu=x,y,z$, $L_{\mu}$ is the length of cell along $\mu$, and $\epsilon \to 0$. To avoid divergence, the charge density must vanish in the region $-\epsilon<x_{\mu}<\epsilon$. Ideally, this region should be set as a vacuum layer. Therefore, the length gauge can only be used for finite systems. The second method is the velocity gauge, which introduces a vector potential $\mathbf{A}(t)$ to simulate the laser field as
\begin{equation}
    \mathbf{A}(t)=-\int \mathbf{E}(t)\,\mathrm{d}t.
\end{equation}
Therefore, the kinetic term of the velocity-gauge Hamiltonian becomes
\begin{equation}
    H_{\mathrm{kinetic}}=\frac{1}{2}\Big[-\mathrm{i}\nabla+\mathbf{A}(t)\Big]^2.
\end{equation}
This method maintains the periodicity of the system. Third, while rt-TDDFT in periodic systems often relies on the velocity gauge to maintain periodicity under external fields, NAO implementations face a critical limitation: the position-dependent phase variations induced by the vector potential are neglected, leading to errors especially for properties like current. To overcome this challenge, Zhao {\it et al.}~\cite{2025-zhao-hybrid} introduces a hybrid gauge in ABACUS that incorporates both the electric field and the vector potential, explicitly preserving phase information within atomic orbitals. Benchmark results confirm that this approach fully rectifies the inaccuracies of the pure velocity gauge in NAO-based rt-TDDFT, delivering reliable and precise simulations of ultrafast dynamics.

\begin{figure}
    \centering
    \includegraphics[width=\linewidth]{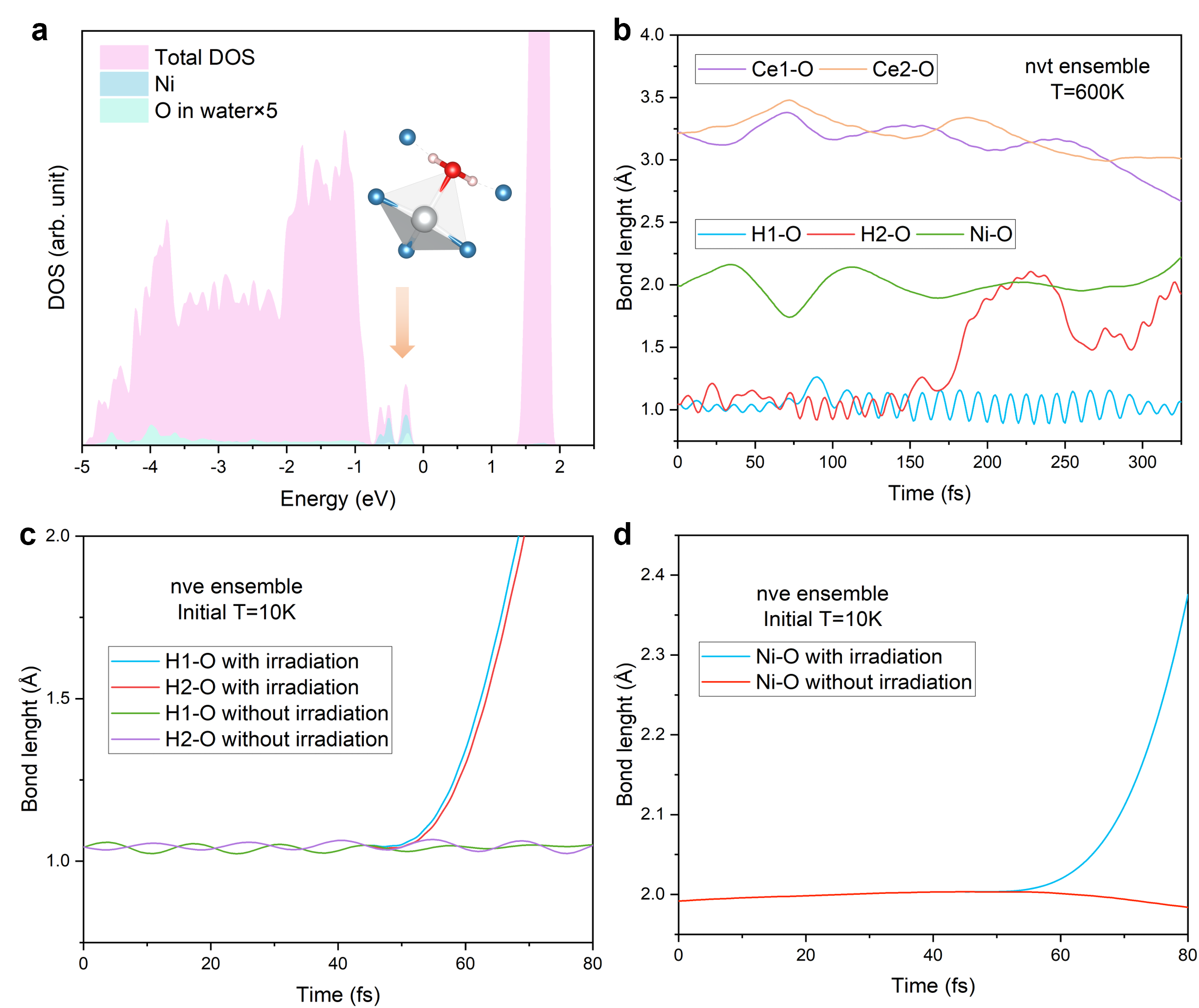}
    \caption{\MC{Real-time TDDFT applications on photocatalysis.~\cite{tddft_application_5_Ren_2024} (a) Projected density of states for H$_2$O on Ni single atom site load on CeO$_2$. (b) Bond length of thermal dissociation of H$_2$O under 600 K. (c-d) Bond length of light-induced dissociation of H$_2$O. (Adapted with permission from Nat. Commun. 15, 4675 (2024). Copyright 2024 Springer Nature.)}}
    \label{fig:tddft_application}
\end{figure}

Based on the rt-TDDFT functionality of ABACUS, Liu {\it et al.}~\cite{tddft_application_5_Ren_2024} studied the dissociation process of water molecules under thermal and photoexcitation conditions using Ni single atoms supported on CeO\textsubscript{2}, as shown in Fig.~\ref{fig:tddft_application}. The ground-state electronic density shows that when water molecules adsorb on Ni single atoms, due to the presence of oxygen-rich defects, the anti-bonding state formed by the hybridization of the Ni atom $d$ orbitals and H\textsubscript{2}O molecular orbitals is occupied (Fig.~\ref{fig:tddft_application}a), leading to weak adsorption strength of H\textsubscript{2}O and a correspondingly low dissociation barrier. Under thermal excitation at 600 K, H\textsubscript{2}O dissociates into H and OH, causing the oxygen vacancies near the Ni exposure site to be covered by OH, which in turn deactivates the catalyst. However, under photoexcitation, the Ni atom site accumulates holes transferred to the H\textsubscript{2}O molecule, weakening the adsorption between Ni and H\textsubscript{2}O. As a result, the dissociation of H\textsubscript{2}O to form O does not fill the oxygen vacancies around the Ni site, ensuring the stability of the catalyst. This mechanism has also been confirmed experimentally. In another study, Liu {\it et al.}~\cite{tddft_application_6_Chen_2024} compared the dissociation of CO$_2$ driven directly by hot carriers generated via plasmonic excitation in metal clusters with the thermochemical reduction of CO$_2$ by H species. They used TDDFT to elucidate the mechanisms and characteristic timescales of both processes and provided experimental evidence from {\it in situ} infrared spectroscopy.

% openlam
\section{ABACUS in the OpenLAM Project} \label{sec:openlam}

As the collection of quantum mechanical data progressively encompasses the entire periodic table, the Deep Potential team has launched an ambitious project named the Large Atomic Model (LAM) based on the practice of DPA pre-training model.~\cite{2022-dpa1,zhang_DPA2_2024} Emphasizing open scientific collaboration, this initiative has been formally designated as the OpenLAM project,~\cite{openlam} reflecting its commitment to transparent development and community-driven progress. ABACUS serves as the cornerstone computational platform for generating cost-effective yet high-fidelity first-principles data to support OpenLAM's development. Through heterogeneous algorithm optimization specifically tailored for plane-wave basis solutions of the Kohn-Sham equations, the software achieves remarkable efficiency gains in data generation.

% APNS
%\section{ABACUS Pseudopotential-Numerical Atomic Orbital Square} \label{sec:apns}
\subsection{APNS Project} \label{sec:apns}

For those DFT codes that require pseudopotentials, several open-source codes for generating pseudopotentials are available, enabling users to tailor pseudopotentials to their specific requirements.~\cite{Schlipf2015,PhysRevB.105.125144} For LCAO calculations using ABACUS, it is crucial to validate both the NAO basis set and the pseudopotentials before practical applications. To address the above issues, we have initiated the "ABACUS Pseudopotential Numerical Atomic Orbital Square" (APNS) project.

Systematic verification of ABACUS remains essential in the APNS project to ensure reliable software development and maintain efficient high-throughput calculations. This aligns with growing research on validating DFT codes.~\cite{Talman1978,Toyoda2010,Wang2019,doi:10.1126/science.aad3000} Major databases like Materials Project,~\cite{jain2020materials} Materials Cloud,~\cite{talirz2020materials} and Computational Materials Repository~\cite{landis2012computational} have improved DFT practices. These datasets help balance accuracy and efficiency in DFT calculations while establishing code implementation standards. The APNS workflow enables automated high-throughput testing with optional user interaction. It evaluates pseudopotentials and NAO basis set across predefined and customizable systems, compatible with both PW and LCAO methods. The system also maintains a results database for test verification and validated pseudopotential/orbital file downloads. Here we list some examples done in the APNS project.

\begin{figure}
    \centering
    \includegraphics[width=0.9\linewidth]{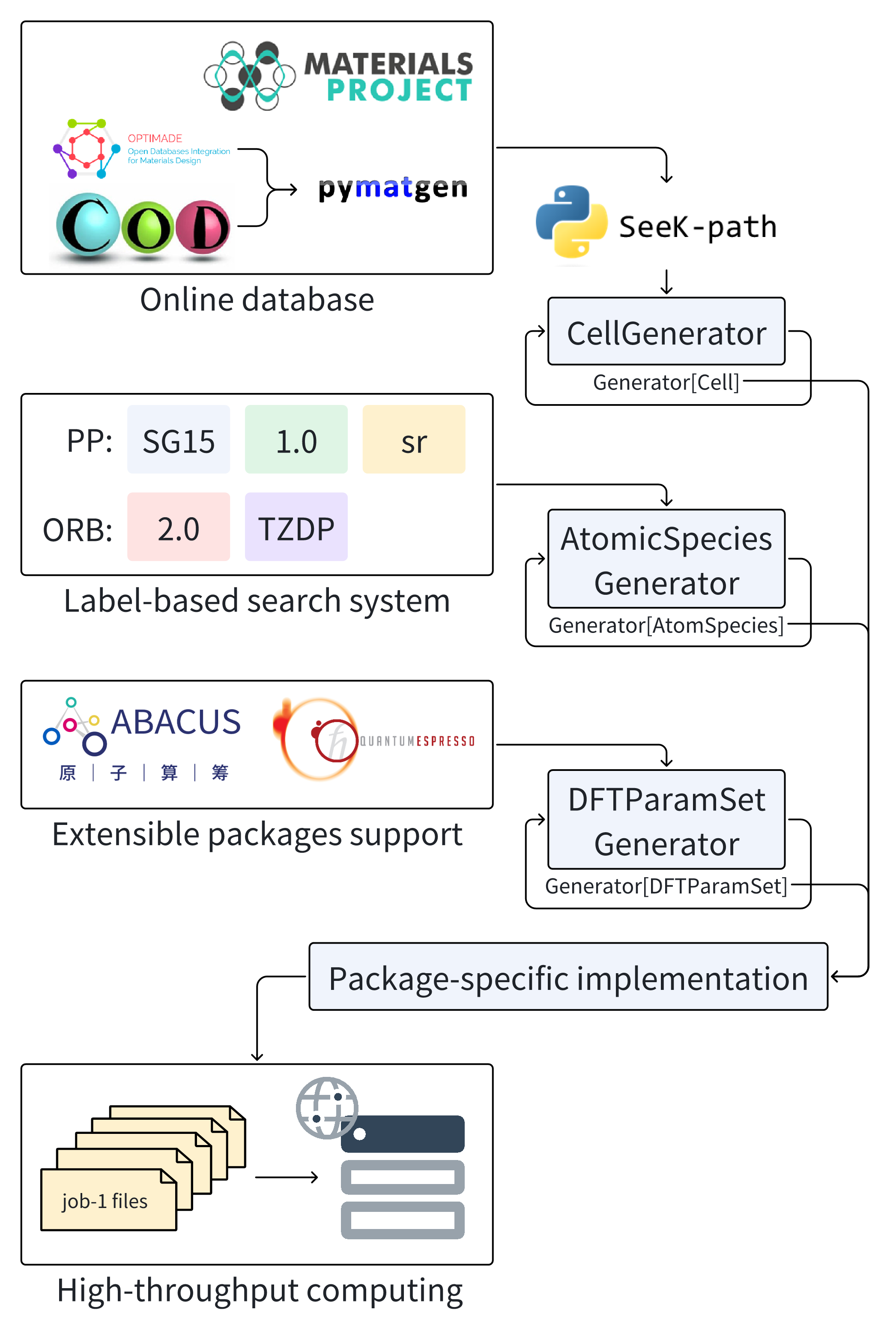}
    \caption{\MC{Workflow-chart of ABACUS Pseudopotential-Numerical atomic orbital Square automatic (APNS), high-throughput workflow incorporating structure generation, download and management (interfaced with online open crystal structure databases Materials Project,~\cite{jain2020materials} Optimade,~\cite{andersen2021optimade} Crystallography Open Database (COD),~\cite{gravzulis2009crystallography} etc.), symmetry analysis (interfaced with SeeK-path~\cite{togo2024spglib,HINUMA2017140}), modular and extensible software input generation (ABACUS and Quantum ESPRESSO~\cite{2020-qe}) and remote high-throughput computing support.}}
    \label{fig:apns1}
\end{figure}

Fig.~\ref{fig:apns1} demonstrates the standardized testing workflow, where the workflow has conducted efficiency and accuracy evaluations for multiple norm-conserving pseudopotential families including SG15,~\cite{SG15} PseudoDojo,~\cite{PseudoDojo} PD03/PD04~\cite{PWmat-pp} (with version variations and semi-core configurations), CP2K's Goedecker-Teter-Hutter pseudopotentials,~\cite{krack2005pseudopotentials} and three ultrasoft types (pslibrary 0.3.1/1.0.0,~\cite{pslibrary-theos} GBRV v1.5~\cite{GARRITY2014446,GBRV}). This workflow has additionally supported ABACUS feature development such as DFT+U implementation and ultrasoft pseudopotential compatibility.

The PW kinetic energy cutoff critically determines calculation accuracy, but excessively high values hinder computational efficiency. To address this trade-off, we established a systematic protocol for optimizing pseudopotentials through convergence testing across key parameters. This process involves two-phase validation. First, convergence testing of cutoff energies (20-300 Ry range) using stable crystal structures from Materials Project,~\cite{talirz2020materials} Crystallography Open Database,~\cite{gravzulis2009crystallography} and Optimade.~\cite{andersen2021optimade} Second, equation of state (EOS) verification with optimized cutoffs. Three physical properties, including Kohn-Sham energy, lattice pressure (stress tensor trace), and band structure similarity metrics~\cite{prandini2018precision} were used as convergence criteria across pseudopotential types.

\begin{figure*}
    \centering
    \includegraphics[width=1.0\textwidth]{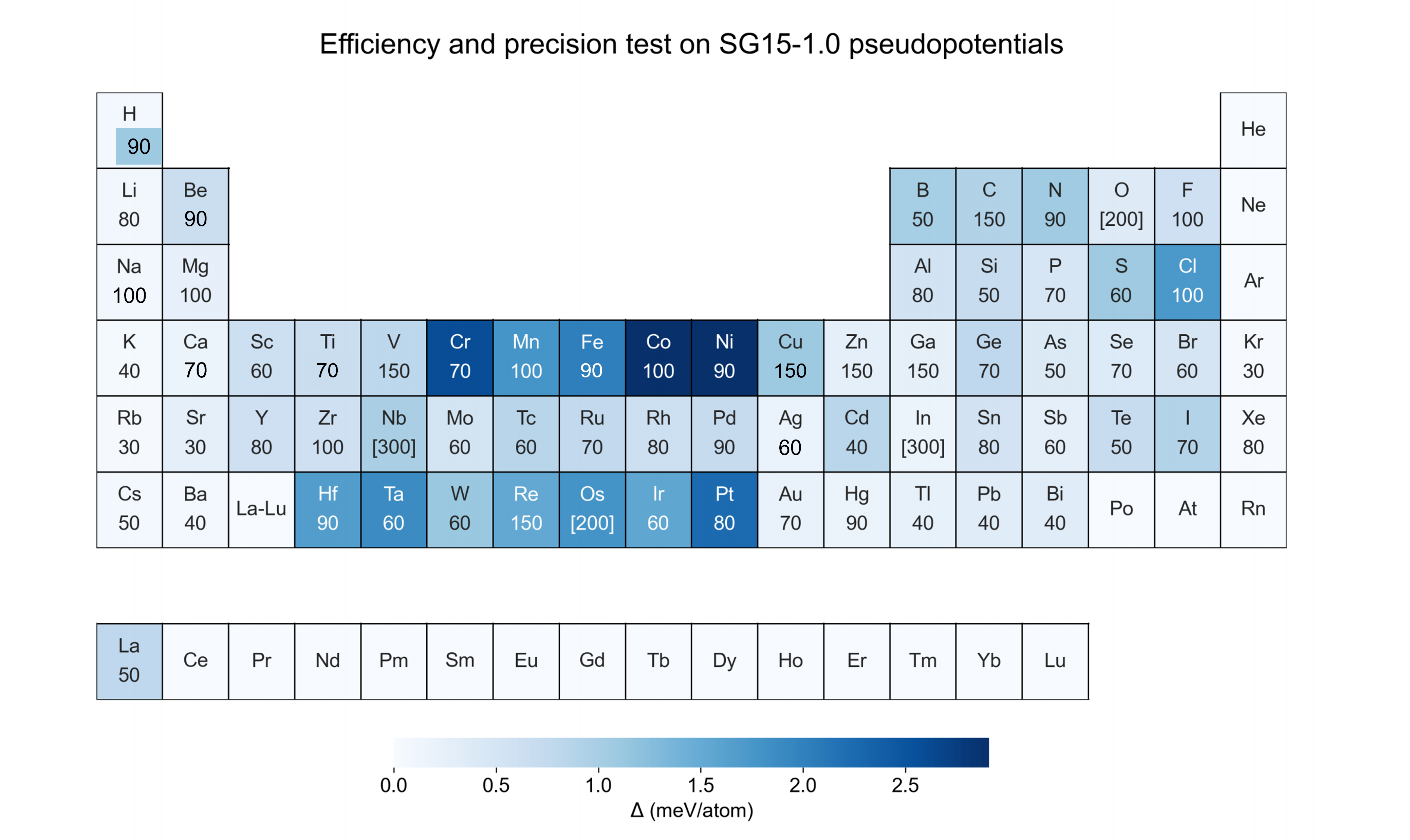}
    \caption{\MC{Efficiency and precision tests on SG15 1.0 pseudopotentials using PW basis of ABACUS. Number below the element symbol labels the converged kinetic energy cutoff ($E^\mathrm{kin}_\mathrm{c}$) under three thresholds: Kohn-Sham energy (< 1 meV/atom), pressure (< 0.1 kbar) and band similarity (< 10 meV). Large $E^\mathrm{kin}_\mathrm{c}$ values which enclosed by brackets are results from the oscillation or slow convergence with respect to pressure. Color of each block indicates the $\Delta$ value calculated with converged $E^\mathrm{kin}_\mathrm{c}$ averaged over BCC, FCC and Diamond structures. Blocks of elements whose pseudopotentials are not available are left empty.}}
    \label{fig:apns2}
\end{figure*}

EOS tests were performed using non-spin-polarized calculations, with $k$-sampled at a spacing of 0.06 Bohr$^{-1}$ in the Brillouin zone for all elements. The crystal volumes $V$ range from 94\% ($V_m$) to 106\% ($V_M$) with a 2\% step size. The kinetic energy cutoff for the plane-wave (PW) basis is set to the maximum of the converged values for all elements in the tested system. The $\Delta$ value,~\cite{lejaeghere2014error} defined as
\begin{equation}
    \Delta \left( a,b \right) =\sqrt{\frac{1}{V_M-V_m}\int_{V_m}^{V_M}{[E_a(V)-E_b(V)]^2\mathrm{d}V}}\label{delta-value-def}
\end{equation}
serves as a scalar indicator to quantify the difference between EOS profiles derived from pseudopotential and all-electron calculations (denoted by $a$ and $b$, respectively). The continuous integral in Eq.~(\ref{delta-value-def}) is evaluated over the curve fitted using the Birch-Murnaghan equation,~\cite{birch1947finite} where $E_a(V)$ and $E_b(V)$ represent the Kohn-Sham energies relative to their equilibrium values.

We constructed three Bravais lattices for each element, including BCC (Body-Centered-Cubic), FCC (Face-Centered-Cubic), and diamond structures. We calculate $E$-$V$ datasets and compare them with the all-electron results reported in Ref.~\onlinecite{Wang2019}. The converged kinetic energy cutoffs and corresponding $\Delta$ values for all available elements in the SG15 1.0 pseudopotential~\cite{Schlipf2015} are shown in Fig.~\ref{fig:apns2}.

For the NAO basis set, we conducted precision tests. Similar to plane-wave (PW) calculations, precision is indicated by the EOS; additionally, the energy difference at the minimum of the EOS profile can serve as an indicator of basis set completeness. For ABACUS LCAO calculations, reliability verification tests for any given pseudopotential are often hindered by the relatively complex and time-consuming NAO generation procedure, which requires sophisticated experience and insights. This makes a rapid, highly automated orbital generation code necessary. Furthermore, as the precision and efficiency of NAOs are expected to improve continuously, a platform for orbital generation algorithm development is needed. Therefore, an open-source subproject, ``ABACUS-ORBGEN'', has been included as a development component of APNS. The current implementation remains based on the algorithms proposed in Refs.~\onlinecite{Chen2010, Lin2021}, but includes features such as automatic setting of orbital configurations for pseudopotentials generated with ONCVPSP codes, automated improvement of orbital transferability through a series of bond length scans and potential curve fitting tasks, and a quick orbital quality estimation function.

\subsection{UniPero Model}

\begin{figure*}
    \centering
    \includegraphics[width=1.0\textwidth]{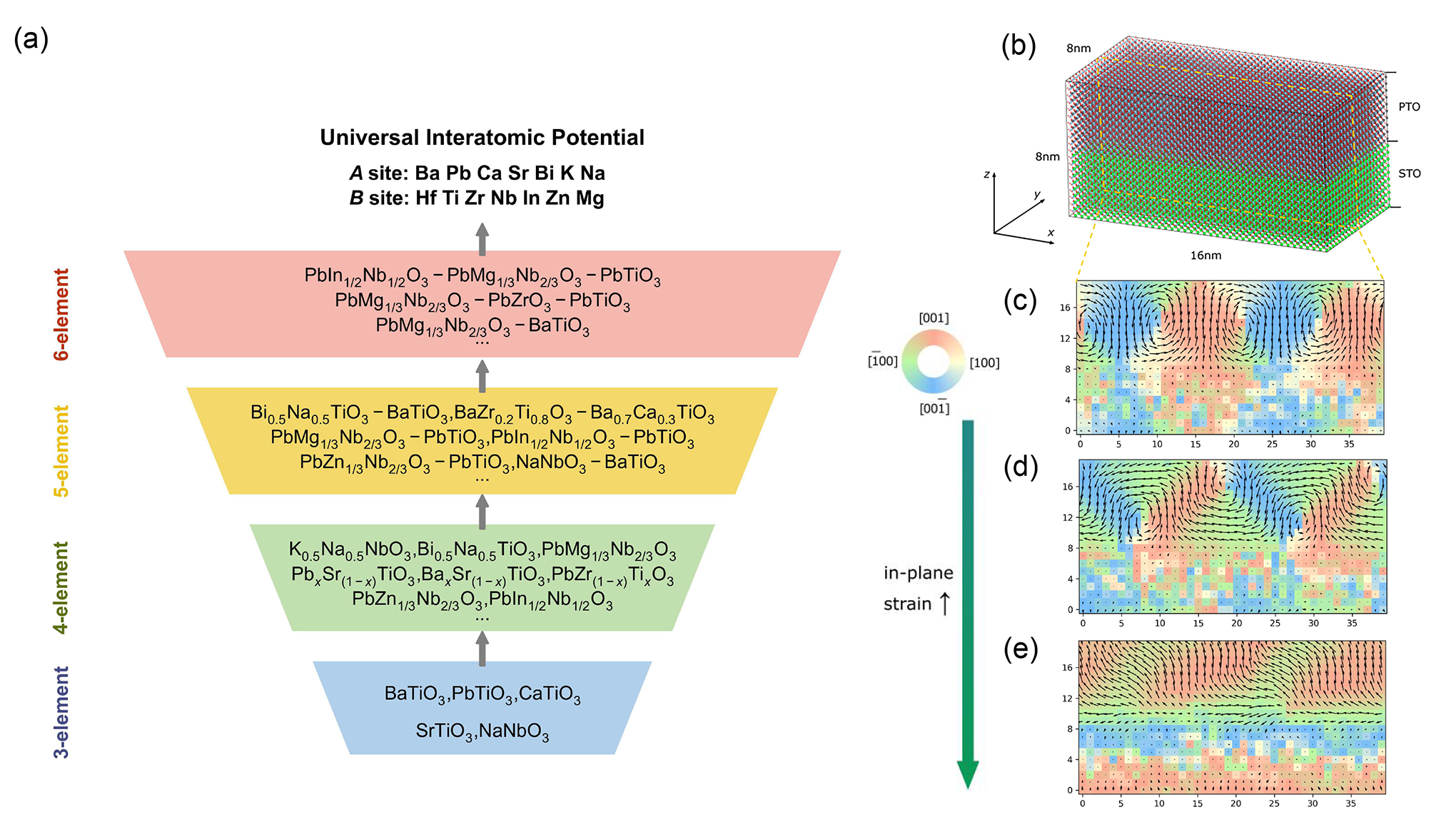}
    \caption{(a) Workflow for developing a universal force field of perovskite oxides. (Adapted with permission from Phys. Rev. B 108, L180104 (2023). Copyright 2023 American Physical Society.) (b) (PTO)$_{10}$/(STO)$_{10}$ superlattices. Unipero predicts an in-plane strain-induced transition from (c) ordered polar vortex lattice to (d) shifted polar vortex lattice, and to (e) electric dipole waves.}
    \label{fig:unipero}
\end{figure*}

Wu {\it et al.}~\cite{wu2023universal} utilized ``modular development of deep potential'' (ModDP)~\cite{Wu23p144102} to create a universal interatomic potential for perovskite oxides (UniPero) using a deep neural network with a self-attention mechanism. This potential spans 26 types perovskite oxides involving 200 components and 14 metal elements. All DFT calculations were performed with ABACUS. As depicted in Fig.~\ref{fig:unipero}(a), they first used DP-GEN to obtain a converged DPA-1 model~\cite{2022-dpa1} for three-element systems such as PbTiO$_3$, SrTiO$_3$. Then, the converged dataset served as a starting point for DP-GEN to improve DPA-1 for four-element perovskite systems such as Pb$_x$Sr$_{1-x}$TiO$_3$ and PbZr$_x$Ti$_{1-x}$O$_3$. Ultimately, the final DPA-1 model (UniPero) can describe six-element perovskite systems, including the ternary solid solution Pb(In$_{1/2}$Nb$_{1/2}$)O$_3$-Pb(Mg$_{1/3}$Nb$_{2/3}$)O$_3$–PbTiO$_3$ (PIN-PMN-PT). UniPero functions as a universal interatomic potential, effectively modeling a wide range of perovskites through molecular dynamics simulations.

The PbTiO$_3$/SrTiO$_3$ (PTO/STO) superlattices serve as a model system for exploring real-space topological textures, including flux closures, vortices, skyrmions, and merons. The authors tested the DPA-1 model by simulating strain-driven topological evolution in the PTO/STO superlattice. Fig.~\ref{fig:unipero}(b) illustrates a 40$\times$20$\times$20 supercell, containing 80,000 atoms, employed to model a (PTO)$_{10}$/(STO)$_{10}$ superlattice. At a strain state where the in-plane lattice constants $a_{IP}$ = 3.937~\AA and $a_{IP}$ = 3.930~\AA, the equilibrium state at 300 K obtained with DPMD simulations adopts an ordered polar vortex lattice with alternating vortex and antivortex (Fig.~\ref{fig:unipero}(c)). As the in-plane strain increases to $a_{IP}$ = 3.950~\AA, the vortex cores shift toward the PTO/STO interfaces (Fig.~\ref{fig:unipero}(b)). Finally, at a large tensile in-plane strain ($a_{IP}$ = 3.955~\AA), it becomes the periodic electric dipole waves characterized by head-to-tail connected electric dipoles in the form of a sine function (Fig.~\ref{fig:unipero}(e)). These results agree with the experimental observations~\cite{Gong21peabg5503} and previous MD simulations.~\cite{Wu23p144102}

\subsection{DPA-Semi Model}

\begin{figure*}
    \centering
    \includegraphics[width=1.0\textwidth]{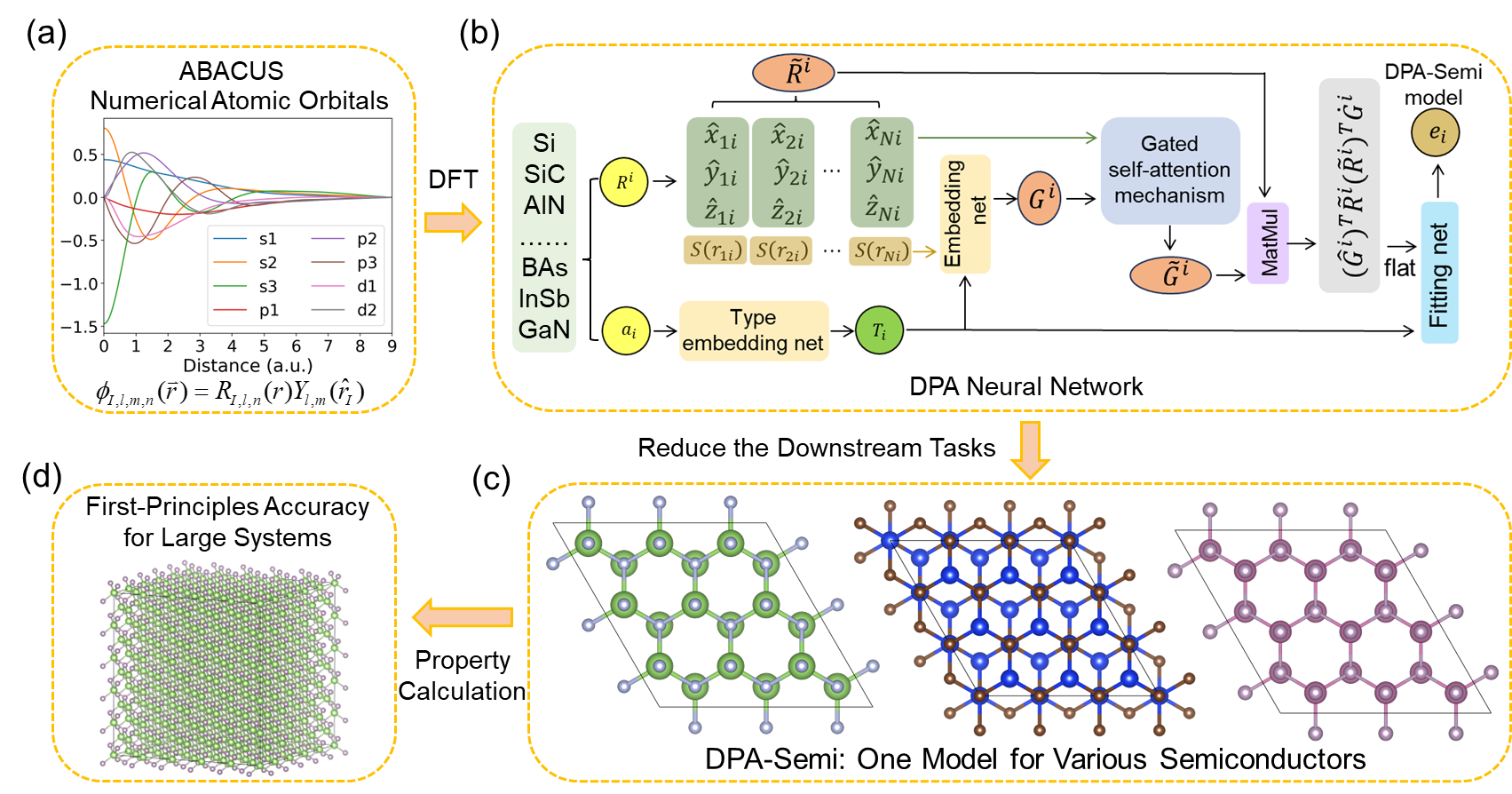}
    \caption{Procedures for developing the DPA-Semi model. (a) Generate atomic datasets using the ABACUS package based on the numerical atomic orbitals as basis set; (b) Generate the DPA-Semi model via the Gated self-attention mechanism based on the DFT atomic datasets; (c) The DPA-Semi model can be used for various kinds of semiconductors, and reduce the computational costs of downstream tasks; (d) The DPA-Semi model is readily applied to calculate properties of large-systems with GGA quality accuracy. (Adapted with permission from J. Chem. Theory Comput. 2024, 20, 13, 5717–5731. Copyright 2024 American Chemical Society.)}
    \label{fig:DPA-Semi-workflow}
\end{figure*}

Liu {\it et al.} \cite{2024-unisemi} generated over 200,000 DFT data for 19 bulk semiconductors ranging from group IIB to VIA, namely, Si, Ge, SiC, BAs, BN, AlN, AlP, AlAs, InP, InAs, InSb, GaN, GaP, GaAs, CdTe, InTe, CdSe, ZnS, CdS. All of the DFT calculations were performed with the ABACUS v3.2 package with the Perdew-Burke-Ernzerhof (PBE)~\cite{1996-pbe} exchange-correlation functional. The triple-zeta plus double polarization (TZDP) NAO basis sets were used for all DFT calculations. The atomic datasets are adopted as training data to generate an attention-based deep potential model using the DPA-1 method,~\cite{2022-dpa1} namely the DPA-Semi model. Fig.~\ref{fig:DPA-Semi-workflow} shows the procedures for developing the DPA-Semi model.

\begin{figure}
    \centering
    \includegraphics[width=0.95\linewidth]{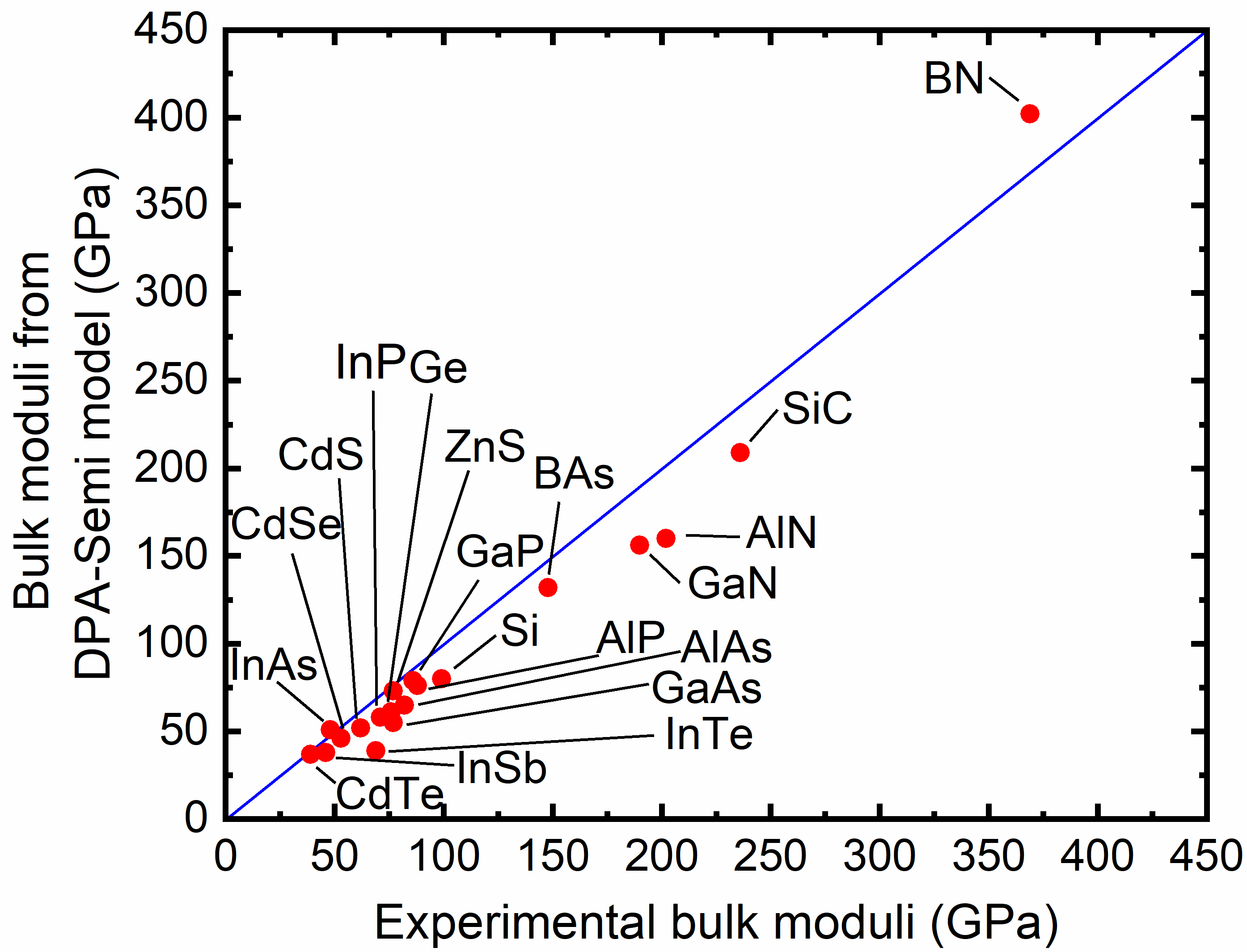}
    \caption{Predicted bulk moduli of various semiconductors by the DPA-Semi model, and available experimental data. (Adapted with permission from J. Chem. Theory Comput. 2024, 20, 13, 5717–5731. Copyright 2024 American Chemical Society.)}
    \label{fig:DPA-Semi-moduli}
\end{figure}

To validate the accuracy of the DPA-Semi model, we find the energy RMSEs of the BAs system are the smallest (0.004 eV/atom), while the force RMSEs of the InSb system are the smallest (0.11 eV/\AA). For the lattice constants, bulk moduli and elastic constants, the results from DPA-Semi are in excellent agreement with the DFT results, as the bulk moduli data are compared in Fig.~\ref{fig:DPA-Semi-moduli}. This work provided reliable evidence that the DPA-Semi model can be readily employed to study several semiconductor systems with {\it ab initio} accuracy.

\subsection{DPA-1, DPA-2 and DPA-3} \label{sec:alloy}

The DPA-1 model~\cite{2022-dpa1} is a universal machine learning interatomic potential that leverages the attention mechanism for molecular and materials modeling. Trained on the OC20 benchmark dataset and specialized materials systems, including aluminum-magnesium-copper alloys, high-entropy alloys, and solid-state electrolytes, this architecture achieves robust transferability across diverse atomic environments. With coverage spanning 56 elements in the periodic table, DPA-1 employs a pretraining-fine-tuning paradigm. The attention-based design supports first-principles accuracy while maintaining computational efficiency comparable to classical force fields.

The DPA-2 model~\cite{zhang_DPA2_2024} employs a multi-task learning framework that effectively integrates datasets from diverse DFT sources. This architecture substantially reduces the model's dependence on data uniformity, enabling utilization of datasets generated by different methods. Notably, the pretraining phase of DPA-2 incorporates systematic learning of chemical compositions and spatial configurations, which dramatically reduces the demand for task-specific fine-tuning data in downstream applications.

The recently developed DPA-3~\cite{2025-dpa3} constructs a hierarchical graph neural network architecture based on line graph transformations, demonstrating substantial improvements in scaling law adherence and cross-dataset generalization capabilities. During the development of both DPA-2 and DPA-3 iterations, the ABACUS software served as a critical computational infrastructure, facilitating DFT calculations across diverse material datasets. Here are two examples.

%\paragraph*{Alloy}
First, based on DFT calculations performed using ABACUS, we are developing general machine-learning-based potential models for alloys, covering 53 elements in the periodic table. The elements are Li, Na, K, Be, \textbf{Mg}, Ca, Sr, Sc, \textbf{Y}, La, \textbf{Ti}, \textbf{Zr}, \textbf{Hf}, \textbf{V}, \textbf{Nb}, \textbf{Ta}, \textbf{Cr}, \textbf{Mo}, \textbf{W}, \textbf{Cu}, Ag, Au, \textbf{Zn}, Cd, \textbf{Mn}, \textbf{Re}, \textbf{Fe}, \textbf{Co}, \textbf{Ni}, Os, Ir, Pt, Rh, Ru, Pd, Ce, Pr, Nd, Sm, Gd, Tb, Dy, Ho, Er, Tm, Lu, \textbf{Al}, Ga, In, Si, Ge, Sn, Pb, where bold ones are commonly used in structural alloys. The specific pseudopotentials utilized for these elements are listed in Ref.~\onlinecite{links-2024} and the calculations were performed efficiently by the ABACUS package on the Deep Computing Unit (DCU) hardware. Specifically, the DP-GEN and APEX workflows generated various structures,~\cite{dpgen,apex} including perfect and perturbed crystal structures, vacancies, interstitials, and surfaces of metals and alloys. Some structures were selected through the concurrent learning workflow in DP-GEN and then labeled by ABACUS, resulting in a training dataset of $\sim$24,000 entries. By integrating previous training datasets from AI Square and the OpenLAM project,~\cite{openlam} the latest general LAM for alloys achieved a root mean square error (RMSE) of $\sim$26 meV/atom for energies and $\sim$0.20 eV/{\AA} for atomic forces across the 53 metals and their alloys. This general LAM for alloys demonstrates superior performance in predicting various properties, including lattice parameters, elastic constants, point defects, and surface formation energies, compared to other LAMs.~\cite{mace} ABACUS has proven to be a stable and reliable DFT workhorse for diverse structures covering these 53 elements.

Second, during the construction of the dataset for the DPA-1 and DPA-2 models of high-pressure SuperHydrides, all labels were calculated using ABACUS with plane-wave basis sets. The thermodynamic stability of the discovered structures was confirmed by their energies above the convex hull, which was determined through structure optimization results from ABACUS. Dynamic stability was assessed using the phonon dispersion spectrum calculated by phonopy,~\cite{23JPCM-phonopy, 23JPSJ-phonopy} with all forces calculated by ABACUS. Our DFT calculations were performed on DCU hardware, significantly enhancing the efficiency of SCF calculations compared to traditional CPUs. We obtained a dataset comprising 218,349 data frames, covering 29 elements and a pressure range of 150-250 GPa. Utilizing this data, the DPA-1 model achieved a training RMSE of 48.1 meV/atom for energy and 334.8 meV/{\AA} for force. The DPA-2 model achieved a training RMSE of 55.2 meV/atom for energy and 298.8 meV/{\AA} for force. On the testing dataset, the DPA-1 model yielded an RMSE of 37.6 meV/atom for energy and 171.4 meV/{\AA} for force, while the DPA-2 model yielded an RMSE of 37.4 meV/atom for energy and 122.1 meV/{\AA} for force. Given the large dataset required for the Superhydrides model, ABACUS proved to be the most economical DFT software available. The high-quality training results further attest to the stable performance of ABACUS.

% interface with other softwares
\section{Interfaces to Other Packages} \label{sec:interface}

\subsection{DeePKS-kit}

DeePKS-kit~\cite{21JCTC-Chen,22CPC-Chen} is an innovative computational framework designed to bridge machine learning with quantum-mechanical simulations. The software integrates deep learning techniques with the Kohn-Sham equations to refine exchange-correlation functionals and can be used with ABACUS or PySCF.~\cite{2020-pyscf} By training neural network on high-fidelity quantum chemistry data, DeePKS-kit dynamically corrects errors in conventional DFT approximations, enabling predictions of electronic properties with near-chemical-accuracy at a fraction of the computational cost.

\begin{figure}
    \centering
    \includegraphics[width=1.0\linewidth]{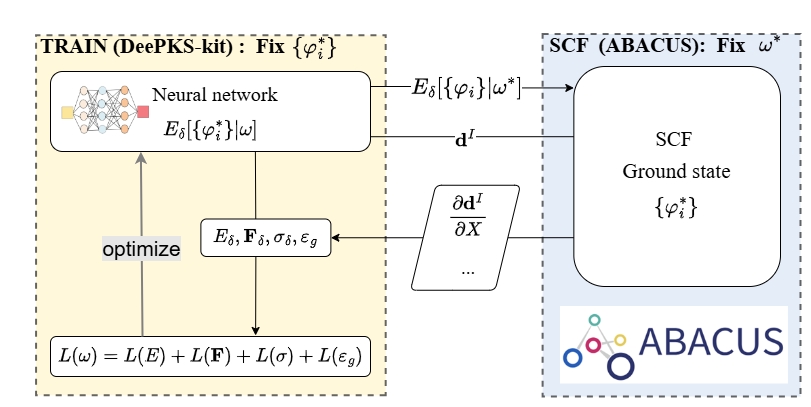}
    \caption{Workflow of training a DeePKS model. The blue and yellow boxes represent the SCF step performed by the ABACUS software and the TRAIN step performed by the DeePKS-kit software, respectively. These two steps iterate over each other until convergence. The neural network can be optimized considering energy $E$, atomic force $\mathbf{F}$, stress $\mathbf{\sigma}$ and band gap $\varepsilon_g$.}
    \label{fig:deepks-workflow}
\end{figure}

By adopting both DeePKS-kit and ABACUS packages, we optimize the DeePKS model in two iterative steps. As shown in Fig.~\ref{fig:deepks-workflow}, $L(\omega)$ is the loss function. Loss function $L(\omega)$ can be composed of various energetic terms including the energy $E$, the atomic force $\mathbf{F}$, the stress tensor $\mathbf{\sigma}$ and the band gap $\varepsilon_g$ of the interested system. The first step is the SCF step, where we fix parameters $\{\omega^{\ast}\}$ in the DeePKS model and solve the Kohn-Sham equation to obtain the ground state electronic wave function. The second step is the TRAIN step, where the electronic wave functions are fixed and the model parameters $\{\omega\}$are optimized. These two steps repeat until convergence is reached.~\cite{21JCTC-Chen}

\subsection{DeePMD-kit} \label{DeePMD}

DeePMD~\cite{18PRL-Zhang,18NIPS-Zhang} is a widely-used~\cite{22MF-Wen} machine-learning molecular dynamics method based on the neural-network potential. ABACUS has an interface with the DeePMD-kit package,~\cite{18CPC-Wang,23JCP-Zeng} allowing its deployment as a molecular dynamics engine in executing DeePMD simulations. Trained on system energy, atomic forces, and lattice stress derived from DFT calculations, this neural network potential enables DeePMD to replicate the potential energy surface with first-principles accuracy. Since the neural network potential is computationally more efficient than the DFT method, DeePMD can simulate large systems at a long time scale, which is barely accessible for AIMD simulations.

Regarding training, DeePMD requires a large number of atomic configurations with system energies, atomic forces, and optionally virial tensors calculated by the DFT software as the training data. Within ABACUS, these requisite data can be obtained either from the self-consistent calculations (section \ref{SCF}) or from the AIMD simulations (section \ref{sec:md}). The Python package \textbf{dpdata}~\cite{dpdata} for atomic simulation data manipulation facilitates convenient transfer of data to the DeePMD training data format.

\subsection{DP-GEN}

To develop the Deep Potential Energy Surface (PES) more efficiently, the DP-GEN (Deep Potential Generator) framework~\cite{19PRM-Zhang,dpgen} employs a recurrent, adaptive learning scheme that systematically refines machine-learned interatomic potentials. The process begins with a minimal initial dataset of atomic configurations labeled via DFT calculations. Through an iterative active learning loop, DP-GEN trains Deep Potential models with varied initializations to explore new regions of the configuration space. During MD simulations, the framework identifies configurations where model predictions diverge significantly and prioritizes these for further DFT validation. This targeted sampling ensures that the training dataset grows strategically, focusing on physically relevant yet under-represented regions of the PES.

The workflow integrates automated DFT computations to validate candidate configurations, DP-GEN has interfaces with first-principles software such as ABACUS, VASP, and Quantum ESPRESSO. As new DFT data points are added, the Deep Potential models undergo retraining, progressively improving their accuracy across diverse atomic environments. The process continues until predefined error thresholds for forces and energies are met. A key innovation of DP-GEN is its parallelized sampling strategy, which enables simultaneous exploration of multiple trajectories to efficiently map high-dimensional configuration spaces while minimizing redundant DFT calculations.

The DP-GEN framework has proven particularly effective for complex systems like multicomponent alloys,~\cite{2021-wu-deep,2024-shi-revisiting} chemical reaction pathways,~\cite{2021-zeng-development,2022-yang-using} and phase transitions under extreme conditions,~\cite{2020-thomas,2023-fu-unraveling} where traditional sampling methods struggle to capture rare events or subtle energy landscapes. Open-source and modular by design, DP-GEN supports customization for specific materials or molecules, offering a versatile tool for accelerating the development of robust, quantum-mechanically informed potentials.

\subsection{DeepH}

The deep-learning density functional theory Hamiltonian (DeepH) method is a neural network approach based on equivariant graph neural networks for modeling the DFT Hamiltonian as a function of material structure.~\cite{2022-deeph} Leveraging the sparsity of the Hamiltonian matrix under NAOs and its compatibility with Walter Kohn's ``quantum nearsightness principle'',~\cite{Kohn1996} the DeepH method can learn from training data of small structures to infer electronic Hamiltonians of large structures, achieving high prediction accuracy with linear-scaling computational cost. The predicted Hamiltonian may be subsequently utilized for post-processing to evaluate properties including band structures, optical properties, and response properties from density functional perturbation theory, etc.~\cite{2022-deeph,2023-deeph-e3,2024-deeph-dfpt}

% Remark: The arxiv version of DeepH is dated back to Apr. 2021
The interface between ABACUS and DeepH was developed in 2022, shortly after the invention of the DeepH approach in 2021. The interface ensures compatibility between ABACUS and various versions of DeepH, including DeepH-pack,~\cite{2022-deeph,deeph-interface} DeepH-E3,~\cite{2023-deeph-e3,deeph-e3} and DeepH-2.~\cite{2024-deeph2} Among these, DeepH-E3 stands out as the most stable open-source implementation to date and has been utilized with ABACUS in several example studies.~\cite{deeph-e3,Tang2024,2024-yang-abacus+deeph} The ABACUS-DeepH interface is versatile, supporting spin-orbit coupling and magnetic systems, and facilitates ABACUS's integration with xDeepH--a specialized variant of DeepH for predicting the electronic structures of magnetic materials.~\cite{2023-xdeeph} To address the modified sparsity patterns in hybrid DFT Hamiltonians, a dedicated toolkit, named DeepH-hybrid, has been developed, which verified DeepH's applicability to hybrid-functional Hamiltonians generated by ABACUS.~\cite{Tang2024,deephhybrid_zenodo} An interface between ABACUS and DeepH-DFPT (a generalization of the DeepH approach for deep-learning density functional perturbation theory) is currently under development to accelerate calculations of electron-phonon coupling.~\cite{2024-deeph-dfpt} In addition, a recently developed ``Hamiltonian Projection and Reconstruction to atomic Orbitals" (H-PRO) method can transform the DFT Hamiltonian of plane-wave basis into localized basis, making the DeepH compatible with the plane-wave mode of ABACUS.~\cite{2024-deeph-pw} Very recently, a universal materials model (UMM) of DeepH, named DeepH-UMM, has been developed, demonstrating exceptional transferability across a wide range of material structures composed of various elements.~\cite{2024-deeph-umm} Given ABACUS's versatility, further development of DeepH-UMM in conjunction with ABACUS holds significant promise for advancing materials discovery.

The current ABACUS-DeepH interface is available on GitHub.~\cite{deeph-interface} For inference with trained DeepH models, ABACUS provides an efficient method for generating the overlap matrix, from which the Hamiltonian can be predicted by DeepH. It is important to note that the overlap and Hamiltonian matrices are assumed to share the same sparsity pattern in DeepH. To ensure clarity, fixing such sparsity pattern with tools provided in DeepH-hybrid is recommended.~\cite{deephhybrid_zenodo}

\subsection{DeePTB}

DeePTB is an open-source package that leverages deep learning to accelerate {\it ab initio} electronic simulations.~\cite{guDeep2024,zhouyinLearning2024} Its integration with ABACUS creates a powerful synergy, combining first-principles calculations with advanced machine learning techniques for efficient large-scale electronic structure predictions. Notably, it has demonstrated the capability to simulate systems containing up to millions of atoms,~\cite{guDeep2024} a scale previously unattainable with traditional methods.

\begin{figure}
\centering
\includegraphics[width=0.95\linewidth]{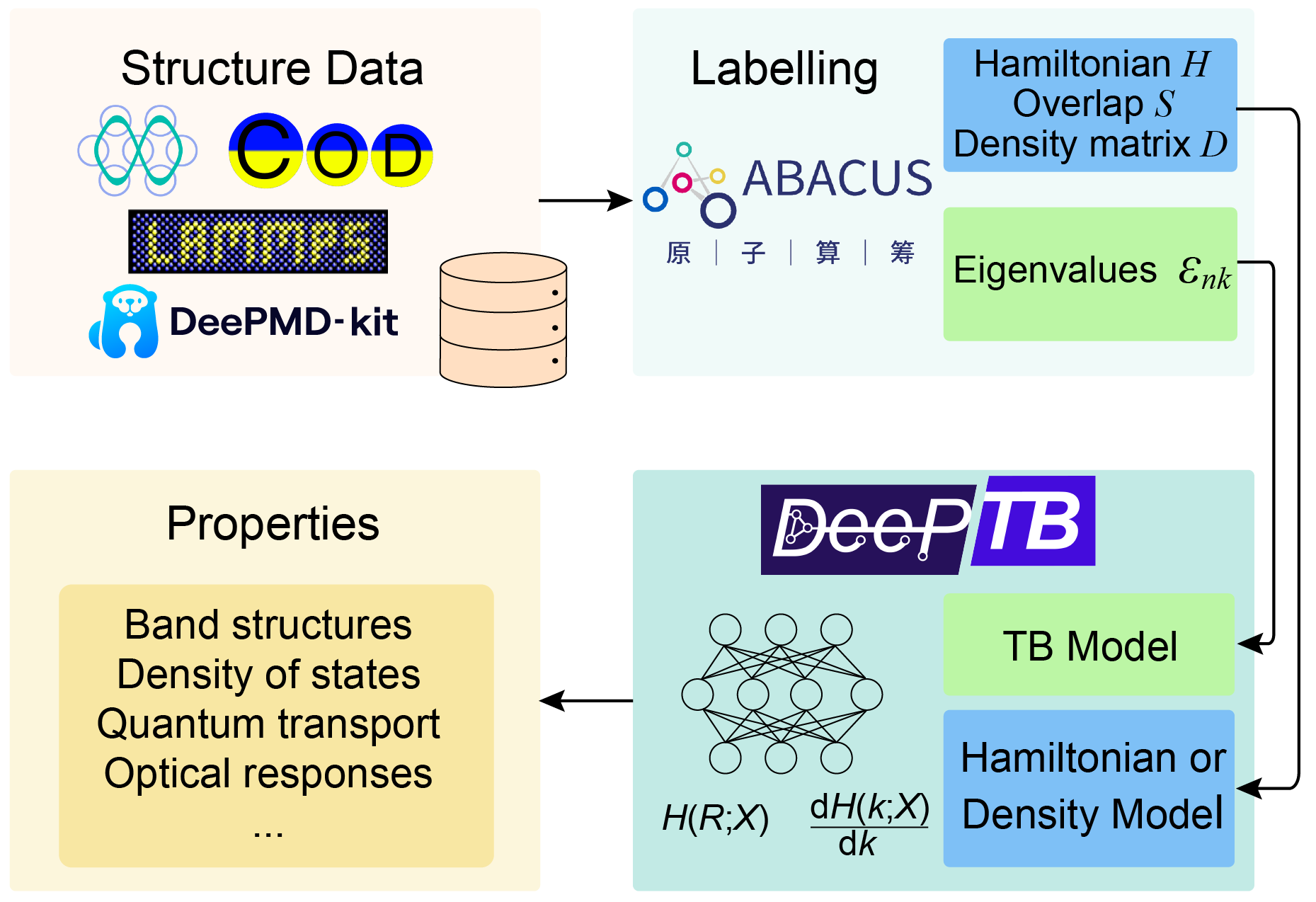}
\caption{Workflow-chart of ABACUS-DeePTB integration for deep learning-based TB and quantum operators (Hamiltonian, overlap, and density matrices) predictions.}
\label{fig:ABACUS-DeePTB}
\end{figure}

Fig.~\ref{fig:ABACUS-DeePTB} illustrates the ABACUS-DeePTB workflow. ABACUS performs DFT calculations on training structural data obtained from crystal databases or molecular dynamics simulations, generating essential quantum mechanical data as labels including energy eigenvalues, NAOs basis Hamiltonian, overlap, and density matrices. DeePTB utilizes this ABACUS-generated data to train two key models: the DeePTB-SK \cite{guDeep2024} and DeePTB-E3 \cite{zhouyinLearning2024} models. The DeePTB-SK model uses energy eigenvalues to develop an improved Slater-Koster TB model for efficient electronic structure prediction, while the DeePTB-E3 model, based on the SLEM (Strictly Localized Equivariant Message-passing) approach, predicts quantum operators including the Hamiltonian, overlap, and density matrices.

This integrated approach significantly enhances the efficiency and scalability of electronic structure simulations, enabling the study of complex and large-scale material systems previously limited by computational constraints. For instance, recent work combining DeePTB with non-equilibrium Green's function (NEGF) methods has demonstrated unprecedented efficiency in quantum transport simulations of large-scale nanodevices.~\cite{zouDeep2024} The ABACUS-DeePTB collaboration demonstrates the potential of combining traditional DFT methods with cutting-edge machine learning in computational materials science, opening new avenues for high-throughput materials discovery and design.

\subsection{HamGNN}

HamGNN (Hamiltonian Graph Neural Network) is an end-to-end model specifically designed for atomic orbital (AO)-based electronic Hamiltonian parameterization.~\cite{2023-zhong-transferable} It inherently satisfies the symmetry requirements of rotation, parity, and translation. HamGNN parameterizes spin-orbit coupling (SOC) and magnetic interactions by incorporating physical principles, thereby accurately capturing the SOC effect and effectively describing the magnetic Hamiltonian across various materials.~\cite{2023-zhong-accelerating} Moreover, HamGNN enhances the accuracy and stability of band structure predictions through a two-stage training strategy: real-space pre-training followed by reciprocal-space fine-tuning, which effectively combines local atomic environments with global electronic behavior within the Brillouin zone. In practical applications—including the modeling of high-order Bi topological insulators,~\cite{2025-zhao-revealing} Bi ferroelectric domain topological interface states,~\cite{2025-luo-topological} and metal-oxide-semiconductor field-effect transistor (MOSFET) device design~\cite{2025-wang-silicon}—HamGNN has demonstrated both high computational efficiency and strong transferability.

The performance of HamGNN relies critically on high-quality training data, where ABACUS plays an essential role. ABACUS stands out as a key tool for building high-precision HSE Hamiltonian training databases, as it is one of the few DFT software packages based on numerical atomic orbitals that support HSE hybrid functionals. When combined with the cross-system transferability of the HamGNN model, this capability enables the development of a universal HSE Hamiltonian model covering the entire periodic table.~\cite{2024-zhong-universal} Recently, the Uni-HamGNN framework achieved universal modeling of SOC effects across the periodic table by decomposing the SOC Hamiltonian into spin-independent terms and SOC correction terms and employing a delta-learning strategy to fit them separately.~\cite{2025-zhong-universal} By incorporating a limited number of SOC Hamiltonian matrices from ABACUS, a universal SOC model with HSE-level accuracy can be further established using the delta-learning strategy. This approach, when applied in the training of the universal HSE SOC model, can significantly reduce the expensive computational cost of HSE SOC calculations.

These advancements in universal Hamiltonian modeling open up numerous practical applications with significant computational advantages. Using the universal HamGNN models trained on HSE Hamiltonian matrices as a computational engine can substantially accelerate the computation of carrier mobility and superconductivity-related electron-phonon coupling properties at the hybrid functional level.~\cite{2024-zhong-accelerating} Furthermore, HamGNN can accurately predict Hamiltonians and non-adiabatic coupling vectors (NACV) that continuously change with atomic displacement, enabling dynamics simulations of excited states at HSE precision.~\cite{2025-zhang-advancing} Looking forward, the deepening integration of ABACUS and HamGNN holds promise for developing even more powerful, universal electronic Hamiltonian prediction models, driving the materials computation field towards new horizons of enhanced efficiency and intelligence. The interface for ABACUS and HamGNN is available online.~\cite{abacus-hamgnn}

\subsection{PYATB}

PYATB (PYthon Ab initio Tight-Binding simulation package) is a Python package based on {\it ab initio} tight-binding Hamiltonian, designed as a tool for calculating and analyzing the electronic structures of materials.~\cite{23CPC-PYATB} It can be viewed as a post-processing program for ABACUS. When ABACUS completes the SCF calculations and generates the tight-binding Hamiltonian, PYATB utilizes this Hamiltonian to perform electronic property calculations. This eliminates the need for the cumbersome construction of \MCC{Maximally localized Wannier functions (MLWFs)} while strictly preserving the Hamiltonian's symmetry, making it particularly well-suited for high-throughput workflows in electronic structure calculations of various materials. Currently, PYATB mainly offers three functional modules, i.e., Bands, Geometric, and Optical. This allows ABACUS, in combination with PYATB, to compute basic band structures, fat bands, and projected density of states (PDOS). Additionally, it can analyze the Berry curvature and Chern number of topological materials,~\cite{10RMP-BerryPhase, 10RMP-TI} as well as compute first- and second-order optical responses such as optical conductivity, shift current, second harmonic generation~\cite{92book-Sturman,00PRB-Sipe,jin_Peculiar_2024} and Berry curvature dipole.~\cite{15PRL-BCD,pang_Tuning_2024a} 

Once the Hamiltonian matrix $\boldsymbol{H}(\mathbf{k})$ and the overlap matrix $\boldsymbol{S}(\mathbf{k})$ are given, the element dipole matrix is given by
\begin{equation}
A_{\mu\nu}^{\mathbf{R}}(\mathbf{k}) = \sum_{\mathbf{R}} \mathrm{e}^{i\mathbf{k}\cdot\mathbf{R}} \langle \phi_{\mu0}|\mathbf{r}|\phi_{\nu\mathbf{R}}\rangle.
\end{equation}
Using these input parameters, PYATB solves the following generalized eigenvalue problem of Eq.~\ref{eq:general-eigenproblem}. Subsequently, based on the wavefunctions, the Berry curvature~\cite{10RMP-BerryPhase, 21JPCM-BerryCurOrb} can be calculated. This allows for the calculation of various topological and optical properties of materials. One can refer to Ref.~\onlinecite{23CPC-PYATB} for the implementation details.

\subsection{Hefei-NAMD}

\begin{figure}
    \centering
    \includegraphics[width=1.0\linewidth]{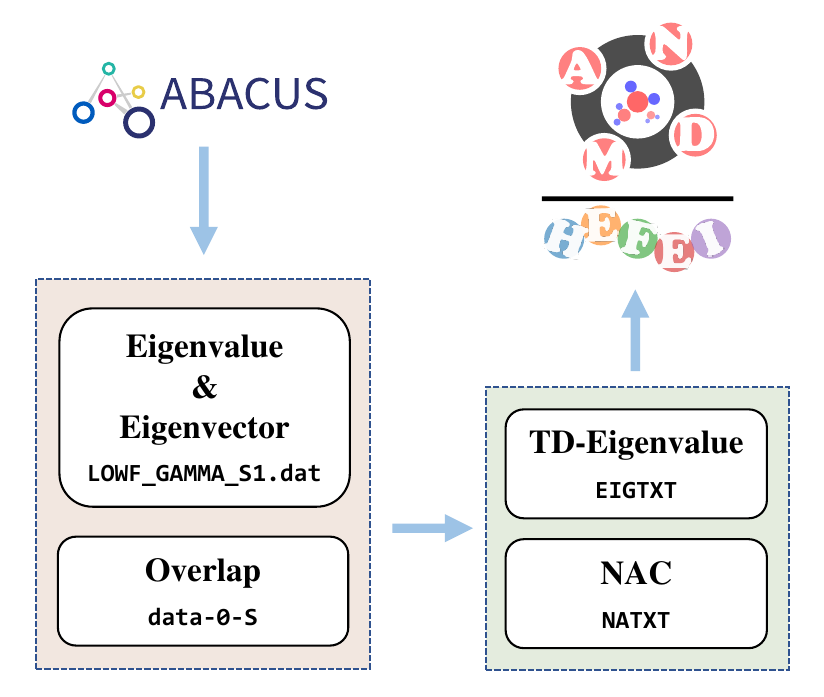}
    \caption{\MC{Workflow-chart of Hefei-NAMD interfaced with ABACUS.}}
    \label{fig:abacus_hnamd}
\end{figure}

Hefei Non-Adiabatic Molecular Dynamics (Hefei-NAMD) is an {\it ab initio} simulation suite for studying excited carrier dynamics in condensed matter systems. It combines real-time time-dependent density functional theory (TDDFT) with the fewest-switches surface hopping scheme and classical-path approximation. Hefei-NAMD has been used to investigate processes such as charge transfer, electron–hole recombination, spin dynamics, and exciton dynamics.~\cite{hnamd2019_zheng, ZZF_PRB_2022, JiangX2021, ZhengZ_2023_NatCompSci} It allows the study of excited carrier dynamics in energy, real, and momentum spaces, while also exploring interactions with phonons, defects, and molecular adsorptions, providing valuable insights into ultrafast dynamics at the atomic scale. Hefei-NAMD works by interfacing with other ab initio codes, e.g. VASP, ABACUS, etc. Fig.~\ref{fig:abacus_hnamd} shows the flowchart of the simulation using Hefei-NAMD interfaced with ABACUS.

\subsection{PEXSI} \label{sec:pexsi}

PEXSI (Pole EXpansion and Selected Inversion)~\cite{CMS2009,JCPM2013} offers an efficient approach for solving the KS equation while circumventing the computational limitations inherent in conventional diagonalization techniques. This method directly computes the real-space density matrix through a $P$-term pole expansion of the Fermi-Dirac function, expressed as
\begin{equation}
     f(\boldsymbol H - \mu \boldsymbol S) \approx \mathrm{Im} \sum_{l=1}^{P} \omega_l^\rho\left[\boldsymbol H - (z_l+\mu)\boldsymbol S\right]^{-1},
\end{equation}
where $\boldsymbol H$ denotes the Hamiltonian matrix, $\boldsymbol S$ is the overlap matrix, $z_l$ and $\omega_l$ correspond to the precomputed poles and weights of the complex contour integration, respectively. The numerical stability of this expansion is ensured through careful selection of these parameters. The computationally intensive matrix inversion operations are efficiently performed using the selected inversion algorithm.

Notably, PEXSI exhibits superior computational scaling compared to traditional diagonalization methods. While standard diagonalization scales cubically as $O(N^3)$, PEXSI achieves at most quadratic scaling ($O(N^2)$), where $N$ is the dimension of the matrix. The method's architecture features a two-level parallelization scheme, incorporating both pole expansion parallelism and selected inversion parallelism, thereby demonstrating exceptional scalability. ABACUS has implemented an interface to the parallel version of PEXSI for solving the KS equation expanded with the NAO basis.

\subsection{Other Packages}

%\subsubsection{ASE}
ASE (Atomic Simulation Environment) comprises a comprehensive suite of tools and Python modules designed for setting up, manipulating, executing, visualizing, and analyzing atomic simulations.~\cite{ASElink} We have developed an ABACUS calculator~\cite{ASE-ABACUSlink} that interfaces with ASE (version 3.23.0b1) and that enables seamless integration of ASE’s robust functionalities for both pre-processing and post-processing tasks. Regarding pre-processing capabilities, the interface facilitates tasks such as converting structural files into various formats, generating $k$-point grids, etc. For post-processing, the integration with ASE empowers ABACUS users with advanced tools for visualizing and analyzing simulation results. Moreover, the self-consistent calculations of ABACUS can be combined with ASE's built-in optimization algorithms, facilitating structural relaxation procedures, conducting MD simulations under varying conditions, performing precise phonon calculations based on the finite displacement method, carrying-out global optimization by genetic algorithm (GA), locating transition states (TS) of reactions by saddle point refinement, etc. For TS search, as an example, the nudged elastic band (NEB)~\cite{NEB-Karsten, IT-NEB-Henkelman, CI-NEB-Henkelman} method and the Dimer~\cite{Dimer-Henkelman, Dimer-Olsen, Dimer-Heyden, Dimer-Kastner} method, which represent double-ended TS searching methods and single-end TS methods respectively, are implemented in ASE. Besides, some other enhanced NEB methods are coded in the ASE package, such as dynamic NEB~\cite{DyNEB-Lindgren} and AutoNEB.~\cite{AutoNEB-Kolsbjerg} Furthermore, there are other saddle point refinement algorithms implemented in ASE, and some of them show better performance, like Sella.~\cite{Sella-Hermes-2019, Sella-Hermes-2021, Sella-Hermes-2022} With the above-mentioned methods available in ASE, ABACUS can be used for searching for transition states.

\begin{figure}
    \centering
    \includegraphics[width=0.9\linewidth]{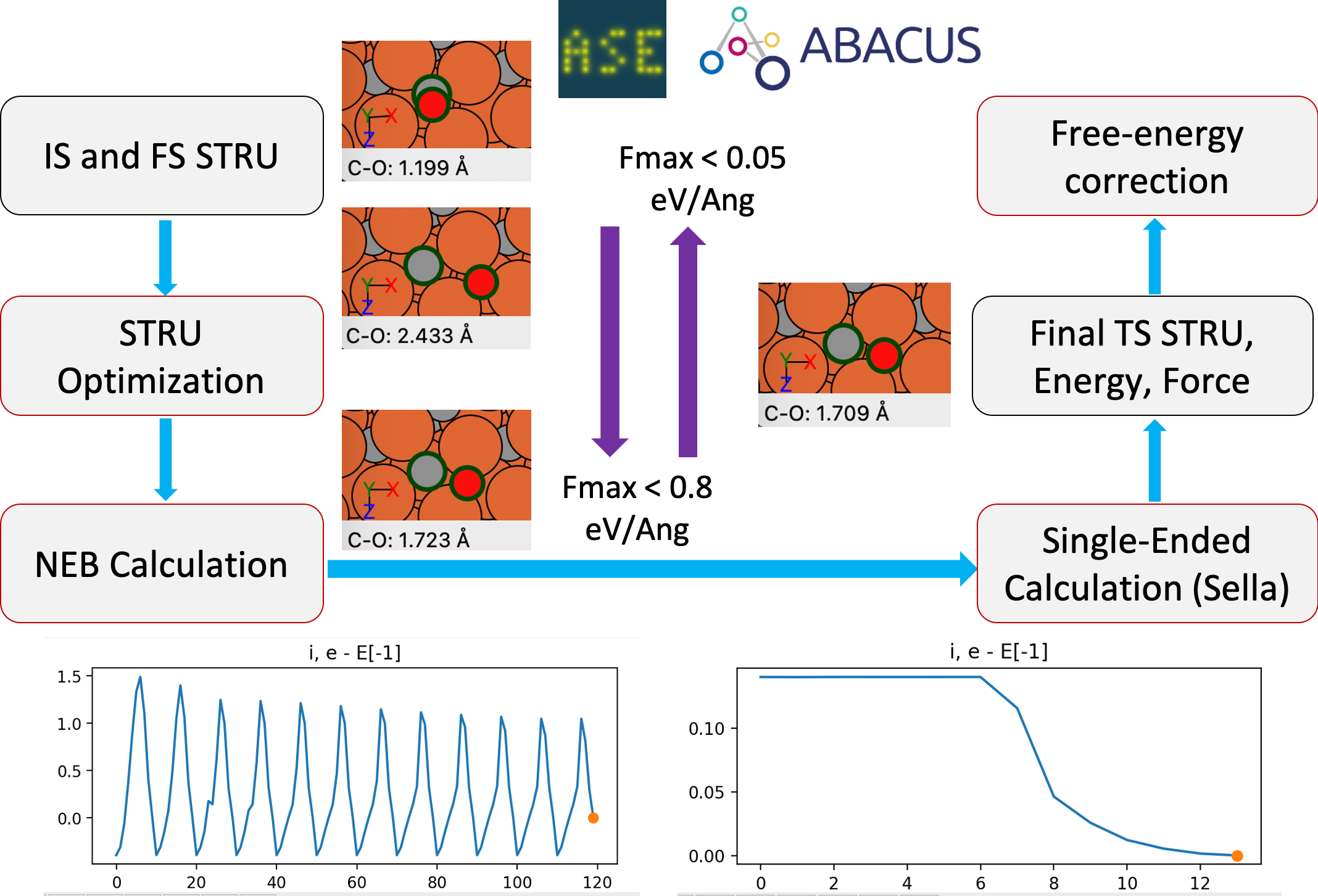}
    \caption{\MC{Workflow-chart of NEB + Sella method for locating TS based on ASE and ABACUS, red block denotes the usage of ABACUS as calculator}}
    \label{fig:ASE-ABACUS-TS}
\end{figure}

ATST-Tools scripts suite~\cite{ATST-Tools-link} is prepared for handy usage of ABACUS and ASE in TS locating jobs (see workflow in Fig.~\ref{fig:ASE-ABACUS-TS}). Apart from directly using TS search tools in ASE, the cooperation of different TS search methods can be done due to ASE's coding flexibility, which leads to better TS locating functionality. One way is to first generate a rough TS by NEB, then utilize a single-ended method like Dimer or Sella to optimize the TS in the target threshold, which incorporates the advantage of both methods.

%\subsubsection{Phonopy}
Phonopy~\cite{23JPCM-phonopy, 23JPSJ-phonopy} is a versatile open-source package designed for calculating phonon and related lattice dynamics properties from first-principles calculations. With the addition of the ABACUS interface in Phonopy version 2.19.1, users can now adopt ABACUS to compute the electronic structures and force constants that Phonopy requires for its phonon calculations.~\cite{20B-Wang, 2024-unisemi}

ShengBTE~\cite{ShengBTE_2014} is an open-source package aimed to solve the linearized Boltzmann Transport Equation (BTE) for phonons, enabling the {\it ab initio} prediction of lattice thermal conductivity in bulk crystalline solids and nanowires. It goes beyond the relaxation-time approximation by explicitly incorporating three-phonon scattering processes and isotopic scattering, yielding predictive results without empirical fitting. ShengBTE requires inputs of second- and third-order interatomic force constants (IFCs), traditionally computed using DFT packages, often aided by preprocessing tools. ABACUS can provide inputs for ShengBTE starting from v3.2.3. 

%\subsubsection{TB2J}
TB2J\cite{he2021tb2j} is an open-source Python package for automatic computation of magnetic interactions between atoms of magnetic crystals from density functional Hamiltonians based on Wannier functions or linear combinations of atomic orbitals. The program is based on Green’s function method with the local rigid spin rotation treated as a perturbation.~\cite{liechtenstein1987local} The ABACUS interface has been added since TB2J version 0.8.0. With the Hamiltonian matrix $\mathbf{H}(\mathbf{k})$ and the overlap matrix $\mathbf{S}(\mathbf{k})$ obtained by an LCAO calculation, TB2J is able to compute the isotropic exchange, symmetric anisotropic exchange, and the Dzyaloshinskii-Moriya interaction (DMI). In 2024, Zhang {\it et al.}~\cite{2024-zhang-abacus+tb2j} study GeFe$_3$N by using ABACUS-TB2J, which help them study the short-range order and strong interplay between local and itinerant magnetism in GeFe$_3$N.

%\subsubsection{Wannier90}
Wannier90~\cite{20JPCM-Wannier90} is a software designed for calculating maximally-localized Wannier functions (MLWFs),~\cite{97PRB-MLWF} which facilitates cost-effective band structure calculations. MLWFs are often used to construct model Hamiltonians. The post-processing capabilities of Wannier90 enable the calculation of various properties, such as Berry curvature, Berry curvature dipole, and shift current.~\cite{20JPCM-Wannier90} Additionally, MLWFs generated by Wannier90 can be analyzed for material surface states and topological properties using the Wannier Tools.~\cite{18CPC-WannierTools} According to the standard scheme for generating maximally localized Wannier functions in Wannier90, ABACUS has implemented an interface compatible with the Wannier90 software. This interface allows ABACUS to generate the necessary files for Wannier90, including overlap files between the periodic parts of Bloch functions at neighboring $\mathbf{k}$-points, projection files of Bloch functions onto trial localized orbitals, eigenvalue files of the energy bands, and real-space distribution files of the periodic parts of Bloch functions.

APEX (Alloy Property EXplorer)~\cite{apex} is versatile and extensible Python package for general alloy property calculations. This package enables users to conveniently establish a wide range of cloud-native property-test workflows by utilizing ABACUS. APEX currently supports the calculations including equation of state (EOS), elastic constants, surface energy, interstitial formation energy, vacancy formation energy, generalized stacking fault energy (Gamma line), and phonon spectra.

CALYPSO (Crystal structure AnaLYsis by Particle Swarm Optimization)~\cite{2012-wang-calypso} is a software designed for prediction of stable and metastable crystal structures. It employs advanced techniques like particle swarm optimization (PSO) to efficiently explore complex potential energy surfaces. CALYPSO has an interface with ABACUS to compute accurate energies and forces during the structure search process. The combined CALYPSO-ABACUS workflow can be utilized for high-throughput structure prediction and inverse design of functional materials under specified chemical compositions and external conditions like pressure.

USPEX (Universal Structure Predictor: Evolutionary Xtallography)~\cite{2006-oganov-crystal} represents a powerful methodology for crystal structure prediction. USPEX employs evolutionary algorithms to efficiently predict stable and metastable structures solely on chemical composition. USPEX can be utilized with ABACUS for accurate energy and force calculations during evolutionary structure searches.

\section{Summary} \label{sec:summary}

In this review, we have detailed the recent advances in the ABACUS package, which is an open-source platform for first-principles electronic structure calculations and molecular dynamics simulations. ABACUS supports both PW and NAO basis sets, and is compatible with a range of electronic structure methods.

On one hand, built on a plane-wave (PW) basis, ABACUS supports iterative methods with parallel algorithms for diagonalizing the Kohn-Sham equation. Additionally, the stochastic DFT method has been implemented, offering an alternative approach to address extremely high-temperature systems with substantially improved computational efficiency. Furthermore, OFDFT has been incorporated, featuring diverse kinetic energy density functionals, including machine learning-based ones, and optimization methods tailored for large-scale materials systems. On the other hand, by using the NAO basis set, ABACUS provides efficient solutions to the Kohn-Sham equation, featuring methods such as hybrid functional, DeePKS, and DFT+U, etc. In addition, real-time TDDFT has been implemented. These methods have been benchmarked and applied to various systems, demonstrating their accuracy and effectiveness.

ABACUS has been used to generate a large amount of first-principles data for training machine-learning-based potentials. The APNS project has been initiated to ensure the reliability and accuracy of ABACUS, where systematic tests for pseudopotentials and numerical atomic orbitals have been conducted, providing a solid foundation for the software's performance. ABACUS interacts with several packages, such as DeePKS-kit, DeePMD-kit, DP-GEN, DeepH, DeePTB, \MCC{HamGNN}, PyATB, Hefei-NAMD, PEXSI, etc., further expanding its functionality and applicability in multi-scale calculations and electronic structure analysis.

In conclusion, with the support of AI techniques, ABACUS has emerged as a robust and flexible platform for electronic structure calculations and analysis, well-positioned to address the challenges in materials science, chemistry, and physics within the AI era. Its ongoing development and integration with AI technologies are expected to further enhance its capabilities, strengthening its role as a valuable tool for researchers and scientists in the field.

\begin{acknowledgments}

We thank Han Wang and Weinan E for many helpful discussions. We gratefully acknowledges supports from AI for Science Institute, Beijing (AISI). M.C. thank funding support from the National Key R\&D Program of China under Grant No.2025YFB3003603, the National Natural Science Foundation of China (No. 12588201, 12588301, 12135002, 12122401,12074007). H.R. is supported by Beijing Natural Science Foundation (No. QY24014). S.X. gratefully acknowledges funding support from the National Natural Science Foundation of China (grant no. 52273223), DP Technology Corporation (grant no. 2021110016001141), School of Materials Science and Engineering at Peking University, and the AI for Science Institute, Beijing (AISI). S.L. acknowledges the supports from National Natural Science Foundation of China (92370104) and  Natural Science Foundation of Zhejiang Province (2022XHSJJ006).  The computational resource is provided by Westlake HPC Center. M.L. gratefully acknowledges the financial support of the National Natural Science Foundation of China (No. 52276212), National Key Research and Development Program of China (No. 2022YFB3803600). Q.O. acknowledges research grants from China Petroleum $\&$ Chemical Corp (funding number 124014). T.W. acknowledges support by The University of Hong Kong (HKU) via seed funds (2201100392, 2409100597). Y.X. acknowledges support by the National Key Basic Research and Development Program of China (grant nos. 2024YFA1409100 and 2023YFA1406400), the National Natural Science Foundation of China (grants nos. 12334003, 12421004, and 12361141826), and the National Science Fund for Distinguished Young Scholars (grant no. 12025405). W.Z. gratefully acknowledges AI for Science Institute, Beijing (AISI), as well as support from the Hongyi postdoctoral fellowship of Wuhan University.

\end{acknowledgments}

\section*{Data Availability Statement}
The data that support the findings of this study are available from the corresponding author upon reasonable request.

\appendix

% Ref
\section*{References}
\hypersetup{hidelinks}
\bibliography{ref}

\end{document}